%% file: 0_main.tex
\documentclass[twocolumn]{aastex63}

\graphicspath{
              {./figures/submit/}
             }

\received{February 3, 2020}
\revised{June 12, 2020}
\accepted{June 12, 2020}
\submitjournal{ApJ}

\shorttitle{MCR-TRGB: A Multiwavelength TRGB Measurement Method}
\shortauthors{Durbin et al.}

\newcommand{\HST}{\textit{HST}}
\newcommand{\Drizzlepac}{\texttt{Drizzlepac}}
\newcommand{\TweakReg}{\texttt{TweakReg}}
\newcommand{\Astrodrizzle}{\texttt{AstroDrizzle}}

\defcitealias{2009ApJS..183...67D}{D09}
\defcitealias{2012ApJS..198....6D}{D12}

\begin{document}

\title{MCR-TRGB: A Multiwavelength-Covariant, Robust Tip of the Red Giant Branch Measurement Method\footnote{Based on observations made with the NASA/ESA Hubble Space Telescope, obtained at the Space Telescope Science Institute, which is operated by the  Association of Universities for Research in Astronomy, Inc., under NASA contract NAS 5-26555.}}

\correspondingauthor{M. Durbin}
\email{mdurbin@uw.edu}

\author[0000-0002-0786-7307]{M.~J. Durbin}
\affiliation{Department of Astronomy, University of Washington, Box 351580, U.W., Seattle, WA 98195-1580, USA}

\author[0000-0002-1691-8217]{R.~L. Beaton}
\affiliation{Department of Astrophysical Sciences, Princeton University, 4 Ivy Lane, Princeton, NJ 08544, USA}

\author[0000-0002-1264-2006]{J.~J. Dalcanton}
\affiliation{Department of Astronomy, University of Washington, Box 351580, U.W., Seattle, WA 98195-1580, USA}

\author[0000-0002-7502-0597]{B.~F. Williams}
\affiliation{Department of Astronomy, University of Washington, Box 351580, U.W., Seattle, WA 98195-1580, USA}

\author[0000-0003-4850-9589]{M.~L. Boyer}
\affiliation{Space Telescope Science Institute, 3700 San Martin Drive, Baltimore, MD 21218, USA}

\begin{abstract}
We present a new method to measure colors and magnitudes of the tip of the red giant branch in multiple bandpasses simultaneously by fitting an $n$-dimensional Gaussian to photometry of candidate tip stars.
We demonstrate that this method has several advantages over traditional edge detection, particularly in regimes where the TRGB magnitude is strongly color-dependent, as is the case in the near-infrared.
We apply this method to a re-reduction of a set of optical and near-IR \HST{} data originally presented in \citet[][D12]{2012ApJS..198....6D}.
The re-reduction takes advantage of the increased depth and accuracy in the NIR photometry enabled by simultaneous reduction with higher resolution optical data in crowded fields \citep{2014ApJS..215....9W}.
We compare three possible absolute calibrations of the resulting apparent TRGB measurements, one adopting the same distance moduli as in \citetalias{2012ApJS..198....6D},
and two based on predicted TRGB absolute magnitudes from two widely-used, modern sets of model isochrones.
We find systematic offsets among the model absolute calibrations at the $\sim\!0.1$ mag level, in line with previous investigations.
The models also have difficulty reproducing the optical-NIR color-magnitude behavior of our measurements, making these observations a useful benchmark for future improvements.
\end{abstract}

\keywords{ distance scale; galaxies: distances and redshifts; galaxies: dwarf; galaxies: halos; galaxies: irregular; galaxies: stellar content; infrared: stars; stars: Population II }


\input{1_introduction.tex}

\input{2_data.tex}

\input{3_measurement.tex}

\input{4_results.tex}

\input{5_discussion.tex}

\input{6_conclusions.tex}

\acknowledgments

We gratefully acknowledge \added{the anonymous referee}, Emily Levesque, \v{Z}eljko Ivezi\'{c}, Anil Seth, and Evan Skillman for helpful feedback on this manuscript, as well as JJ Eldridge, Olivia Jones, and several members of the Carnegie-Chicago Hubble Program team for illuminating discussions. 
We also thank the GalRead group at Princeton University, in particular Andy Goulding, for directing us to Extreme Deconvolution.

\added{We acknowledge the people of the Dkhw'Duw'Absh, the Duwamish Tribe, the Muckleshoot Tribe, the Lenape, and other tribes on whose traditional lands we have performed this work.}

Support for this work was provided by NASA through grant \#AR-15016, and through Hubble Fellowship grant \#51386.01 awarded to R.L.B., from the Space Telescope Science Institute, which is operated by AURA, Inc., under NASA contract NAS 5-26555.

This research has made use of ``Aladin sky atlas" developed at CDS, Strasbourg Observatory, France, and of the NASA/IPAC Extragalactic Database (NED), which is operated by the Jet Propulsion Laboratory, California Institute of Technology, under contract with the National Aeronautics and Space Administration. 

\vspace{5mm}
\facilities{HST(ACS/WFC), HST(WFC3/IR)}

\software{AstroML \citep{astroMLpaper, 2014ascl.soft07018V},
          Astropy \citep{2013A&A...558A..33A, 2018AJ....156..123A},
          Astroquery \citep{2017ascl.soft08004G, 2019AJ....157...98G},
          Dask \citep{rocklin2015, dask},
          DOLPHOT \citep{2000PASP..112.1383D, 2016ascl.soft08013D},
          Drizzlepac \citep{2012ascl.soft12011S, 2013ASPC..475...49H, 2015ASPC..495..281A},
          KDEpy \citep{KDEpy},
          Matplotlib \citep{2007CSE.....9...90H},
          NumPy \citep{numpy},
          Pandas \citep{pandas, mckinney2011},
          Seaborn \citep{seabornv090},
          SciPy \citep{scipy},
          Scikit-learn \citep{sklearn},
          SEP \citep{2016JOSS....1...58B, 2018ascl.soft11004B},
          Vaex \citep{2018ascl.soft10004B, 2018A&A...618A..13B}
          }


\bibliographystyle{aasjournal}

\input{bibliography.tex}

\clearpage %

\appendix \label{app}

\input{app_tests.tex}
\onecolumngrid %

\end{document}

%% file: 1_introduction.tex
\section{Introduction} \label{sec:intro}

The tip of the red giant branch (TRGB) is defined as the truncation of the RGB phase of stellar evolution. The TRGB is reached when the \replaced{degeneracy in the core is lifted}{helium flash ignites, terminating the expansion and cooling of the outer layers} \citep{2006essp.book.....S}. 
\replaced{Breaking of degeneracy}{Helium ignition} occurs at a more or less fixed \deleted{core} temperature, and thus the maximum bolometric luminosity ($L_{\rm bol}$) produced by the core is \replaced{nearly constant}{well-constrained} \citep[see e.g., ][]{1978ApJS...36..405S, 2000ApJ...532..430V, 2006essp.book.....S, 2017A&A...606A..33S}.
However, \replaced{while the bolometric luminosity may be similar for all TRGB stars,}{both the bolometric luminosity and} the observed luminosity in a given bandpass will vary from star to star depending on the \added{effective temperature,} atmospheric chemistry, and on which elements and molecules selectively absorb and emit flux.
While the TRGB can be used as a ``standardizable candle", care must be taken to understand the wavelength dependence of the \added{observed} TRGB luminosity \citep[see][for \added{a discussion of} additional physical details]{2017A&A...606A..33S}. 

\citet{1944ApJ...100..137B}, when first resolving M\,31 into stars, noticed a field of red-stars of roughly equal brightness, which we now associate with the TRGB of ``old'' stellar populations. 
However, the optical TRGB (OPT-TRGB) was not used as a distance indicator until \citet{1993ApJ...417..553L}, \replaced{which}{who} leveraged precise color-magnitude diagrams (CMDs) of globular clusters from \citet{1990AJ....100..162D} to demonstrate an effective technique to ``detect'' the truncation of the RGB sequence observationally, and thereby determine a distance to the host system. 

The \citet{1993ApJ...417..553L} methods are conceptually simple; to detect the truncation of the RGB sequence, one \replaced{just needs to identify}{identifies} the magnitude at which there is a sharp jump in star counts, as expected for the edge of the RGB sequence. 
\citet{1993ApJ...417..553L} applied an edge-detection algorithm that approximates the first-derivative of a discrete function \citep[a Sobel filter;][]{Sobel1968} to measure the point of greatest change in the RGB luminosity function, which they identified as the apparent magnitude of the TRGB. 
Since \citet{1993ApJ...417..553L}, algorithms to detect the TRGB and calibrations of the absolute TRGB have evolved  \citep[a review and comparison is given in][]{2018SSRv..214..113B}, but the core of the technique has stayed the same. In general, the OPT-TRGB employed in the $I$ filter is thought to have a \added{near-}constant magnitude M$_{\rm I}\sim-4$ mag for most old ($t > 5$ Gyr) and metal-poor ([Fe/H$] < -0.5$ dex) stellar populations -- populations that are nearly ubiquitous in galaxies of all Hubble types and luminosity classes \citep{2018SSRv..214...90K}. These properties have made the detection of the OPT-TRGB an effective distance determination method out to $\sim$31 Mpc \citep{2017ApJ...836...74J}. 

While the OPT-TRGB is a powerful tool with several key science drivers \citep[for example,][among others]{2013AJ....146...86T, 2019MNRAS.486.1192T, 2019ApJ...880...52A, 2019ApJ...882...34F}, extending this method to the near-infrared (IR-TRGB, hereafter) has several advantages: 
(i) the stars themselves are $\sim 1-1.5$ mag brighter and comparable in luminosity to $P\sim$ 10 day Cepheids \citep[see Fig.~30 in][]{2018SSRv..214..113B};
(ii) the impact of extinction is reduced by \added{up to} a factor of 6 \citep{2005ApJ...619..931I} \added{\citep{2014MNRAS.444..392C}}, permitting exploration of galaxies behind high extinction \citep[see e.g.,][]{2019ApJ...872L...4A} and reducing any dust-based systematics; and 
(iii) the next generation of astronomical facilities, whether 30-m class telescopes on the ground, wide-field telescopes in space, or large-aperture telescopes in space, are likely to realize their highest efficiency in the near- to mid-infrared. 
Thus, there is enormous potential for the IR-TRGB, although there remain challenges to its implementation at high precision. 

The first detailed characterization of the IR-TRGB in \added{the Hubble Space Telescope (HST)'s} WFC3/IR filters was presented in \citet[][D12 hereafter]{2012ApJS..198....6D} in which 23 galaxies with optical imaging from the ACS Nearby Galaxy Survey Treasury \citep[ANGST,][hereafter D09]{2009ApJS..183...67D} were supplemented with WFC3/IR imaging in the F110W and F160W filters \citep[][GO-11719]{2009hst..prop11719D}.
\citetalias{2012ApJS..198....6D} detected the IR-TRGB applying a Sobel filter to F110W--F160W color-magnitude diagrams, and then converted the dust-corrected apparent magnitudes to an absolute scale via distances derived in \citetalias{2009ApJS..183...67D}, using the OPT-TRGB calibrated to models described in \citet{2008PASP..120..583G}. 
\citetalias{2012ApJS..198....6D} found a strong correlation between the absolute F160W magnitude of the IR-TRGB and the F110W-F160W color, such that redder TRGB stars had a brighter absolute magnitude.
The correlation was expected due to metallicity variations among the sample, such that more metal rich stars had redder colors, pushing a larger fraction of their bolometric flux into the NIR. However, the \citetalias{2012ApJS..198....6D} IR-TRGB was brighter than contemporaneous theoretical models by 0.05 to 0.10 mag and, generally, was notably different from globular cluster observations that had been converted from 2MASS into the WFC3/IR system. 
The general conclusion from this paper was that while the IR-TRGB was promising, there were significant unresolved issues. 
A subsequent and similar analysis by \citet{2014AJ....148....7W}, however, essentially found the same underlying mag-color relationship for the IR-TRGB, albeit these authors argued for a break in the slope at F110W--F160W = 0.95 mag.

\begin{figure*}[ht]
\epsscale{1.15}
\plottwo{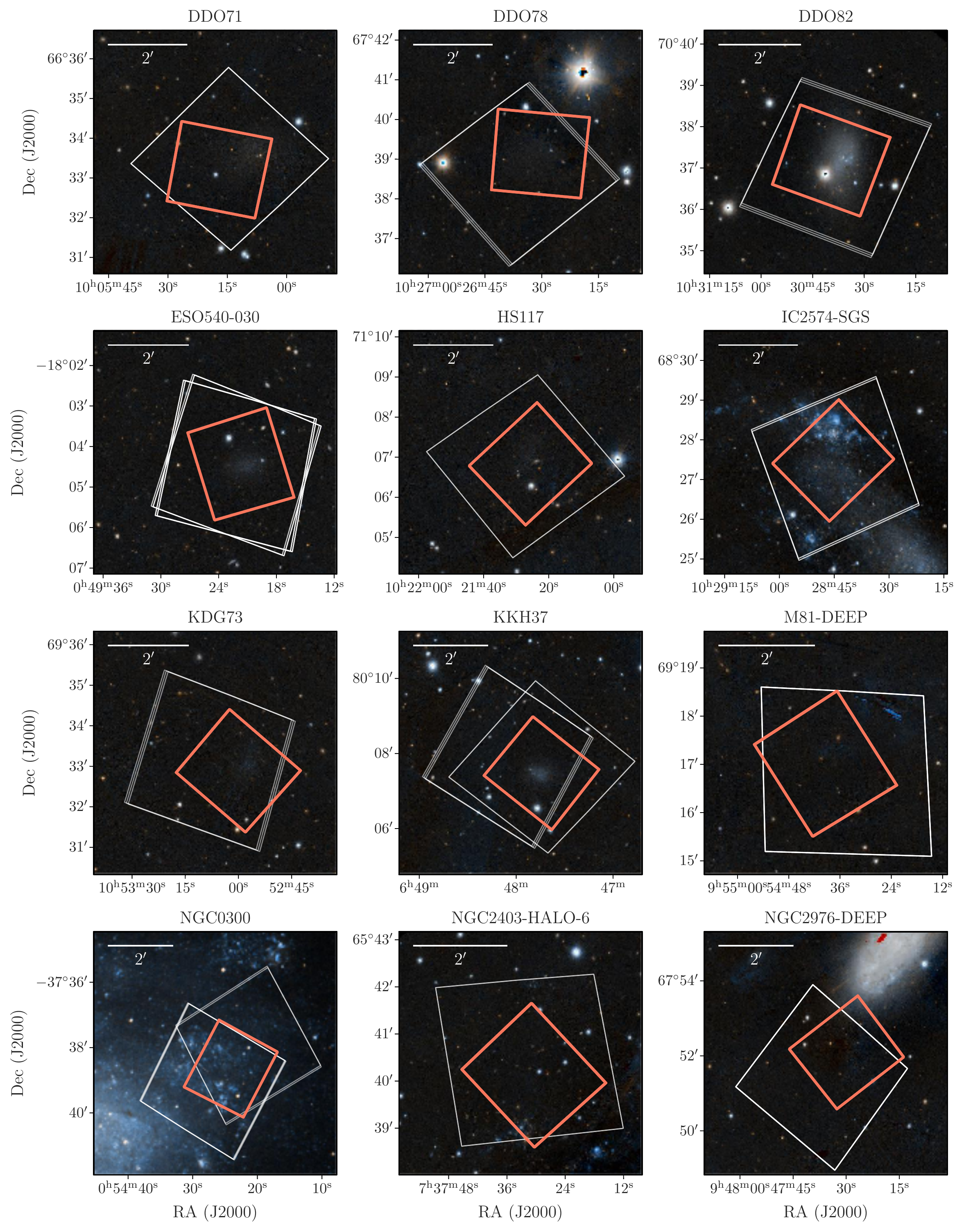}{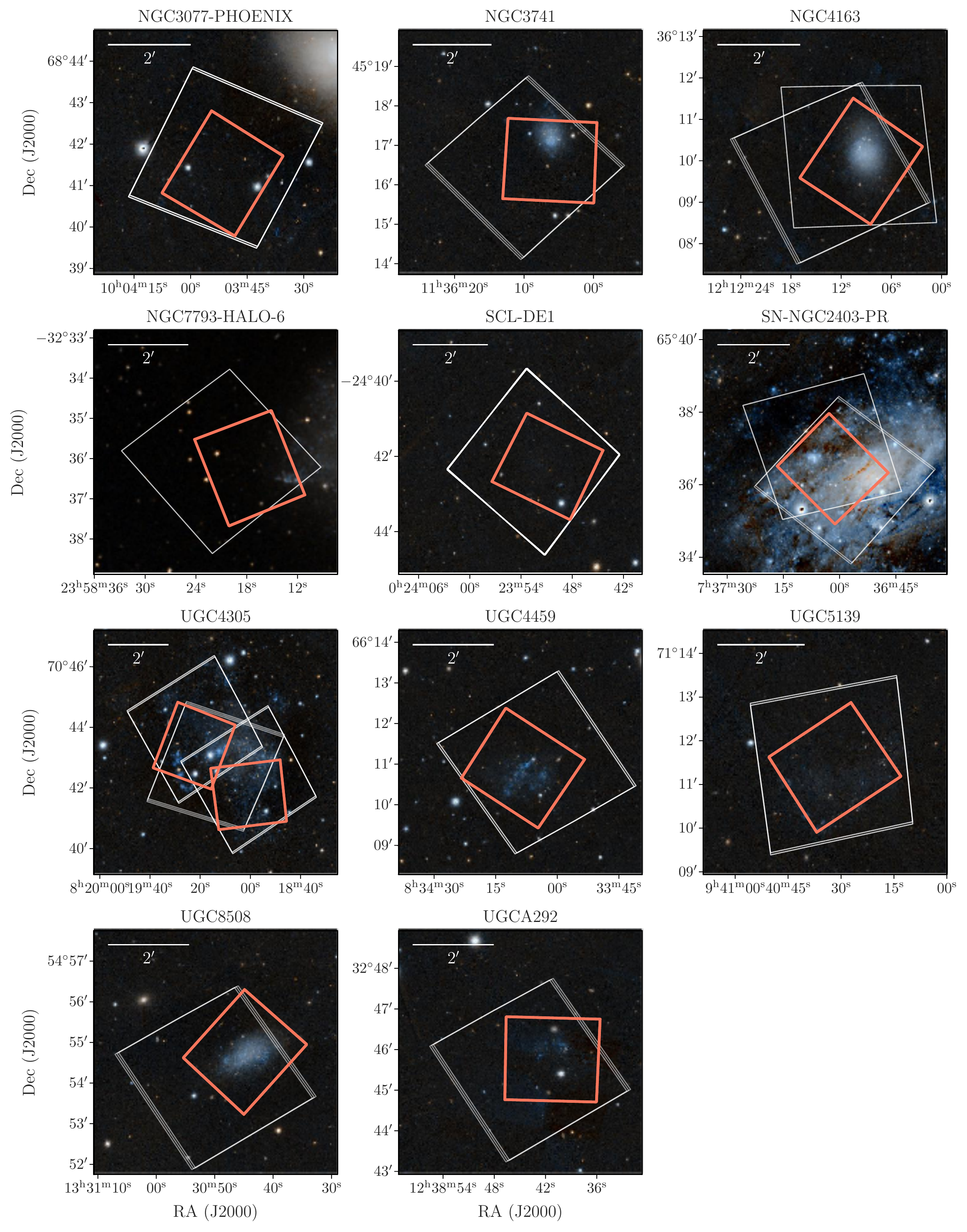}
\caption{Footprints of the HST observations originally presented in \citetalias{2009ApJS..183...67D} and \citetalias{2012ApJS..198....6D}, which we reanalyze in this work. ACS/WFC footprints are shown by thin white lines, and WFC3/IR footprints are in thick orange. Background images are PanSTARRs $z+g$ for all targets except NGC\,300 \& NGC\,7793-HALO-6, which use DSS2. All background images were retrieved through the HiPS thumbnail service provided by the Universit{\'e} de Strasbourg.} 
\label{fig:footprints}
\end{figure*}

More recent, ground-based work in the 2MASS filter system by \citet{2018ApJ...858...12H}, \citet{2018ApJ...858...11M}, and \cite{2018AJ....156..278G} produced empirical color-magnitude relations for the IR-TRGB.
These, however, are significantly different from those determined for WFC3/IR \added{on HST}. 
In their review, \citet{2018SSRv..214..113B} compared the WFC3/IR and 2MASS IR-TRGB slopes to demonstrate that these independent WFC3/IR and 2MASS calibrations largely agree when considered within a given filter system and that the apparent differences are morely likely due to inherent differences between the filter systems. 
As a result, calibrations from the ground-based 2MASS systems are likely inapplicable to the space-based WFC3/IR system. 

In addition to advancing empirical measurements of the IR-TRGB, recent papers have also updated theoretical relationships derived from stellar models. 
A key work is that of \citet{2017A&A...606A..33S}, which directly compared the theoretical IR-TRGB for a range of metallicities and ages in the BaSTI \citep{2013A&A...558A..46P} model suite.
\cite{2017A&A...606A..33S} report both physical and color-magnitude relationships for the IR-TRGB, \replaced{though the authors}{but} note that uncertainties in the bolometric corrections \added{and stellar $T_{\rm{eff}}$ scale} make direct use of these relationships challenging \citep[as discussed further in][]{2018SSRv..214..113B}. 
\citet{2019ApJ...880...63M} studied the variation in the TRGB with age and metallicity from the optical to the mid-IR using simulated photometry based on the PARSEC \citep{2012MNRAS.427..127B, 2017ApJ...835...77M} model suite, and found that rectifying the photometry to a fiducial tip reduced the range of variations in the measured F160W TRGB to 0.04 mag. 
Thus, while the potential for the IR-TRGB is well-recognized \citep[see e.g.,][among others]{2018SSRv..214..113B}, the empirical evidence for its reliability \replaced{is less clear}{remains unclear}.

\citetalias{2012ApJS..198....6D} presented a number of concerns regarding their analysis that ranged from the relatively new data processing and calibration of WFC3/IR data, to crowding in the images (for which the higher-resolution optical images are clearly deeper and more complete). 
However, since \citetalias{2012ApJS..198....6D}, major large-scale projects like the Panchromatic Hubble Andromeda Treasury \citep[PHAT;][]{2012ApJS..200...18D, 2014ApJS..215....9W}, the Cosmic Assembly Near Infrared Deep Extragalactic Legacy Survey \citep[CANDELS;][]{2011ApJS..197...35G, 2011ApJS..197...36K}, and the Ultra Deep Field \citep{2013ApJS..209....3K, 2019A&A...621A.133B}, have led to substantial improvement both in our technical knowledge of \added{the} WFC3/IR \added{camera} and in the development of multiwavelength data-processing techniques that significantly improve the WFC3/IR photometric quality. Additionally, there have been multiple internal efforts to improve WFC3/IR calibration and data products \citep[for a comprehensive overview see][]{2018wfc..rept...12M}.
It is the purpose of this work to apply these techniques \deleted{along with a new method to self-consistently measure the TRGB across multiple filters} to the \citetalias{2012ApJS..198....6D} dataset and revisit the discrepancies identified in \citetalias{2012ApJS..198....6D} regarding the IR-TRGB \citep{2017hst..prop15016D}. 
\added{We also take advantage of and expand upon recent works \citep[e.g.][]{2018ApJ...858...12H, 2018ApJ...858...11M, 2020ApJ...891...57F} that have demonstrated the effectiveness of calibrating the TRGB in multiple bandpasses by selecting a set of fiducial ``tip stars" and fitting their multiwavelength behavior; we present a generalized version of this method here.}

The outline of the paper is as follows.
\autoref{sec:data} describes the observations, image processing, photometry, and artificial star tests. 
\autoref{sec:measurement} presents techniques to isolate the RGB, identify candidate TRGB stars, and trace their multiwavelength behavior.
\autoref{sec:results} presents the final measured TRGB apparent magnitudes and colors, and compares the absolute magnitudes and distance moduli obtained from previously published distances and then from calibration to two sets of theoretical isochrones.
\autoref{sec:discussion} presents a discussion of our results, attempts to resolve concerns from \citetalias{2012ApJS..198....6D}, and discusses lingering concerns regarding the full realization of the IR-TRGB.
\autoref{sec:conclusions} presents a summary of our work and discusses directions of future research.
Throughout the main text, we limit visualizations to a representative set of galaxies; figures for the full sample are given as figure sets.

%% file: 2_data.tex
\section{Data} \label{sec:data}

\subsection{Observations}

\input{table_sample.tex}

We re-reduced the optical and near-infrared \HST{} imaging data described in \citetalias{2009ApJS..183...67D} and \citetalias{2012ApJS..198....6D}.
The \citetalias{2012ApJS..198....6D} observations were a WFC3/IR imaging follow-up (SNAP-11719) to the optical ACS/WFC data presented in \citetalias{2009ApJS..183...67D}. The F110W+F160W observations cover 26 pointings in 22 Local Volume galaxies with a range of star-formation histories. The majority of the galaxies are low-metallicity dwarfs, with the exception of M81. \autoref{tab:sample}, reproduced from \citetalias{2012ApJS..198....6D}, presents summary information about the galaxies in our sample, including coordinates, angular diameter, apparent $B$ magnitude, foreground reddening, T-type, H\textsc{I} line widths, and group membership. We note that not all of these galaxies have the purely old stellar populations that are considered optimal for measuring the TRGB.

We analyzed 24 of the 26 datasets that were included in \citetalias{2012ApJS..198....6D}. 
To maintain uniformity in the final dataset and analyses, we excluded two targets (NGC404 and NGC2403-DEEP) because their optical data were taken by WFPC2 rather than ACS. Additionally, we combined the two pointings of Holmberg II (UGC4305-1 and UGC4305-2 in \citetalias{2012ApJS..198....6D}) into a single target UGC4305 here, as they have slight overlap in the NIR and substantial overlap in the optical. All targets have ACS imaging in F814W (comparable to Johnson-Cousins $I$) and at least one of the F475W, F555W, and F606W filters (comparable to SDSS $g$, Johnson-Cousins $V$, and broad Johnson-Cousins $V$ respectively).
\autoref{fig:footprints} shows the footprints of the ACS/WFC (white) and WFC3/IR (orange) on either PanSTARRS or DSS2 imaging for each of the 23 distinct targets used here. 

\input{table_obs.tex}

\autoref{tab:obs} describes the ACS/WFC and WFC3/IR observations used for this work including references to the original proposals, total F814W exposure time, and offsets of the observation from the galaxy center.

We retrieved all data in the form of calibrated individual exposures (\texttt{*flt} files for WFC3/IR and CTE-corrected \texttt{*flc} files for ACS/WFC) from the Mikulski Archive for Space Telescopes (MAST) with Astroquery \citep{2017ascl.soft08004G, 2019AJ....157...98G} on January 28, 2019, and obtained up-to-date reference files with the HST CRDS {\tt bestref} tool \citep{2004ASPC..314..824S}.

\subsection{Alignment \& Photometry}

\begin{figure}[ht]
    \plotone{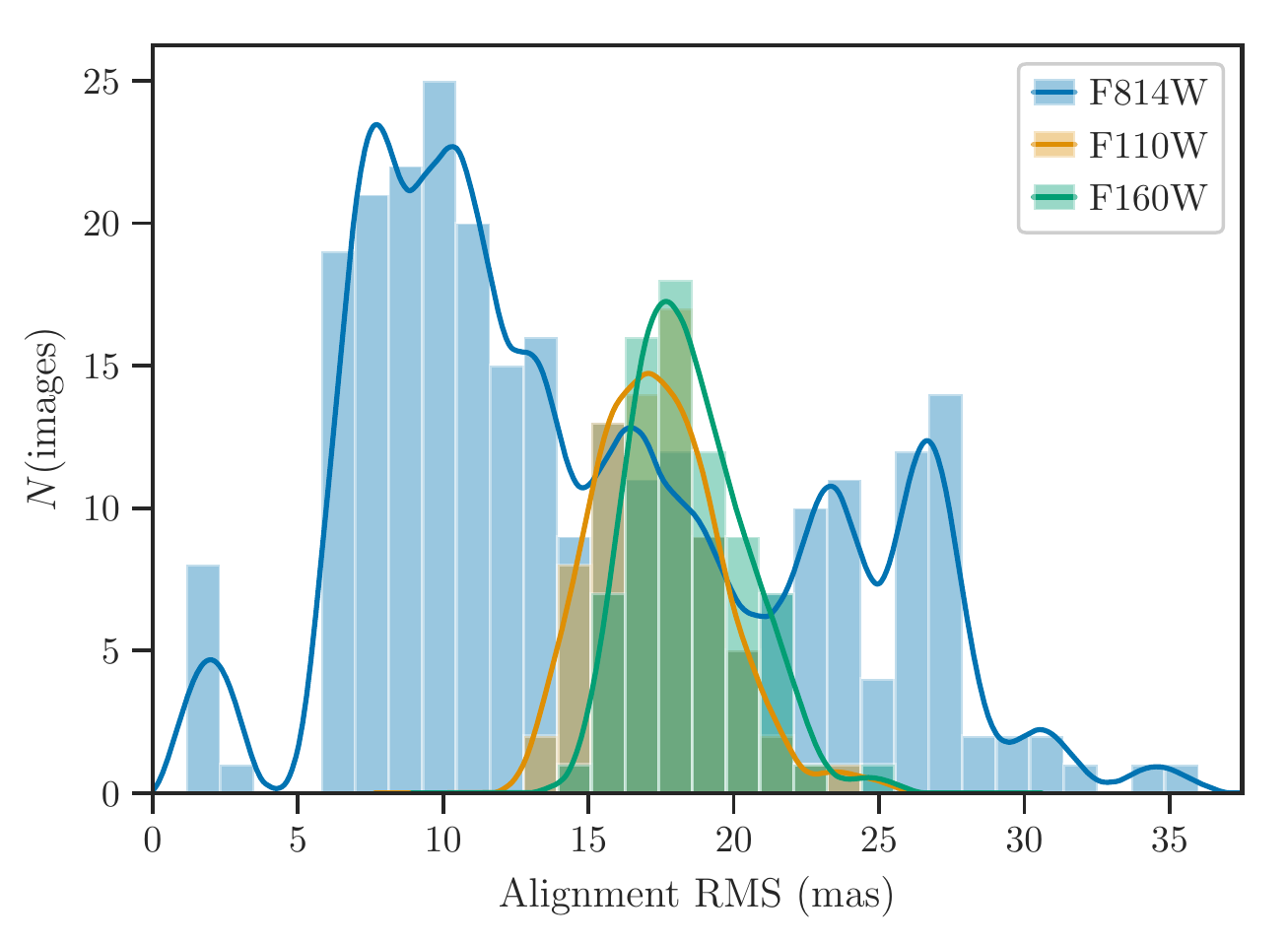}
    \caption{Distributions of the RMS scatter of alignment residuals for F814W, F110W, and F160W. Both near-IR filters have a residual scatter on the order of 0.025\farcs, or $\sim$0.2 WFC3/IR pixels. In F814W the alignment RMS has a peak closer to 0.01\farcs, but there is a long tail of images with higher scatter, likely due to variations in exposure depth and source densities.}
    \label{fig:alignment_rms}
\end{figure}

We aligned all exposures using \TweakReg{} and \Drizzlepac{} 2.0 \citep{2013ASPC..475...49H, 2015ASPC..495..281A}. \TweakReg{} aligns images by calculating an affine transform (shifts, rotation, and scale) that best describes the transformation between astrometric catalogs from two images, one of which is treated as the fiducial ``reference" image. It then calculates an updated WCS solution for the non-reference image using the affine transform.

By default, \TweakReg{} extracts astrometric source catalogs from input images with a point source extraction routine based on DAOFIND \citep{1987PASP...99..191S}, which is optimized for point source detection. 
However, many of our exposures are too sparsely populated with bright stars to produce a reliable cross-filter alignment solution from point sources alone, requiring the addition of background galaxies to the astrometric source catalogs.\footnote{Although it is true that extended sources are less optimal for alignment, as their morphologies may vary across filters affecting their calculated centroids, they are nonetheless useful in the absence of sufficient point sources.} 
We therefore followed the procedure described in \citet{2015acs..rept....5L} to align images on Source Extractor \citep{1996A&AS..117..393B} catalogs rather than \TweakReg{}-produced catalogs. 
Source Extractor's detection algorithm is largely morphology-agnostic, which enables the robust detection of both point and extended sources.
We used SEP \citep{2016JOSS....1...58B}, a Python and C reimplementation of core Source Extractor algorithms, to derive all catalogs used in alignment.

We chose ACS/WFC F814W as our ``reference" filter for all targets, as it is the only optical filter common to all targets, and in most cases it is the deepest and most likely to contain sources that are detected across multiple filters.
We aligned all frames for each target with the following steps:

\begin{enumerate}
    \item Extract initial source catalogs from all F814W exposures with SEP and align these with \TweakReg{};
    \item Combine all aligned F814W exposures into a single distortion-corrected reference image with \Astrodrizzle{}, and extract a deep reference catalog from the drizzled image;
    \item Realign all F814W exposures to the reference image using catalogs from the cosmic ray cleaned (\texttt{*crclean}) images produced by \Astrodrizzle{};
    \item Align all other exposures to the reference image with \TweakReg{}.
\end{enumerate}

We did not attempt to derive an absolute astrometric solution for any of our targets, as the majority are severely limited by the $\sim\!2\arcmin\times2\arcmin$ WFC3/IR field of view and do not have enough bright sources to reliably match against external astrometric catalogs such as \emph{Gaia}. For the purposes of this work, internally consistent alignment on a per-target basis is sufficient.

\autoref{fig:alignment_rms} compares the RMS scatter of the alignment residuals for the common filters of F814W, F110W, and F160W.
The residuals for the two WFC3/IR filters are very similar, with a residual scatter of $\sim$0\farcs025 or 0.2 WFC3/IR pixels. 
The residuals for F814W are more scattered, with a peak at 0\farcs01 (0.2 ACS/WFC pixels) and a long tail, likely due to differences in the underlying image datasets themselves (e.g., different exposure depths and source densities).

We carried out photometry on the aligned images with the pipeline described in \citet{2014ApJS..215....9W}, which wraps the \HST{} photometry package DOLPHOT \citep{2000PASP..112.1383D}.
Briefly, DOLPHOT uses a set of fiducial PSF models that are empirically scaled for each frame to account for frame-to-frame differences, such as those induced by ``breathing" \citep{1994rhis.conf..157H}.
The cross-camera wrapper utilizes a single underlying source list such that DOLPHOT can iteratively measure each individual source simultaneously across frames employing techniques optimized for crowded fields. 
As described in \citet{2014ApJS..215....9W}, the output photometry for each source requires additional characterization to have realistic uncertainties incorporating all concerns; these are derived via artificial star tests that are described in the following subsection. 

The key difference in the procedure adopted here compared to that of \citetalias{2012ApJS..198....6D} is that we perform {\it simultaneous cross-camera photometry} rather than reducing the datasets independently and then matching catalogued sources. Due to the differences in the native angular resolution between ACS/WFC and WFC3/IR (0\farcs05/pixel vs.\ 0\farcs13/pixel respectively), the simultanous procedure should produce a more complete and robust WFC3/IR dataset due to improved deblending and more complete source lists.

We rejected large contaminating sources, such as bright foreground stars and background galaxies, by convolving the images with a 2D Gaussian kernel with width 0\farcs75 (15 WFC3/IR pixels) and extracting sources from the convolved images with SEP. We used the ellipse parameters $a$, $b$, and $\theta$ of the sources to mask potentially contaminated pixels, with $a$ and $b$ multiplied by 5 to ensure that a sufficient fraction of the contaminating flux was masked.

\subsection{Artificial Star Tests} \label{ssec:ASTs}

The primary sources of photometric uncertainty in these data are total exposure depth, which determines the Poisson noise of photon counts, and stellar surface density, which affects the likelihood of a star being blended with surrounding sources. The former are well-captured by DOLPHOT's accounting of photon-counting uncertainties. The latter, however, require additional tests to fully characterize, especially given that blending is typically the dominant source of bias and uncertainty in crowding-limited data.

\begin{figure}[ht]
    \plotone{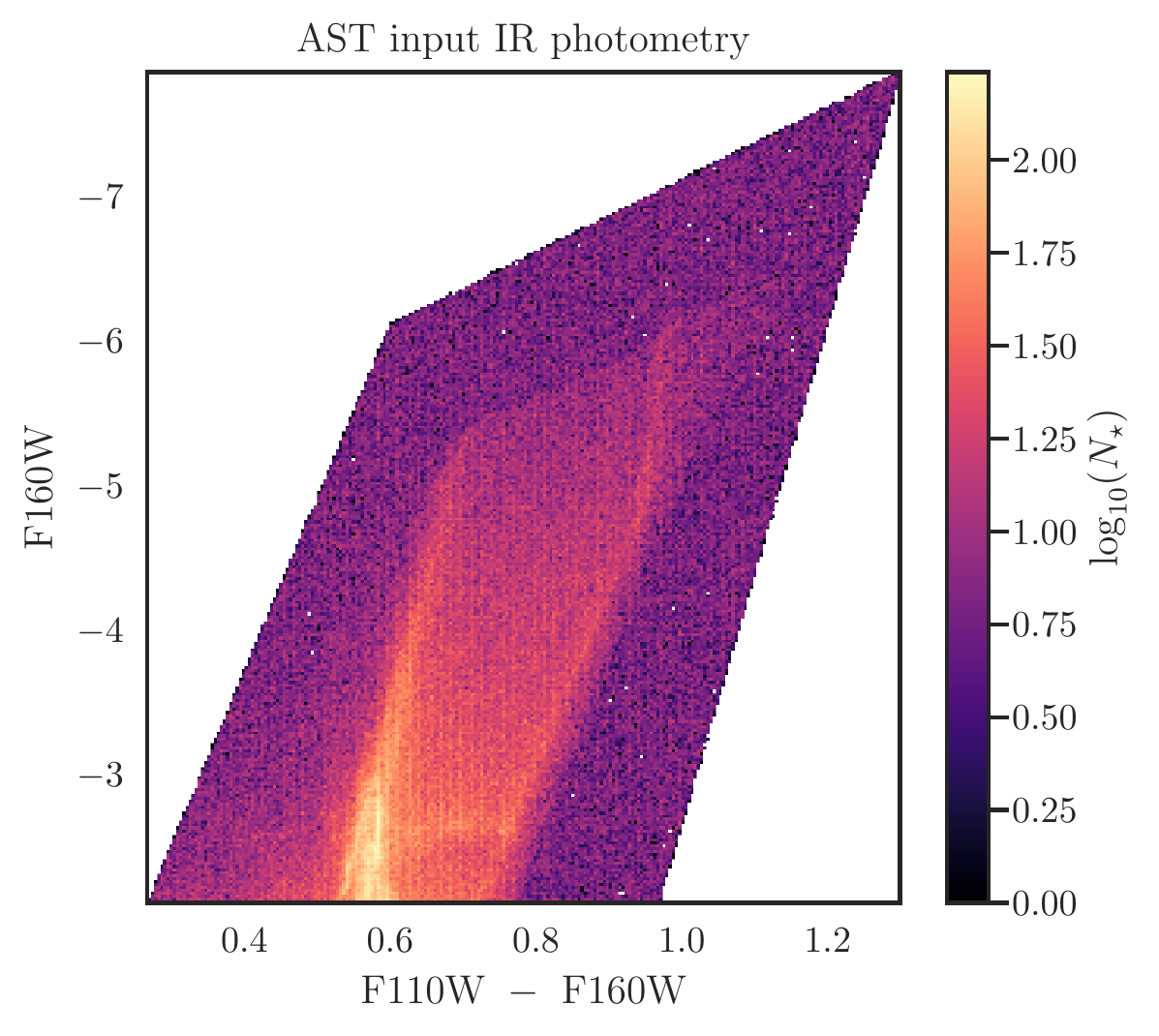}
    \caption{Hess diagram of input AST photometry in the near-IR. The densest portions (orange to yellow) are from the CMDs, whereas the uniform sampling is purple. }
    \label{fig:ast_inputs}
\end{figure}

We evaluated the photometric biases, scatter, and completeness of our data with a series of artificial star tests (ASTs). We generated 20,000 artificial stars to be injected into the image stack for each target, for a total of 460,000 ASTs. We prioritized the near-IR RGB when selecting artificial star magnitudes. Half were drawn directly from simulated absolute photometry generated with MATCH \citep{2002MNRAS.332...91D} from PARSEC models. The other half were assigned random magnitudes within our F110W-F160W selection box, with optical magnitudes taken from simulated stars with comparable near-IR photometry. \autoref{fig:ast_inputs} shows a CMD of the full set of input NIR photometry. All absolute input magnitudes were then adjusted by the per-filter foreground reddening and \citetalias{2012ApJS..198....6D} distance modulus for each target and assigned random pixel coordinates within the NIR image footprints, excluding the locations of masked contaminating sources such as extended background galaxies and bright foreground stars. These input stars were then inserted into the image stack in batches of 1000 at a time and processed identically to the original photometry.

As the AST input locations were assigned at random, they do not necessarily reflect the true distributions of density and depth for any single target. We therefore resampled the full set of AST results to match the distribution of these quantities for each target as closely as possible, as follows.

\begin{figure}[ht]
    \epsscale{1.1}
    \plotone{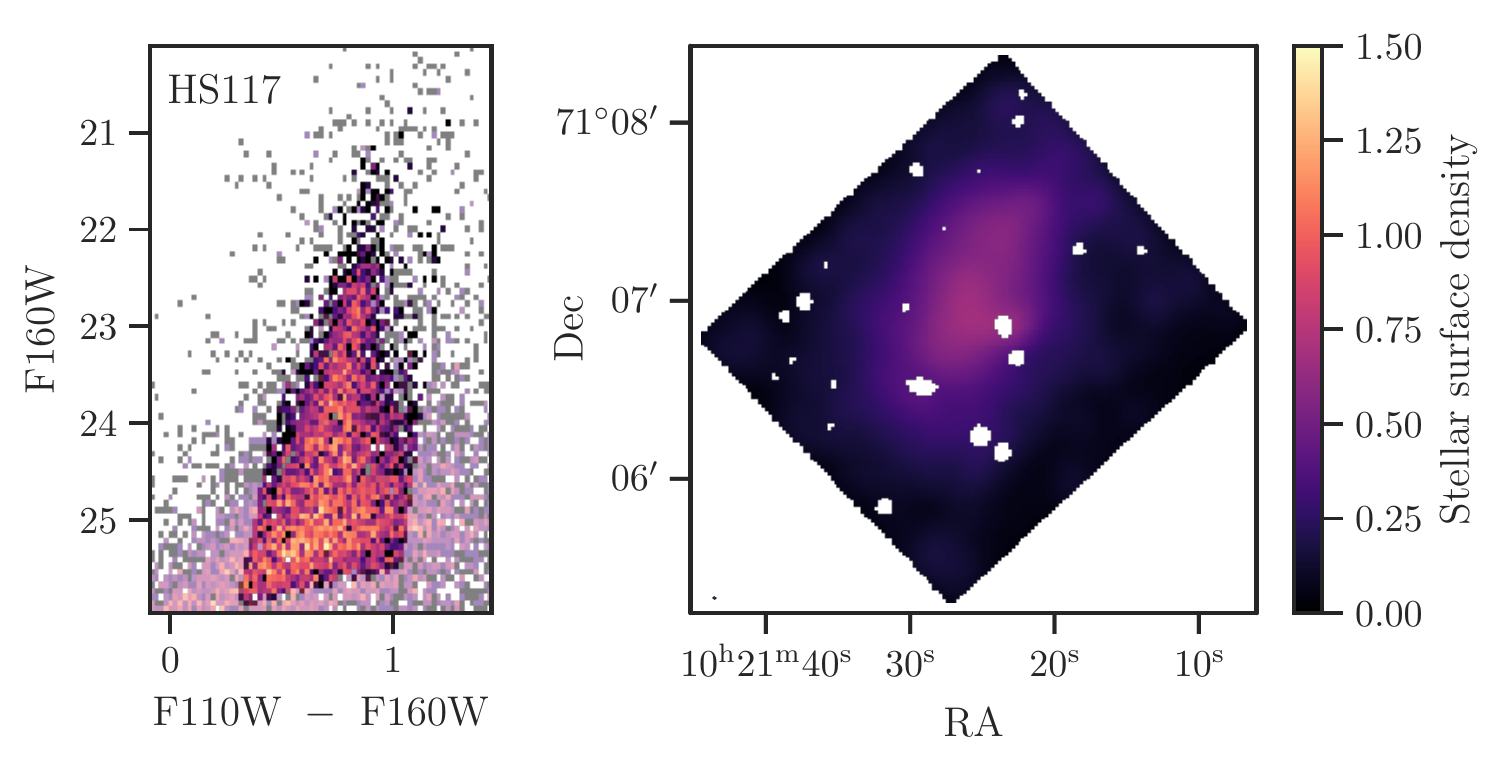}
    \plotone{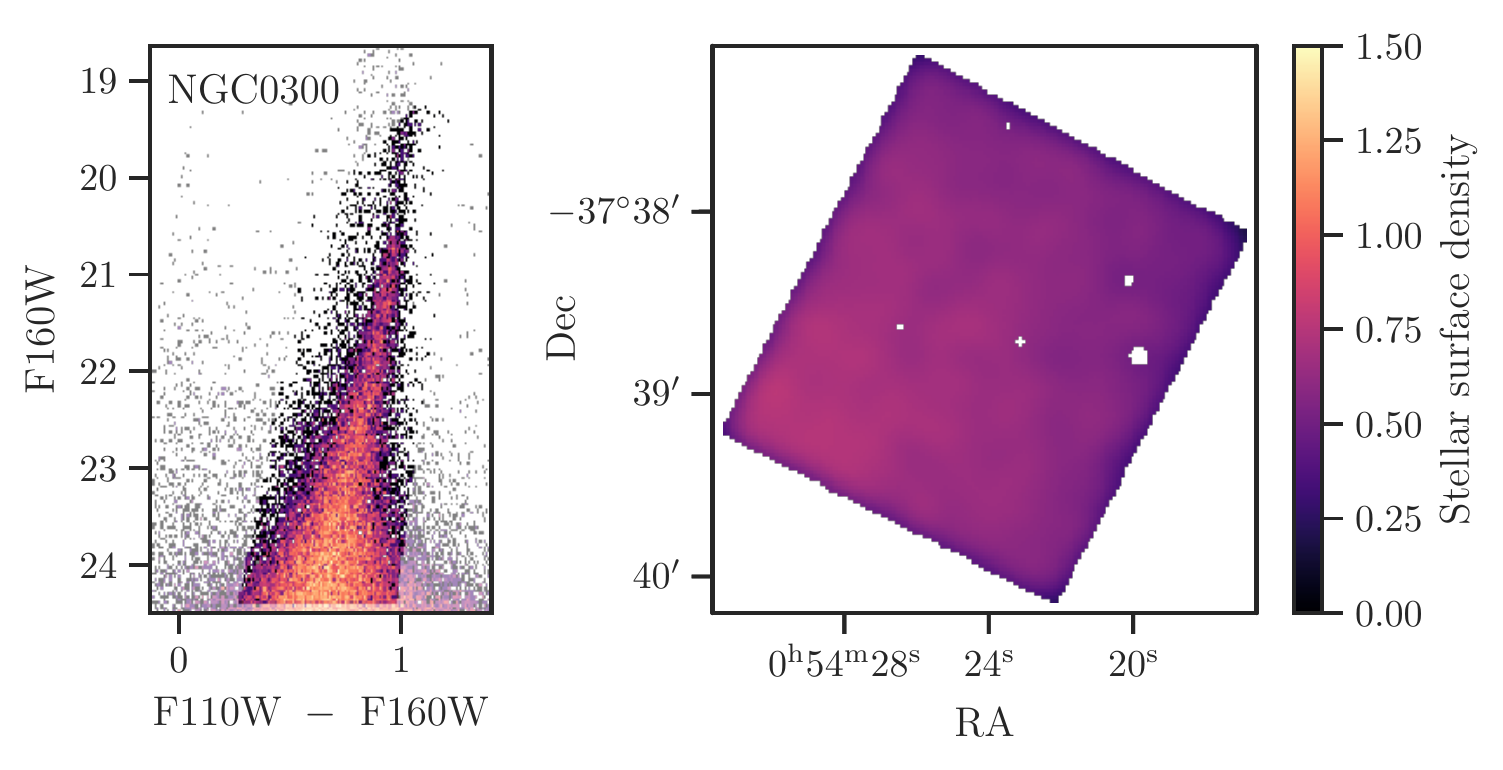}
    \plotone{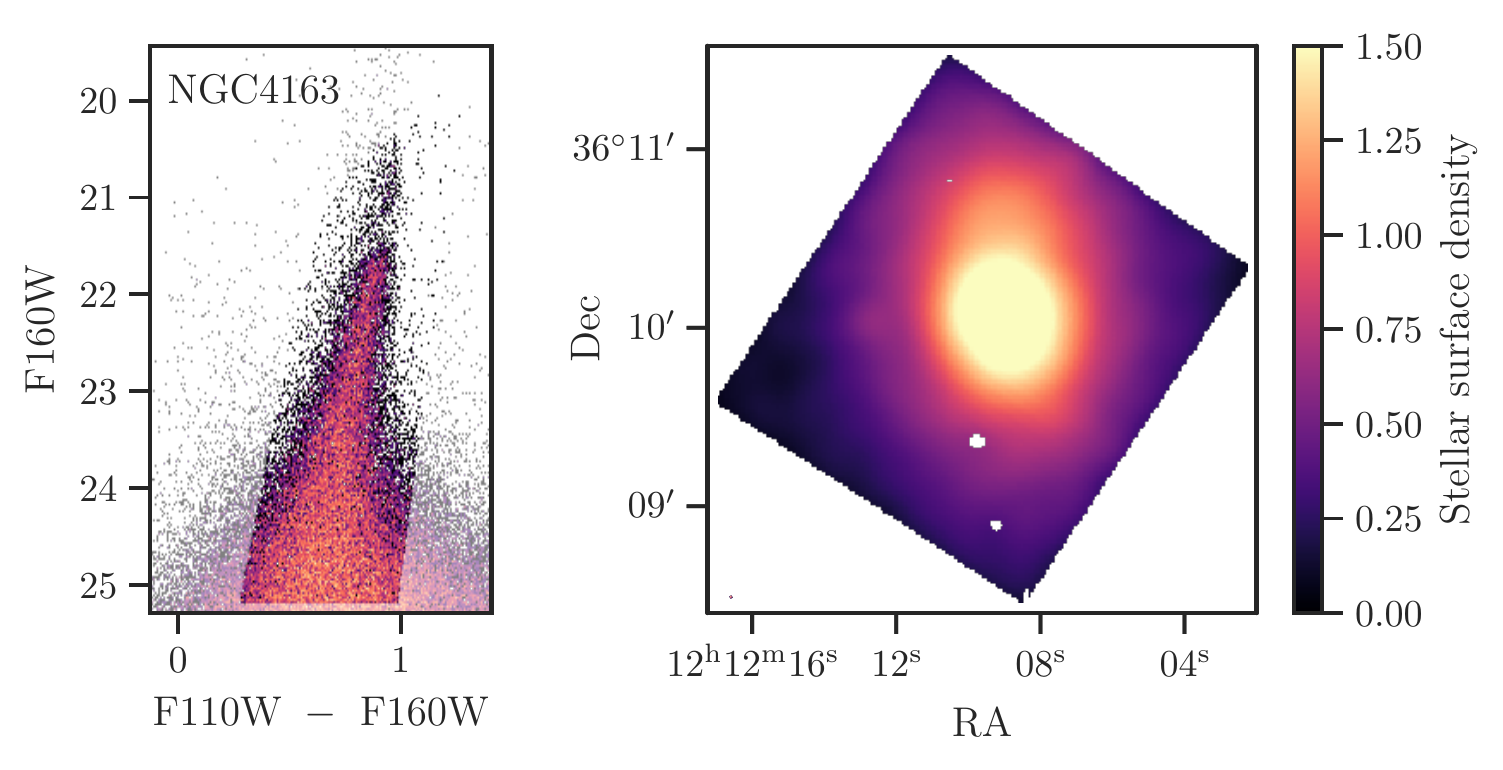}
    \caption{Left: NIR Hess diagrams of three galaxies in our sample, with the selections of stars included in our surface density calculations highlighted. Right: corresponding stellar surface density maps for each target. All density maps are scaled to the same limits (0 to 1.5 RGB stars per square arcsecond) to highlight the range of stellar densities in our sample. Gaps in the density images show where contaminating sources such as foreground stars and background galaxies were rejected.
    The complete figure set (23 images) is available in the online journal.
    }
    \label{fig:density_maps}
\end{figure}

We evaluated stellar surface densities using kernel density estimation \citep{rosenblatt1956, parzen1962} with the Python package \texttt{KDEpy} \citep{KDEpy}. We selected the photometry to be used for density estimation using the same near-IR selection box as in the ASTs, with the additional criteria of having a mean near-IR signal-to-noise greater than 3. We then constructed stellar surface density maps by convolving source coordinates with a Gaussian kernel with a width of 5\arcsec, and tagged all photometry with their local densities. Density maps for three example targets are shown in \autoref{fig:density_maps}. In the analysis presented in \autoref{sec:measurement} we used only photometry with local densities less than 1.5 stars/$\square \arcsec$, except for the high-density target SN-NGC2403-PR, where we used a maximum local density of 3 stars/$\square \arcsec$.

\input{figset_density.tex}

While all near-IR exposures were taken with identical exposure times and are therefore of comparable depth, there is considerable variation in the optical exposure depths, which in turn may affect DOLPHOT's source detection and subsequent deblending of near-IR sources. To characterize exposure depth, we use the weight maps generated by \texttt{Astrodrizzle} for the combined F814W reference images to assign fiducial total exposure times to the locations of each source.

For each target we separated to the photometry into 10 bins according to density vs.\ depth using K-means clustering \citep{kmeans, minibatchkmeans}, and resampled the full set of ASTs to match the observed distributions of densities and depths.

We then use the resampled ASTs to assign fiducial photometric uncertainties, biases, and completenesses to all of our photometry. We define the photometric bias to be the median of the differences between observed and input AST magnitudes, the photometric error to be the interquartile range of the same, and the completeness to be the fraction of stars with non-null observed magnitudes. We calculate these quantities as a function of AST input magnitudes in each filter.

We subtract filter-appropriate foreground extinctions from all photometry, with values obtained from \citet{2011ApJ...737..103S}; the corresponding $V$-band extinctions are listed in \autoref{tab:sample}. We assume negligible internal extinction for all targets, as the majority of our targets are either low-metallicity dwarfs or halos of spiral galaxies. Target SN-NGC2403-PR is an exception, but in that case we find that the photometric uncertainties due to crowding are large enough that an attempt to analyze or correct for internal extinction would likely not be productive.

\subsection{Comparison to D12 photometry}

\begin{figure}[ht]
    \plotone{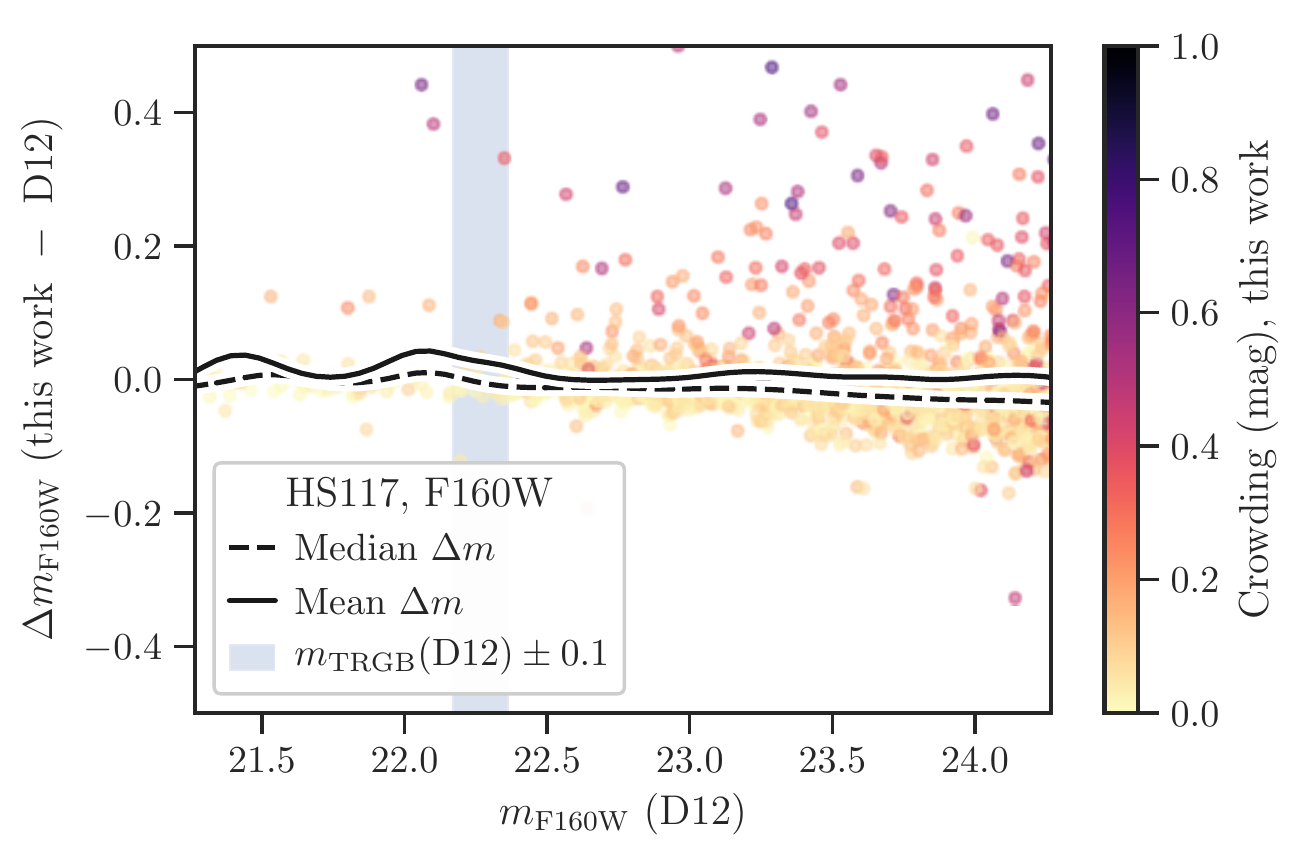}
    \plotone{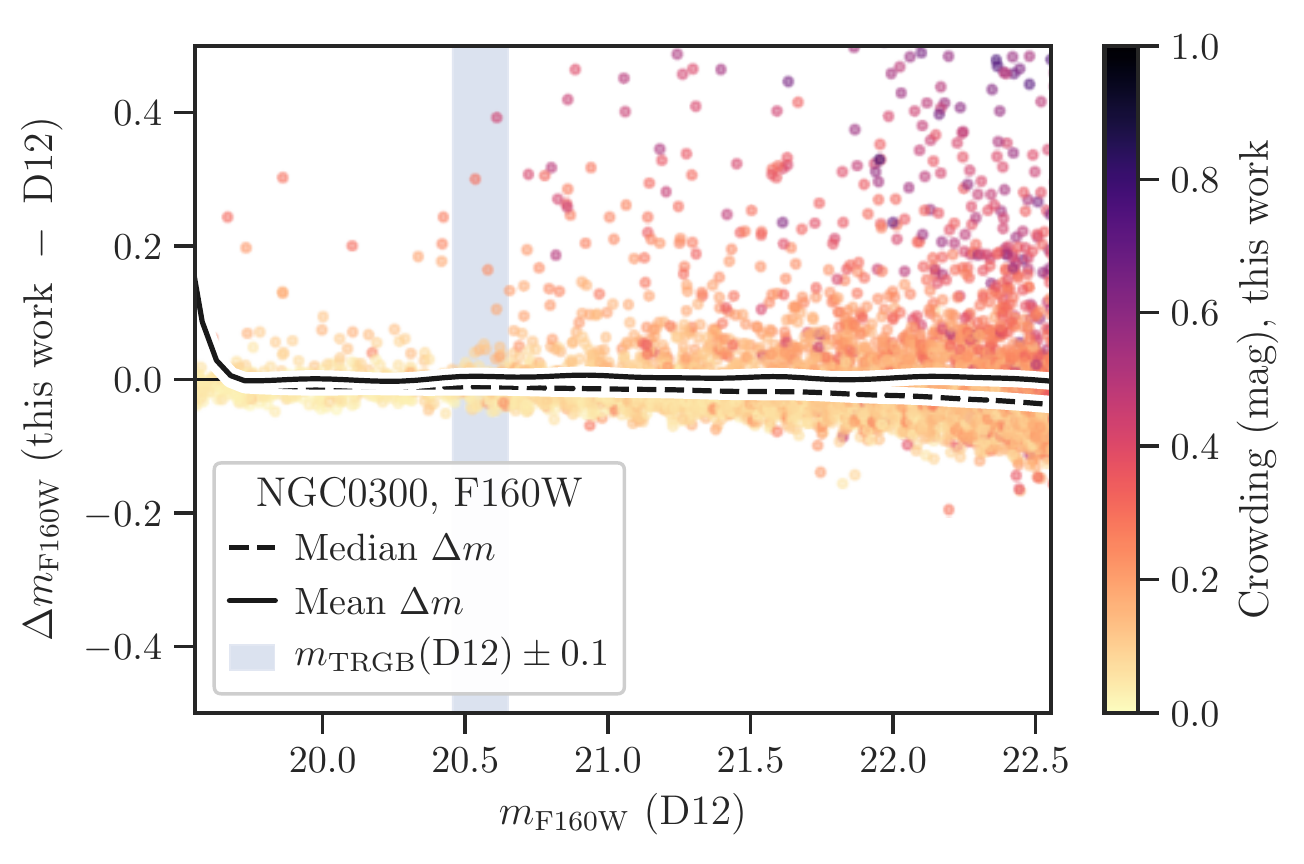}
    \plotone{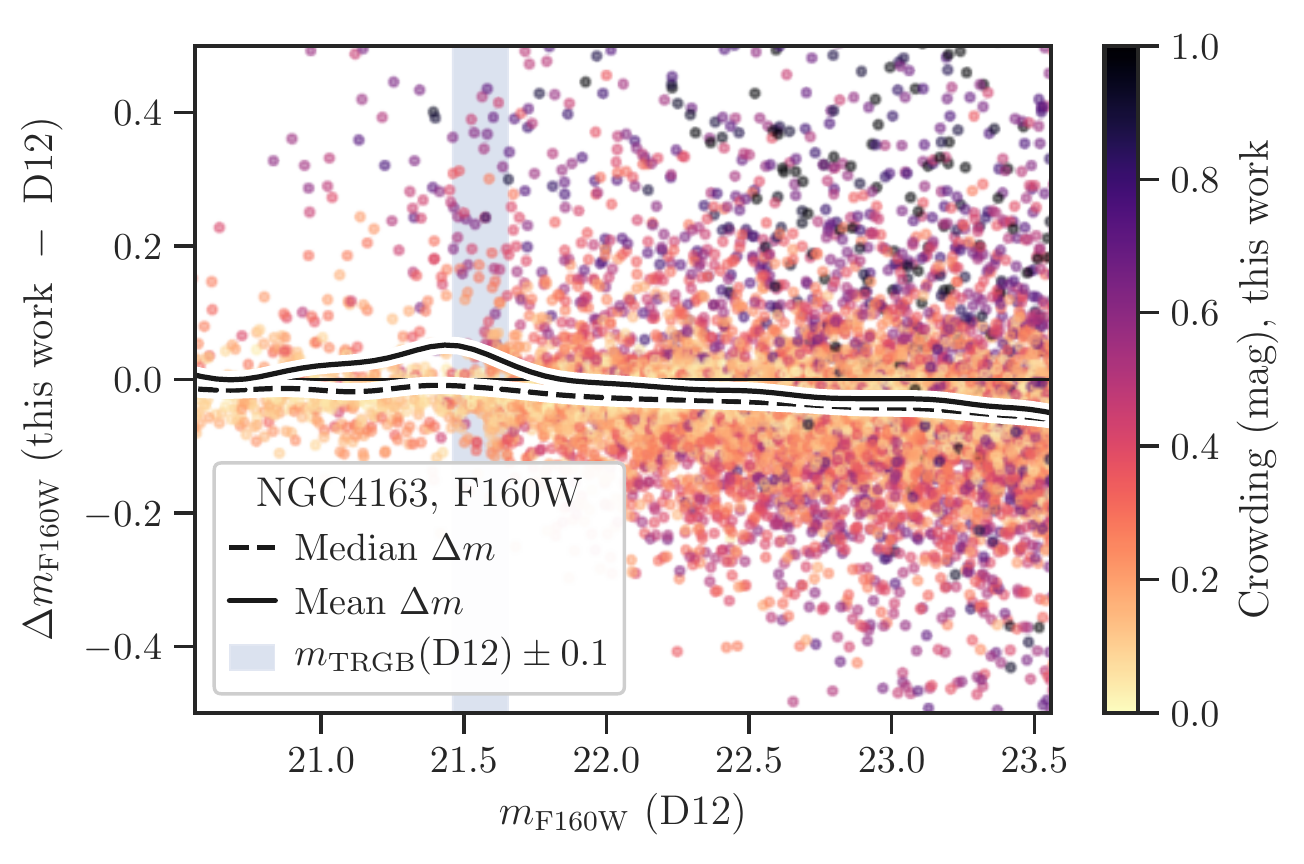}
    \caption{Changes in photometry between \citetalias{2012ApJS..198....6D} and this work for HS117 (top), NGC\,300 (middle), and NGC\,4163 (bottom), with the D12 magnitude on the x-axis and $\Delta m$ on the y-axis. The color-coding indicates the DOLPHOT crowding parameter of the new photometry, which is the number of magnitudes subtracted from the initial measurement due to neighboring sources. The rolling mean and median are shown by the solid and dashed lines respectively.
    The complete figure set for both F110W and F160W (46 images) is available in the online journal.
    }
    \label{fig:old_new}
\end{figure}

Here we directly compare this generation of photometry to that of \citetalias{2012ApJS..198....6D} by crossmatching individual stars. We first convert the IR pixel coordinates of the original photometry to the WCS defined by our realigned images. We select an initial sample of stars within 1 mag of the \citetalias{2012ApJS..198....6D} TRGB values and maximum per-filter old-to-new magnitude differences of 0.5 mag, and match on RA and Dec using a kd-tree \citep{KDTree} with a maximum distance of 2\arcsec. We then find the robust coordinate transformation parameters between the new and old photometry using RANSAC regression \citep{RANSAC} on the matched initial sample with a maximum residual value of 0\farcs1. We apply this transformation to the full set of old photometry coordinates and match the transformed coordinates again with a kd-tree, this time with a maximum distance of 0\farcs1. \autoref{fig:old_new} shows the changes in magnitude as a function of the original \citetalias{2012ApJS..198....6D} magnitudes in F160W, with the \citetalias{2012ApJS..198....6D} TRGB $\pm 0.1$ mag highlighted.

\input{figset_old_new.tex}

Interestingly, we find that near the tip, the median magnitude difference is typically very small (on the order of 0.01 mag) but negative for uncrowded stars, indicating that this generation of photometry is slightly brighter than the previous. However, even the sparsest fields show a population of high-crowding stars that are several tenths of a magnitude dimmer than their \citetalias{2012ApJS..198....6D} counterparts.

%% file: table_sample.tex
\begin{table*}[!ht]
\centering
\caption{Sample galaxies}

\begin{tabular}{llrrrrrrrrl}
\hline
\hline
Galaxy &          Alt. &        RA &      Dec & Diam. &    $B_T$ &    $A_V$ &    $m-M$ &     $T$ &        $W_{50}$ &  Group \\
       &         Names &     (J2000) &    (J2000) &   (\arcmin) &        &        &        &       &  ( $\mathrm{km}~\mathrm{s}^{-1}$) &        \\
\hline
 DDO53 &         U4459 &  08:34:06.5 &   66:10:45 &   1.6 &  14.55 &  0.104 &  27.79 &  10.0 &          25 &    M81 \\
 DDO78 &               &  10:26:27.9 &   67:39:24 &   2.0 &  15.80 &  0.058 &  28.18 &  -3.0 &             &    M81 \\
 DDO82 &         U5692 &  10:30:35.0 &   70:37:10 &   3.4 &  13.57 &  0.112 &  27.90 &   9.0 &             &    M81 \\
   HoI &   U5139,DDO63 &  09:40:28.2 &   71:11:11 &   3.6 &  13.64 &  0.137 &  27.95 &  10.0 &          29 &    M81 \\
  HoII &         U4305 &  08:19:05.9 &   70:42:51 &   7.9 &  11.09 &  0.087 &  27.65 &  10.0 &          66 &    M81 \\
 HS117 &               &  10:21:25.2 &   71:06:58 &   1.5 &  16.50 &  0.316 &  27.91 &  10.0 &          13 &    M81 \\
 I2574 &   U5666,DDO81 &  10:28:22.4 &   68:24:58 &  13.2 &  10.84 &  0.100 &  27.90 &   9.0 &         115 &    M81 \\
  KDG2 &  E540-030,KK9 &  00:49:21.1 &  -18:04:28 &   1.2 &  16.37 &  0.064 &  27.61 &  -1.0 &             &    Scl \\
 KDG63 &   U5428,DDO71 &  10:05:07.3 &   66:33:18 &   1.7 &  16.01 &  0.270 &  27.74 &  -3.0 &          19 &    M81 \\
 KDG73 &               &  10:52:55.3 &   69:32:45 &   0.6 &  17.09 &  0.052 &  28.03 &  10.0 &          18 &    M81 \\
 KKH37 &               &  06:47:45.8 &   80:07:26 &   1.2 &  16.40 &  0.204 &  27.56 &  10.0 &          20 &        \\
   M81 &   N3031,U5318 &  09:55:33.5 &   69:04:00 &  26.9 &   7.69 &  0.232 &  27.77 &   3.0 &         422 &    M81 \\
  N300 &               &  00:54:53.5 &  -37:40:57 &  21.9 &   8.95 &  0.034 &  26.50 &   7.0 &         149 &  14+13 \\
 N2403 &         U3918 &  07:36:54.4 &   65:35:58 &  21.9 &   8.82 &  0.110 &  27.50 &   6.0 &         231 &    M81 \\
 N2976 &         U5221 &  09:47:15.6 &   67:54:49 &   5.9 &  11.01 &  0.241 &  27.76 &   5.0 &          97 &    M81 \\
 N3077 &         U5398 &  10:03:21.0 &   68:44:02 &   5.4 &  10.46 &  0.188 &  27.92 &  10.0 &          65 &    M81 \\
 N3741 &         U6572 &  11:36:06.4 &   45:17:07 &   2.0 &  14.38 &  0.066 &  27.55 &  10.0 &          81 &  14+07 \\
 N4163 &         U7199 &  12:12:08.9 &   36:10:10 &   1.9 &  13.63 &  0.055 &  27.29 &  10.0 &          18 &  14+07 \\
 N7793 &               &  23:57:49.4 &  -32:35:24 &   9.3 &   9.70 &  0.054 &  27.96 &   7.0 &         174 &    Scl \\
  Sc22 &        Sc-dE1 &  00:23:51.7 &  -24:42:18 &   0.9 &  17.73 &  0.042 &  28.11 &  10.0 &             &    Scl \\
 U8508 &         IZw60 &  13:30:44.4 &   54:54:36 &   1.7 &  14.12 &  0.042 &  27.06 &  10.0 &          49 &  14+07 \\
 UA292 &      CVnI-dwA &  12:38:40.0 &   32:46:00 &   1.0 &  16.10 &  0.043 &  27.79 &  10.0 &          27 &        \\
\hline
\end{tabular}

\tablecomments{Reproduced from \citetalias{2012ApJS..198....6D}, with updates to $A_V$ from \citet{2011ApJ...737..103S}. Name, position, diameter, $B_{\rm{T}}$, $W_{50}$, and T-type taken from \citet{2004AJ....127.2031K}. $m-M$ from \citetalias{2009ApJS..183...67D} and \citet{2003A&A...404...93K} for NGC\,7793; Group membership from \citet{2005AJ....129..178K} or \citet{2006AJ....132..729T}.}

\label{tab:sample}
\end{table*}

%% file: table_obs.tex
\begin{table*}[!ht]
\centering

\caption{Observations}
\footnotesize
\begin{tabular}{l l l >{\raggedleft}p{0.8cm} >{\raggedleft}p{1.2cm} r r r >{\raggedright}p{1.5cm} >{\raggedright\arraybackslash}p{2.2cm} }
\hline
\hline
Galaxy & Target name & Date obs. &  Offset (\arcmin)  &  Exptime (s) & $\Sigma_\mathrm{min}$ & $\Sigma_\mathrm{max}$ &  $N_\star$ &     Opt. propid &                Opt. filters \\
\hline
KDG63   &            DDO71 &  2010-04-21 16:33:04 &    0.97 &           9000 &   0.00 &   0.86 &   68477 &             GO-9884 &                F606W, F814W \\
DDO78   &            DDO78 &  2010-04-20 15:13:25 &    0.34 &           2292 &   0.02 &   0.54 &   56458 &            GO-10915 &                F475W, F814W \\
DDO82   &            DDO82 &  2010-05-07 07:27:41 &    0.38 &           2442 &   0.01 &   2.99 &  187699 &            GO-10915 &         F475W, F606W, F814W \\
KDG2    &       ESO540-030 &  2009-12-17 12:32:10 &    0.15 &           7840 &   0.00 &   0.60 &   28087 &            GO-10503 &                F606W, F814W \\
HS117   &            HS117 &  2010-02-24 02:35:38 &    0.13 &            900 &   0.00 &   0.46 &   13011 &             GO-9771 &                F606W, F814W \\
I2574   &       IC2574-SGS &  2010-02-25 03:34:37 &    3.28 &           6400 &   0.10 &   1.40 &  286852 &             GO-9755 &                F555W, F814W \\
KDG73   &            KDG73 &  2010-06-09 18:17:42 &    0.43 &           2274 &   0.00 &   0.22 &   12721 &            GO-10915 &                F475W, F814W \\
KKH37   &            KKH37 &  2009-09-29 11:12:38 &    0.09 &           3441 &   0.00 &   1.36 &   30966 &   GO-10915, GO-9771 &         F475W, F606W, F814W \\
M81     &         M81-DEEP &  2010-06-13 01:26:19 &   13.88 &          29953 &   0.02 &   0.23 &   63093 &            GO-10915 &         F475W, F606W, F814W \\
N300    &          NGC0300 &  2010-04-19 18:17:32 &    6.26 &           2982 &   0.05 &   0.59 &  197750 &   GO-10915, GO-9492 &  F475W, F555W, F606W, F814W \\
N2403   &   NGC2403-HALO-6 &  2010-04-25 04:57:54 &    5.58 &            720 &   0.01 &   0.46 &   20441 &            GO-10523 &                F606W, F814W \\
N2976   &     NGC2976-DEEP &  2010-02-25 02:34:59 &    3.03 &          27191 &   0.02 &   1.50 &   96662 &            GO-10915 &         F475W, F606W, F814W \\
N3077   &  NGC3077-PHOENIX &  2010-02-21 23:20:39 &    3.89 &          19200 &   0.02 &   0.38 &   70482 &             GO-9381 &                F555W, F814W \\
N3741   &          NGC3741 &  2009-11-07 02:03:02 &    0.51 &           2331 &   0.00 &   2.14 &   48369 &            GO-10915 &                F475W, F814W \\
N4163   &          NGC4163 &  2010-03-23 18:11:32 &    0.23 &           3150 &   0.01 &   3.89 &  153523 &   GO-10915, GO-9771 &         F475W, F606W, F814W \\
N7793   &   NGC7793-HALO-6 &  2010-06-14 19:43:15 &    6.02 &            740 &   0.01 &   0.45 &   20079 &            GO-10523 &                F606W, F814W \\
Sc22    &          SCL-DE1 &  2009-09-08 01:16:49 &    0.18 &          17920 &   0.00 &   0.24 &   18967 &            GO-10503 &                F606W, F814W \\
N2403   &    SN-NGC2403-PR &  2010-04-22 08:27:47 &    0.90 &           1450 &   0.37 &   7.59 &  433196 &  GO-10182, GO-10402 &         F475W, F606W, F814W \\
HoII    &          UGC4305 &  2010-02-26 10:10:22 &    0.54 &           9920 &   0.02 &   1.65 &  327523 &  GO-10605, GO-10522 &                F555W, F814W \\
DDO53   &          UGC4459 &  2010-04-23 11:46:34 &    0.25 &           4768 &   0.02 &   0.47 &   63451 &            GO-10605 &                F555W, F814W \\
HoI     &          UGC5139 &  2009-08-21 23:26:49 &    0.35 &           5936 &   0.04 &   0.52 &  105305 &            GO-10605 &                F555W, F814W \\
U8508   &          UGC8508 &  2009-10-14 20:11:32 &    0.18 &           2349 &   0.00 &   1.57 &   73755 &            GO-10915 &                F475W, F814W \\
UA292   &          UGCA292 &  2010-05-18 13:08:17 &    0.35 &           2274 &   0.00 &   0.21 &   17668 &  GO-10915, GO-10905 &         F475W, F606W, F814W \\
\hline
\end{tabular}

\tablecomments{Here the date observed is the date of the last IR exposure; the offset is the distance between the center of the IR footprint and the galaxy coordinates as given in \autoref{tab:sample}; the exposure time is the total F814W exposure time (all IR observations have uniform exposure times of 600s in F110W and 900s in F160W); and $N_\star$ is the number of stars that were detected in all of F814W, F110W, and F160W.}

\label{tab:obs}
\end{table*}

%% file: figset_density.tex
\figsetstart
\figsetnum{4}
\figsettitle{Surface density maps}

    \figsetgrpstart
    \figsetgrpnum{4.1}
    \figsetgrptitle{DDO71}
    \figsetplot{f4_1.pdf}
    \figsetgrpnote{
        Stellar surface density map for target DDO71.
    }
    \figsetgrpend

    \figsetgrpstart
    \figsetgrpnum{4.2}
    \figsetgrptitle{DDO78}
    \figsetplot{f4_2.pdf}
    \figsetgrpnote{
        Stellar surface density map for target DDO78.
    }
    \figsetgrpend

    \figsetgrpstart
    \figsetgrpnum{4.3}
    \figsetgrptitle{DDO82}
    \figsetplot{f4_3.pdf}
    \figsetgrpnote{
        Stellar surface density map for target DDO82.
    }
    \figsetgrpend

    \figsetgrpstart
    \figsetgrpnum{4.4}
    \figsetgrptitle{ESO540-030}
    \figsetplot{f4_4.pdf}
    \figsetgrpnote{
        Stellar surface density map for target ESO540-030.
    }
    \figsetgrpend

    \figsetgrpstart
    \figsetgrpnum{4.5}
    \figsetgrptitle{HS117}
    \figsetplot{f4_5.pdf}
    \figsetgrpnote{
        Stellar surface density map for target HS117.
    }
    \figsetgrpend

    \figsetgrpstart
    \figsetgrpnum{4.6}
    \figsetgrptitle{IC2574-SGS}
    \figsetplot{f4_6.pdf}
    \figsetgrpnote{
        Stellar surface density map for target IC2574-SGS.
    }
    \figsetgrpend

    \figsetgrpstart
    \figsetgrpnum{4.7}
    \figsetgrptitle{KDG73}
    \figsetplot{f4_7.pdf}
    \figsetgrpnote{
        Stellar surface density map for target KDG73.
    }
    \figsetgrpend

    \figsetgrpstart
    \figsetgrpnum{4.8}
    \figsetgrptitle{KKH37}
    \figsetplot{f4_8.pdf}
    \figsetgrpnote{
        Stellar surface density map for target KKH37.
    }
    \figsetgrpend

    \figsetgrpstart
    \figsetgrpnum{4.9}
    \figsetgrptitle{M81-DEEP}
    \figsetplot{f4_9.pdf}
    \figsetgrpnote{
        Stellar surface density map for target M81-DEEP.
    }
    \figsetgrpend

    \figsetgrpstart
    \figsetgrpnum{4.10}
    \figsetgrptitle{NGC0300}
    \figsetplot{f4_10.pdf}
    \figsetgrpnote{
        Stellar surface density map for target NGC0300.
    }
    \figsetgrpend

    \figsetgrpstart
    \figsetgrpnum{4.11}
    \figsetgrptitle{NGC2403-HALO-6}
    \figsetplot{f4_11.pdf}
    \figsetgrpnote{
        Stellar surface density map for target NGC2403-HALO-6.
    }
    \figsetgrpend

    \figsetgrpstart
    \figsetgrpnum{4.12}
    \figsetgrptitle{NGC2976-DEEP}
    \figsetplot{f4_12.pdf}
    \figsetgrpnote{
        Stellar surface density map for target NGC2976-DEEP.
    }
    \figsetgrpend

    \figsetgrpstart
    \figsetgrpnum{4.13}
    \figsetgrptitle{NGC3077-PHOENIX}
    \figsetplot{f4_13.pdf}
    \figsetgrpnote{
        Stellar surface density map for target NGC3077-PHOENIX.
    }
    \figsetgrpend

    \figsetgrpstart
    \figsetgrpnum{4.14}
    \figsetgrptitle{NGC3741}
    \figsetplot{f4_14.pdf}
    \figsetgrpnote{
        Stellar surface density map for target NGC3741.
    }
    \figsetgrpend

    \figsetgrpstart
    \figsetgrpnum{4.15}
    \figsetgrptitle{NGC4163}
    \figsetplot{f4_15.pdf}
    \figsetgrpnote{
        Stellar surface density map for target NGC4163.
    }
    \figsetgrpend

    \figsetgrpstart
    \figsetgrpnum{4.16}
    \figsetgrptitle{NGC7793-HALO-6}
    \figsetplot{f4_16.pdf}
    \figsetgrpnote{
        Stellar surface density map for target NGC7793-HALO-6.
    }
    \figsetgrpend

    \figsetgrpstart
    \figsetgrpnum{4.17}
    \figsetgrptitle{SCL-DE1}
    \figsetplot{f4_17.pdf}
    \figsetgrpnote{
        Stellar surface density map for target SCL-DE1.
    }
    \figsetgrpend

    \figsetgrpstart
    \figsetgrpnum{4.18}
    \figsetgrptitle{SN-NGC2403-PR}
    \figsetplot{f4_18.pdf}
    \figsetgrpnote{
        Stellar surface density map for target SN-NGC2403-PR.
    }
    \figsetgrpend

    \figsetgrpstart
    \figsetgrpnum{4.19}
    \figsetgrptitle{UGC4305}
    \figsetplot{f4_19.pdf}
    \figsetgrpnote{
        Stellar surface density map for target UGC4305.
    }
    \figsetgrpend

    \figsetgrpstart
    \figsetgrpnum{4.20}
    \figsetgrptitle{UGC4459}
    \figsetplot{f4_20.pdf}
    \figsetgrpnote{
        Stellar surface density map for target UGC4459.
    }
    \figsetgrpend

    \figsetgrpstart
    \figsetgrpnum{4.21}
    \figsetgrptitle{UGC5139}
    \figsetplot{f4_21.pdf}
    \figsetgrpnote{
        Stellar surface density map for target UGC5139.
    }
    \figsetgrpend

    \figsetgrpstart
    \figsetgrpnum{4.22}
    \figsetgrptitle{UGC8508}
    \figsetplot{f4_22.pdf}
    \figsetgrpnote{
        Stellar surface density map for target UGC8508.
    }
    \figsetgrpend

    \figsetgrpstart
    \figsetgrpnum{4.23}
    \figsetgrptitle{UGCA292}
    \figsetplot{f4_23.pdf}
    \figsetgrpnote{
        Stellar surface density map for target UGCA292.
    }
    \figsetgrpend

\figsetend

%% file: figset_old_new.tex
\figsetstart
\figsetnum{5}
\figsettitle{Photometry comparisons}

    \figsetgrpstart
    \figsetgrpnum{5.10}
    \figsetgrptitle{HS117 F160W}
    \figsetplot{f5_10.pdf}
    \figsetgrpnote{
        Comparison between IR-only and full-stack photometry for target HS117, filter F160W.
    }
    \figsetgrpend

    \figsetgrpstart
    \figsetgrpnum{5.20}
    \figsetgrptitle{NGC0300 F160W}
    \figsetplot{f5_20.pdf}
    \figsetgrpnote{
        Comparison between IR-only and full-stack photometry for target NGC0300, filter F160W.
    }
    \figsetgrpend

    \figsetgrpstart
    \figsetgrpnum{5.30}
    \figsetgrptitle{NGC4163 F160W}
    \figsetplot{f5_30.pdf}
    \figsetgrpnote{
        Comparison between IR-only and full-stack photometry for target NGC4163, filter F160W.
    }
    \figsetgrpend

\figsetend

%% file: 3_measurement.tex
\section{TRGB Measurement} \label{sec:measurement}

In this section we describe the steps we use to measure the apparent magnitudes and colors of the IR-TRGB.
We adopt a multiwavelength approach, which we call ``MCR-TRGB", that we summarize for the reader in advance of detailed descriptions.
First, we isolate the RGB sequence from the other stellar populations.
From the RGB sample, we do a tip detection to select stars in the vicinity of the TRGB.
This initial sample is then separated into potential sub-populations to isolate those that have colors and magnitudes consistent with being TRGB stars.
The color and magnitude distributions of the candidate tip stars are then fitted jointly for all applicable color-magnitude spaces to build the final color-magnitude calibrations.
This approach has several advantages over traditional Sobel edge-detection for the purpose of this work, which we discuss in detail in \autoref{sec:discussion}.

Throughout this section, the methods are demonstrated using galaxies that span a range in metallicity and RGB shape. Identical figures for each of the 23 galaxies in the sample are provided as figure sets.

\subsection{Initial RGB Star Selection} \label{ssec:RGB_Selection}

\input{figset_rgb.tex}
\begin{figure*}[ht]
    \plotone{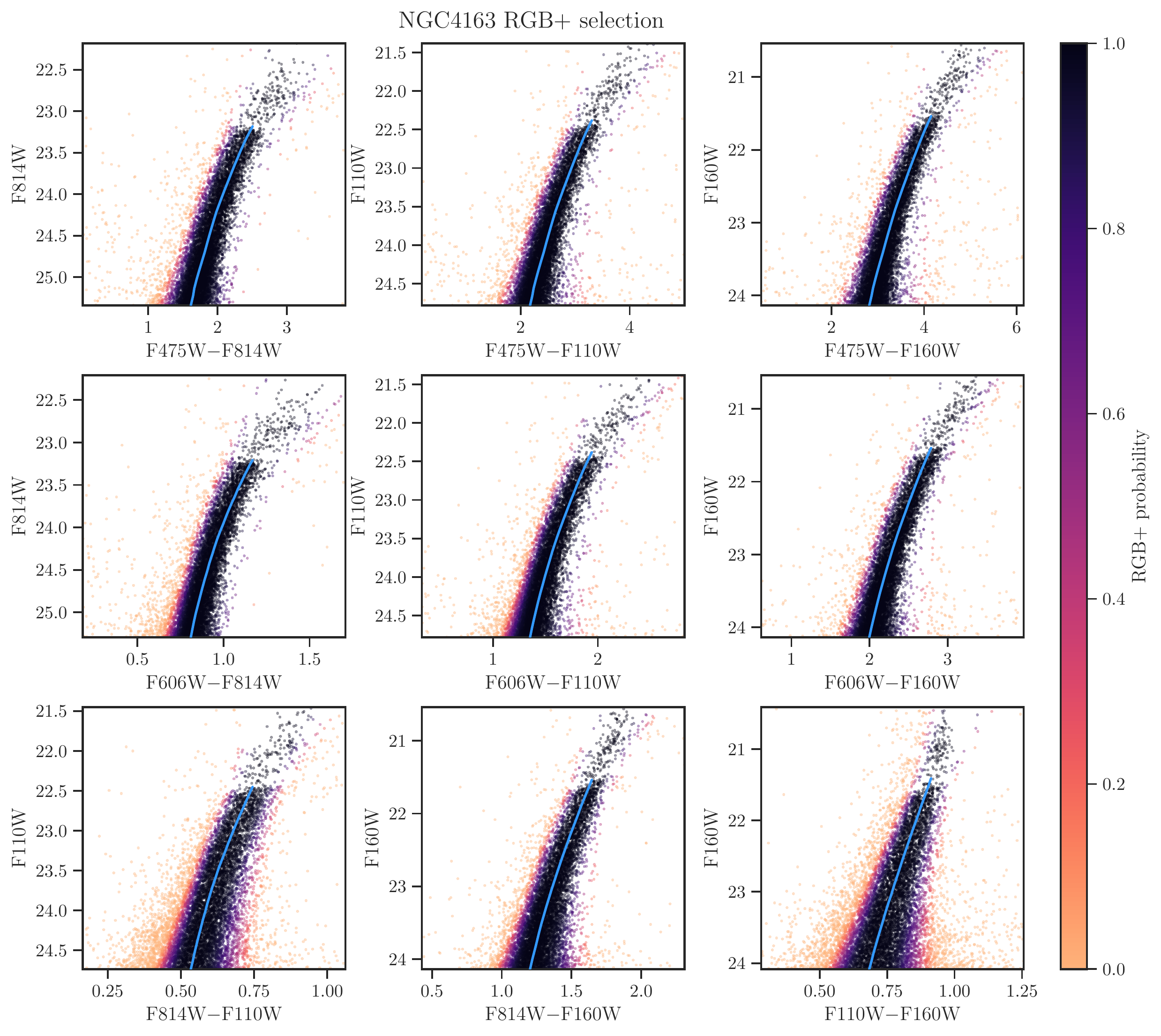}
    \caption{A demonstration of filter-by-filter RGB selections for NGC4163. Each of the 9 panels contains a color-magnitude combination used to determine $P({\rm RGB})+$ for stars in the color-magnitude range.
    The points in each panel are color-coded by the final probability as indicated on the color-bar.
    The best-fit PARSEC \replaced{isochrone}{synthetic RGB} in each combination is shown\deleted{, with its age and metallicity labelled for each panel}. 
    \deleted{While the best-fit isochrones are largely of consistent age and metallicity for this target, this is not necessarily the case for all targets.
    We also emphasize the discrepancy between the F110W--F160W best-fit isochrone and the rest, which suggests that forcing a single set of physical parameters to fit all filter combinations may induce bias in RGB+ star selection.}
    The complete figure set (23 images) is available in the online journal.
}
\label{fig:rgb_selection}
\end{figure*}

A maximally complete and minimally contaminated sample of RGB stars is essential for characterizing the TRGB and the RGB luminosity function near the tip. Unfortunately there are many stars that have colors and magnitudes similar to RGB stars, such as red helium-burning (RHeB) and asymptotic giant branch (AGB) stars. These ``contaminant" populations can blur the TRGB edge or distort its measured magnitude (see discussion in \citetalias{2012ApJS..198....6D}).

Typically, RGB stars are selected using strict \added{binary} color-magnitude cuts; we describe two particular examples. \citetalias{2012ApJS..198....6D} initially select stars with colors in the range $0.6 < {\rm F110W}-{\rm F160W} < 1.1$ mag and magnitudes brighter than 1 mag below their initial TRGB estimate, and then make further rejections based on the standard deviations of a linear fit to the remaining stars in color-magnitude space.
The Carnegie-Chicago Hubble Program \citep{2017ApJ...845..146H, 2018ApJ...852...60J, 2018ApJ...861..104H, 2018ApJ...866..145H, 2019ApJ...882...34F} makes color cuts with a fiducial RGB slope and a color width chosen visually to encompass the edges of the RGB near the tip.

Here, we leverage the multiwavelength information available for our targets to probabilistically identify stars that fall along characteristic RGB color-magnitude sequences.
For each target we construct a set of \added{{\it{red}} vs.\ {\it{blue--red}}} CMDs using F814W, F110W, and F160W as the red filters, and using all available optical filters other than F814W as the blue. We also construct CMDs in F814W vs.\ F814W--F160W, F110W vs.\ F814W--F110W, and F110W vs.\ F110W--F160W for all targets.
The number of unique color-magnitude combinations varies from 6 to 12 depending on the number of available optical filters for each target. 
We apply broad initial color and magnitude cuts based on the \citetalias{2012ApJS..198....6D} TRGB measurements. 
\autoref{fig:rgb_selection} provides example CMDs after cuts for NGC\,4163 in the color-magnitude combinations used for this analysis.

Next, we define an RGB locus in each filter combination by fitting \replaced{an isochrone}{a predicted RGB color-magnitude sequence} to \added{the photometry in} each color-magnitude \replaced{sequence}{combination} independently.
We minimize the median distance between the observed photometry and a grid of \added{synthetic photometry} derived from PARSEC \added{\citep{2017ApJ...835...77M}} isochrones of ages 4 to 14 Gyr and [Fe/H] $-3.0$ to $-0.2$ dex, which have been limited to RGB stars brighter than $M_{\rm F110W} = -2$ mag \added{and converted to apparent magnitudes using the distance moduli from \citetalias{2012ApJS..198....6D}}. 
The panels of \autoref{fig:rgb_selection} have their best-fit \replaced{isochrones}{isochrone-predicted RGB sequences} overlaid in blue\deleted{and both its age and metallicity are labelled for reference}.
\deleted{However, }As our goal is to trace the RGB \added{color-magnitude} locus \added{across all available bandpasses} rather than to measure any underlying properties of the stellar populations, we do not force a single age and metallicity combination to fit all color-magnitude combinations.\added{(We note that the metallicities of the ``best-fit" isochrones for a single target can vary filter-to-filter by up to nearly a full dex, especially in the case of low-metallicity targets where the upper RGB color only weakly depends on metallicity; see \autoref{ssec:model_discussion} for further discussion of filter-to-filter differences between observed and predicted photometry at the TRGB.)}

An initial ``RGB-sequence probability" is then assigned to each star based on the distance between its observed position in color-magnitude space and the nearest point on the \replaced{interpolated isochrone}{predicted RGB} for each color-magnitude combination. 
The points in the panels of \autoref{fig:rgb_selection} are color-coded by these probabilities. 

The purpose of this process is to construct a luminosity function with which to make an initial TRGB estimate, as described in \autoref{ssec:edge_detection}.
We therefore extrapolate the fitted RGB sequences out to at least 1.5 mag brighter than the measured \citetalias{2012ApJS..198....6D} TRGB apparent magnitudes in all filters.
As a result, stars brighter than the TRGB that fall along the \added{predicted} color-magnitude \replaced{sequences defined by the isochrones}{loci} will be assigned high RGB-sequence probabilities.
These probabilities should be understood as estimates of a star's proximity to the color-magnitude relations characteristic of each target's RGB sequence, rather than as identifications of only the stars that are truly on the RGB.

The individual RGB-sequence probabilities are then averaged across all color-magnitude combinations to produce \replaced{final}{global} RGB-sequence probabilities, which we call $P({\rm RGB})+$.

\subsection{Edge Detection} \label{ssec:edge_detection}

\begin{figure}[ht]
    \epsscale{1.1}
    \plotone{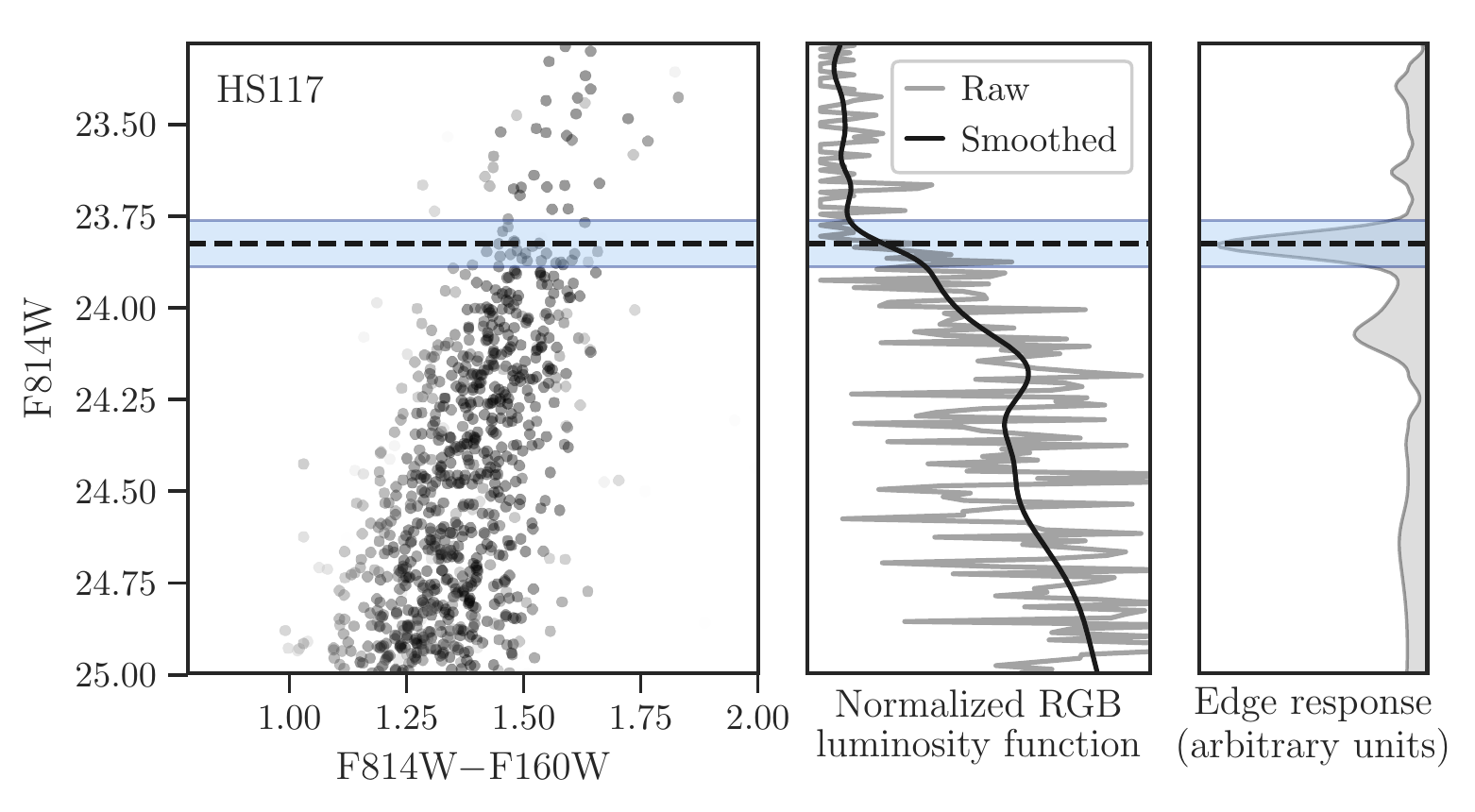}
    \plotone{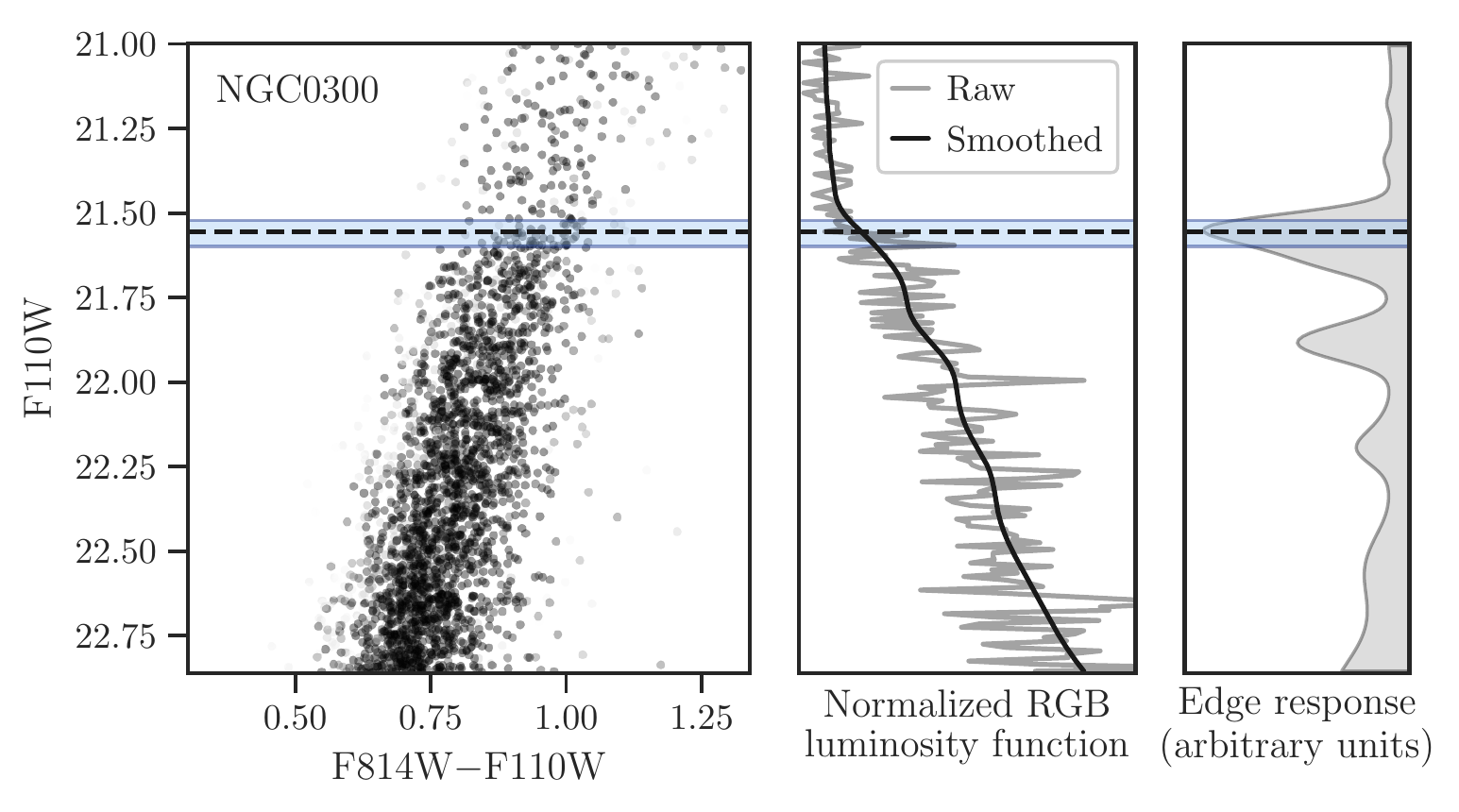}
    \plotone{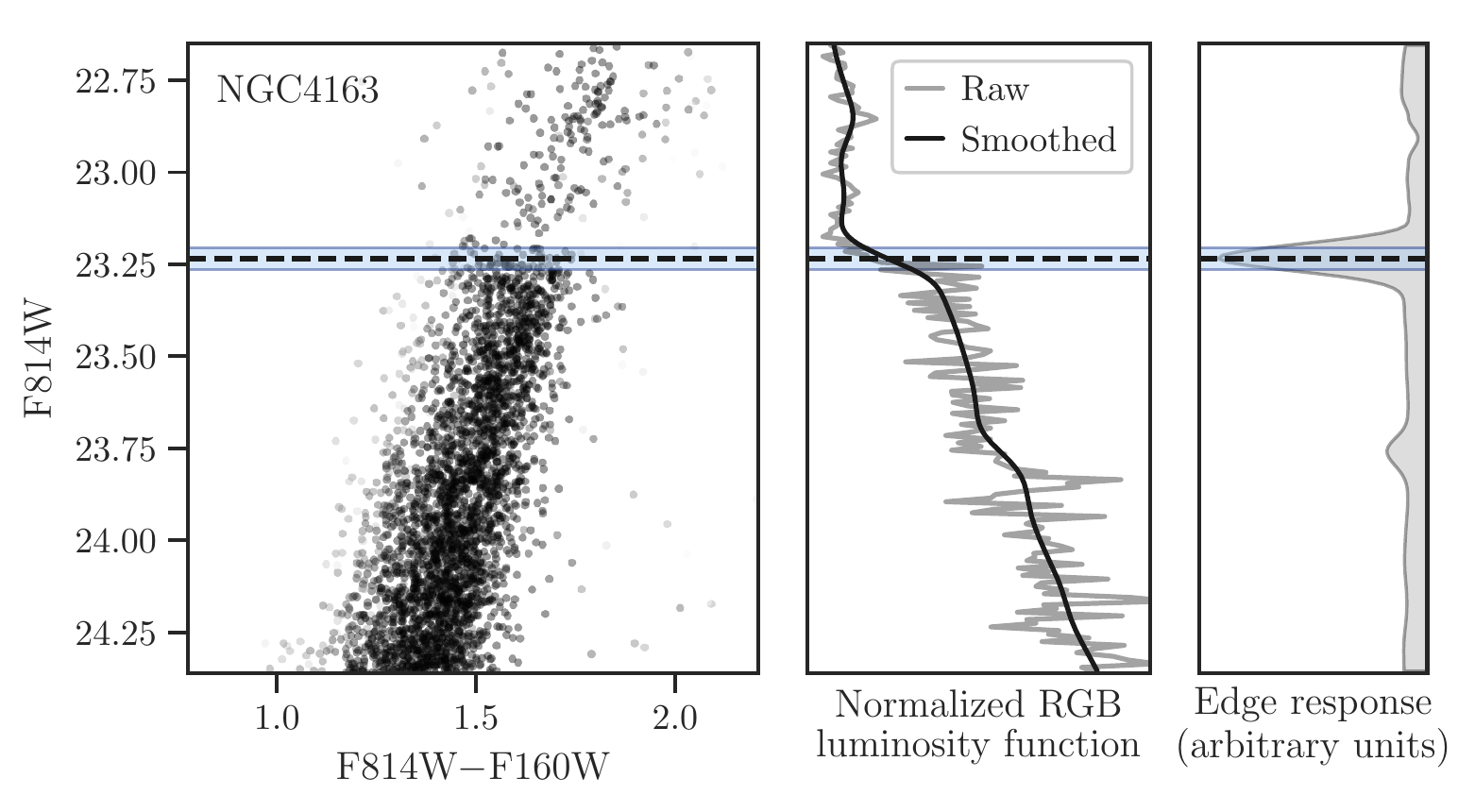}
    \caption{Tip star selection with edge detection for three demonstrative galaxies in our sample. From top to bottom, HS117, NGC\,300, and NGC\,4163. 
    For each galaxy, three panels are shown, from left to right, the CMD of high-probability RGB stars, the raw (gray) and smoothed (black) luminosity function, and the Sobel edge response ($\eta$). 
    The initial magnitude of the TRGB is identified as the magnitude at $\eta_{\rm max}$, which is identified as the dashed line in each panel.
    TRGB candidate stars are selected within the blue band, the width of which is deteremined by the photometric uncertainty at the tip and by the number of stars on the upper RGB as described in the text. 
    The complete figure set (23 images) is available in the online journal.
}
\label{fig:edge_detection}
\end{figure}

We make an initial selection of candidate tip stars by applying a Sobel edge detection to the RGB-weighted luminosity function (LF). For each target we choose the filter with the sharpest LF; that is, the filter in which the tip magnitude is least dependent on color. This is F814W for most targets, and F110W for targets with ${\rm F110W}-{\rm F160W} > 0.95$ mag, as measured in \citetalias{2012ApJS..198....6D}). We first construct a luminosity function (shown in the middle column of \autoref{fig:edge_detection}) by marginalizing $P({\rm RGB})+$ over color as a function of magnitude. We use a bin size of 0.01~mag, which is a factor of $\sim$5 smaller than the typical magnitude uncertainty.

For each galaxy, the middle panels of \autoref{fig:edge_detection} show the raw LF (blue), where the noise is consistent with Poisson fluctuations.
We first smooth the LF with a Savitzky-Golay filter \citep{savgol}, a low-pass filter originally developed to suppress noise in spectroscopic data by fitting a polynomial within a rolling window. This technique effectively removes Poisson noise spikes while preserving sharp features such as the TRGB edge.
However, there may be remaining spurious edges from photometric variance or stochastic sampling of the luminosity function, particularly in sparse data.
To reduce the impact of these false edges, we smooth the LF once more using GLOESS (Gaussian-windowed, Locally Weighted Scatterplot Smoothing); for an in-depth description see \citet{2017ApJ...845..146H} and references therein. Briefly, GLOESS is an implementation of one-dimensional Gaussian kernel density estimation, which we have modified to accept a variable kernel width.
We select a fiducial kernel width using the {\tt KDEpy} implementation of the Improved Sheather-Jones algorithm \citep{botev2010}, which chooses an optimal kernel width based on the overall density of the data.
We then multiply this fiducial width by the square of the photometric uncertainties scaled by their median value as a function of magnitude, which de-emphasizes LF variation fainter than the TRGB, where photometric uncertainties are higher.
The final smoothed LF, shown overlaid in black on the raw LFs in \autoref{fig:edge_detection}, is then used for the initial TRGB detection.

To detect the TRGB, we begin by applying a Sobel filter, which is one of the most widely used means of finding the tip \citep[see summary and comparisons in][]{2018SSRv..214..113B}. The Sobel filter approximates the first derivative of a discrete dataset via convolution with a kernel.
In its simplest form, this kernel is $[-1,0,1]$, which effectively subtracts counts in the $i-1$ bin from the $i+1$ bin to determine the edge-response, $\eta$, for bin $i$.
This kernel is applied to the smoothed LF, and the response is shown for each galaxy in the right panels of \autoref{fig:edge_detection}.
In \autoref{fig:edge_detection}, the magnitude of maximum Sobel response, $m(\eta_{\rm max})$, is indicated by the dashed line across all panels.

We then select candidate tip stars near $m(\eta_{\rm max})$ within a range we call $\Delta \eta$. The value of $\Delta \eta$ is determined using two quantities: i) the median photometric error within $\pm 0.1$ mag of $m(\eta_{\rm max})$, $\sigma_{\rm phot}^{\eta_{\rm max}}$, and ii), a minimum number of tip candidate stars $N_{\star}^{\rm min}$.
We define $N_{\star}^{\rm min}$ as the square root of the number of stars 1 magnitude below $m(\eta_{\rm max})$, with a hard minimum of 30 stars.
For each target, we make an initial selection of stars within $\pm 1\sigma_{\rm phot}^{\rm TRGB}$, and then iteratively expand the selection range by $0.5 \sigma_{\rm phot}^{\rm TRGB}$ on each side until either $N_{\star}^{\rm min}$ is reached or $\Delta \eta$ is over 0.2 mag. For the majority of our targets, the initial selection window of $\pm 1\sigma_{\rm phot}^{\rm TRGB}$ is enough to meet $N_{\star}^{\rm min}$. 
Our final $\Delta \eta$ is shown by the blue band in the panels of \autoref{fig:edge_detection} for our example pointings.

Out of the stars that fall within the fiducial tip magnitude range, we first select likely RGB stars as those with $P({\rm RGB})+ > 0.6$, which roughly corresponds to stars that were identified as RGB+ sequence candidates with over 90\% probability in at least two-thirds of the filter combinations we used to assign RGB probabilities. We then reject stars with anomalous magnitudes in at least one filter with Local Outlier Factor outlier detection \citep{LOF}, which evaluates the relative isolation of points using $k$-nearest neighbors. We take this trimmed sample of stars to be our final set of tip star candidates, which we then use to measure tip magnitudes and colors as described in the following section.

\input{figset_edge.tex}

We note that this selection of likely RGB tip stars is performed based on the resuts of applying the Sobel filter to the filter where the tip is ``flat'' with color. The Sobel filter, by design, looks for an sharp edge in a one-dimensional distribution. Two dimensional implementations of the Sobel Filter exist, but still require conversion of our CMDs into a binned form. Thus, application of the one-dimensional Sobel filter to a distribution that has magnitude-color behavior may not fully detect the true edge in the distribution. Lastly, where there is strong magnitude-color trend, because our colors are more imprecise than our magnitudes, the intrinsic slope can be distorted by the color-spread in our data. Thus, in the next section, we develop a method to utilize the tip stars we have just identified to trace the intrinsic TRGB slope across our set of \replaced{filter}{color}-magnitude combinations.

\subsection{Multiwavelength tip fitting} \label{ssec:tip_fitting}

We characterize the color and magnitude distributions of our candidate tip stars using Extreme Deconvolution \citep[XDGMM,][]{XDGMM}, a modification of Gaussian mixture modeling that accounts for uncertainties in the input data. Specifically, we use XDGMM to fit a single six-dimensional Gaussian to the F814W, F110W, and F160W magnitudes and the F814W--F160W, F814W--F110W, and F110W--F160W colors of the tip star candidates. 
Although the underlying distribution of tip stars in this parameter space is not intrinsically Gaussian, we find that a single Gaussian is a reasonable approximation for the majority of our tip star samples. Additionally, for the faintest and sparsest of our targets, low star counts and photometric uncertainties on the same order as the width of the tip star selection windows do not allow us to place reasonable constraints on more complex models, such as multi-component Gaussian mixtures.  We discuss potential alternative modeling approaches in \autoref{ssec:cons}.

For the uncertainties we use as inputs to XDGMM, we divide each star's individual photometric uncertainties by $P({\rm RGB}+)$, effectively weighting the input points by $P({\rm RGB}+)$. 
We emphasize that XDGMM, as a tool, allows us to take into account these uncertainties and weights on the RGB+ likelihood to trace the tip in filters where the Sobel edge is less effective due to color-magnitude slopes.

\begin{figure*}[ht]
    \plotone{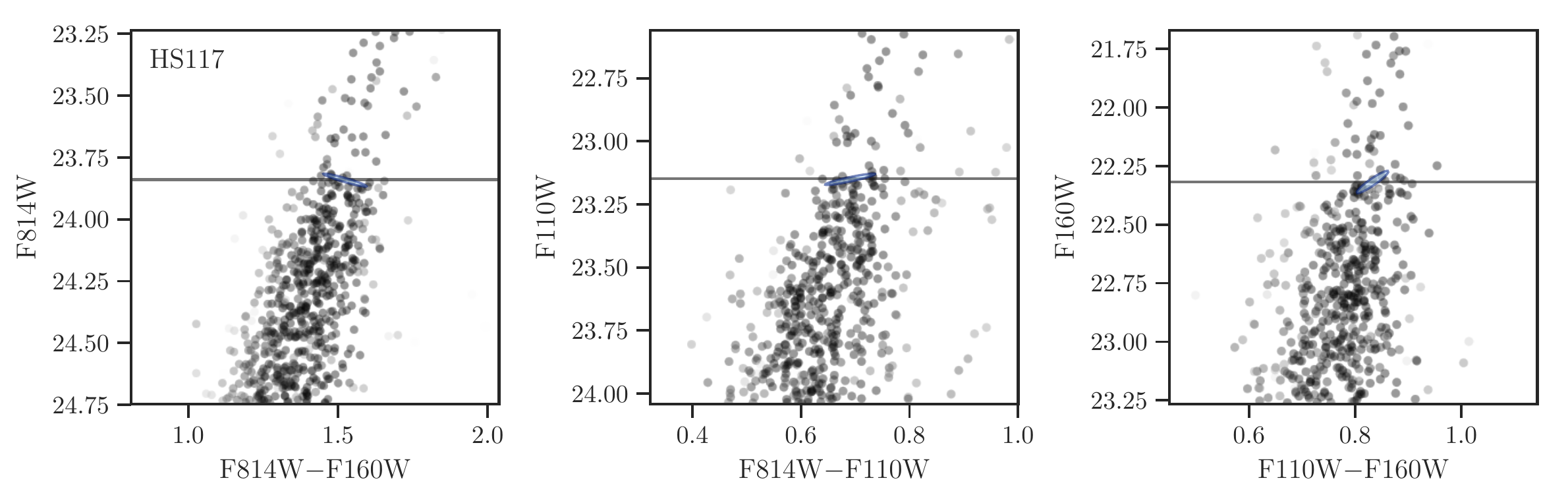}
    \plotone{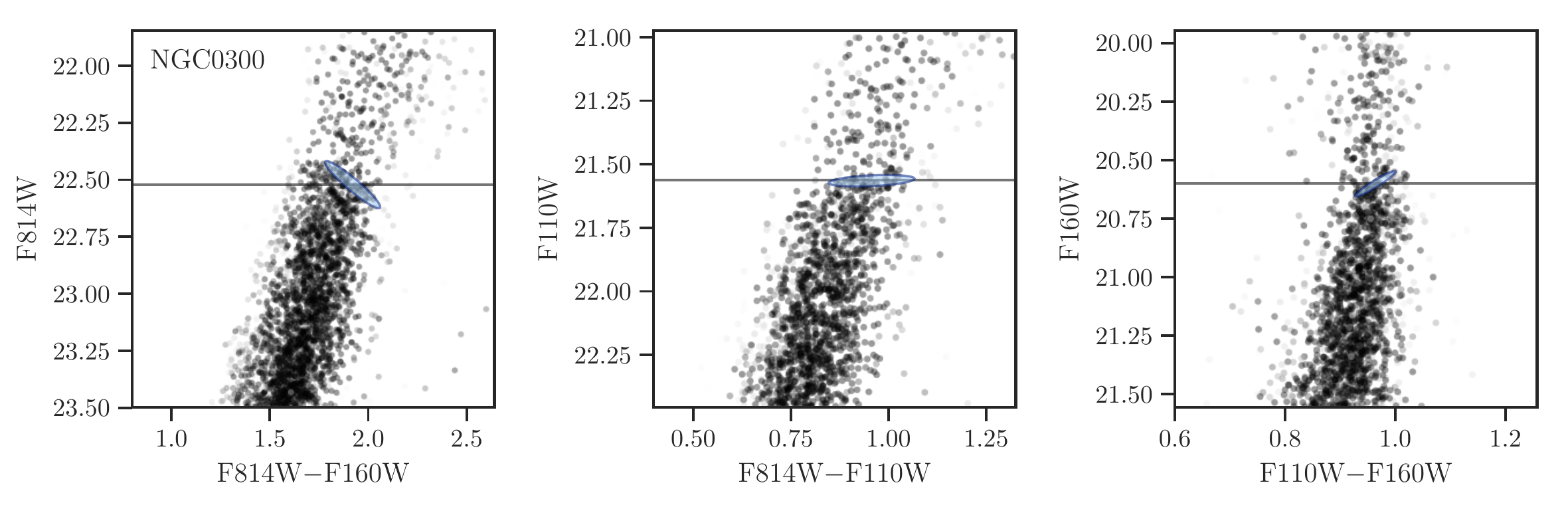}
    \plotone{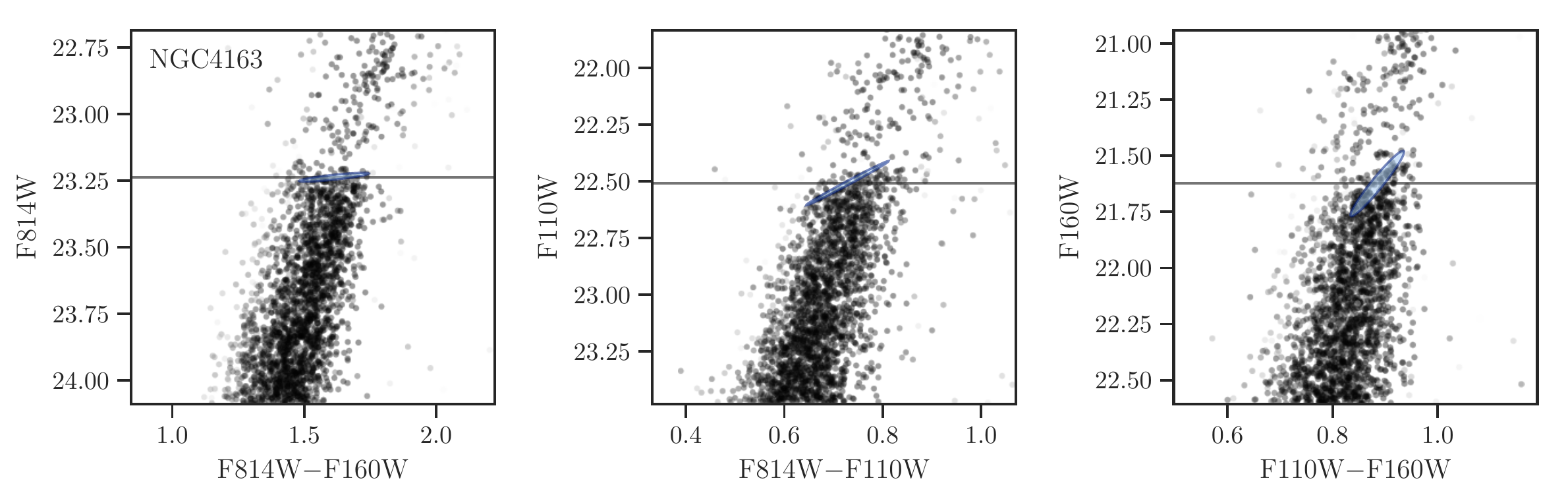}
    \caption{Results of XDGMM fits to candidate TRGB stars for HS117 (top), NGC\,300 (middle), and NGC\,4163 (bottom). For each galaxy the fits are shown for the following color-magnitude combinations: F814W, F814W--F160W (left), F110W, F814W--F110W (center), and F160W, F110W--F160W (right). 
    The solid horizontal lines in each panel identify the mean magnitudes of the tip from XDGMM (which are typically very close to the Sobel-detected edge), and the overplotted ellipses show the 95\% color-magnitude confidence regions of the two-dimensional fitted tip.  
    We note that because this method uses a set of candidate tip stars selected based on their magnitudes in a single band (either F814W or F110W), the 2-D ellipses may not follow the visual impression of the tip in other bandpasses. This is especially apparent when the color-width of the RGB is of order the color uncertainties, which is typical in the NIR.
    The complete figure set (23 images) is available in the online journal.
}
\label{fig:tip_fitting}
\end{figure*}

We take the means of the fitted distributions to be our final apparent tip magnitudes and colors. Results of these fits are shown for our sample galaxies in \autoref{fig:tip_fitting}, where we plot ellipses showing the 95\% confidence regions of the XDGMM fits in three color-magnitude combinations. The width, height, and position angle of each ellipse are derived from two-dimensional slices of the full six-dimensional covariance matrix.

Potential systematic and statistical biases of this method are discussed in \autoref{sec:artdata_tests}; overall, we find that the results are comparable to those of edge detection in most cases.

\input{figset_tip.tex}

%% file: figset_rgb.tex
\figsetstart
\figsetnum{6}
\figsettitle{RGB selection CMDs}

    \figsetgrpstart
    \figsetgrpnum{6.1}
    \figsetgrptitle{DDO71}
    \figsetplot{f6_1.pdf}
    \figsetgrpnote{
        CMDs showing individual RGB-sequence probabilities for target DDO71.
    }
    \figsetgrpend

    \figsetgrpstart
    \figsetgrpnum{6.2}
    \figsetgrptitle{DDO78}
    \figsetplot{f6_2.pdf}
    \figsetgrpnote{
        CMDs showing individual RGB-sequence probabilities for target DDO78.
    }
    \figsetgrpend

    \figsetgrpstart
    \figsetgrpnum{6.3}
    \figsetgrptitle{DDO82}
    \figsetplot{f6_3.pdf}
    \figsetgrpnote{
        CMDs showing individual RGB-sequence probabilities for target DDO83.
    }
    \figsetgrpend

    \figsetgrpstart
    \figsetgrpnum{6.4}
    \figsetgrptitle{ESO540-030}
    \figsetplot{f6_4.pdf}
    \figsetgrpnote{
        CMDs showing individual RGB-sequence probabilities for target ESO540-030.
    }
    \figsetgrpend

    \figsetgrpstart
    \figsetgrpnum{6.5}
    \figsetgrptitle{HS117}
    \figsetplot{f6_5.pdf}
    \figsetgrpnote{
        CMDs showing individual RGB-sequence probabilities for target HS117.
    }
    \figsetgrpend

    \figsetgrpstart
    \figsetgrpnum{6.6}
    \figsetgrptitle{IC2574-SGS}
    \figsetplot{f6_6.pdf}
    \figsetgrpnote{
        CMDs showing individual RGB-sequence probabilities for target IC2574-SGS.
    }
    \figsetgrpend

    \figsetgrpstart
    \figsetgrpnum{6.7}
    \figsetgrptitle{KDG73}
    \figsetplot{f6_7.pdf}
    \figsetgrpnote{
        CMDs showing individual RGB-sequence probabilities for target KDG73.
    }
    \figsetgrpend

    \figsetgrpstart
    \figsetgrpnum{6.8}
    \figsetgrptitle{KKH37}
    \figsetplot{f6_8.pdf}
    \figsetgrpnote{
        CMDs showing individual RGB-sequence probabilities for target KKH37.
    }
    \figsetgrpend

    \figsetgrpstart
    \figsetgrpnum{6.9}
    \figsetgrptitle{M81-DEEP}
    \figsetplot{f6_9.pdf}
    \figsetgrpnote{
        CMDs showing individual RGB-sequence probabilities for target M81-DEEP.
    }
    \figsetgrpend

    \figsetgrpstart
    \figsetgrpnum{6.10}
    \figsetgrptitle{NGC0300}
    \figsetplot{f6_10.pdf}
    \figsetgrpnote{
        CMDs showing individual RGB-sequence probabilities for target NGC0300.
    }
    \figsetgrpend

    \figsetgrpstart
    \figsetgrpnum{6.11}
    \figsetgrptitle{NGC2403-HALO-6}
    \figsetplot{f6_11.pdf}
    \figsetgrpnote{
        CMDs showing individual RGB-sequence probabilities for target NGC2403-HALO-5.
    }
    \figsetgrpend

    \figsetgrpstart
    \figsetgrpnum{6.12}
    \figsetgrptitle{NGC2976-DEEP}
    \figsetplot{f6_12.pdf}
    \figsetgrpnote{
        CMDs showing individual RGB-sequence probabilities for target NGC2976-DEEP.
    }
    \figsetgrpend

    \figsetgrpstart
    \figsetgrpnum{6.13}
    \figsetgrptitle{NGC3077-PHOENIX}
    \figsetplot{f6_13.pdf}
    \figsetgrpnote{
        CMDs showing individual RGB-sequence probabilities for target NGC3077-PHOENIX.
    }
    \figsetgrpend

    \figsetgrpstart
    \figsetgrpnum{6.14}
    \figsetgrptitle{NGC3741}
    \figsetplot{f6_14.pdf}
    \figsetgrpnote{
        CMDs showing individual RGB-sequence probabilities for target NGC3741.
    }
    \figsetgrpend

    \figsetgrpstart
    \figsetgrpnum{6.15}
    \figsetgrptitle{NGC4163}
    \figsetplot{f6_15.pdf}
    \figsetgrpnote{
        CMDs showing individual RGB-sequence probabilities for target NGC4163.
    }
    \figsetgrpend

\figsetend

%% file: figset_edge.tex
\figsetstart
\figsetnum{7}
\figsettitle{Initial TRGB star selection}

    \figsetgrpstart
    \figsetgrpnum{7.1}
    \figsetgrptitle{DDO71}
    \figsetplot{f7_1.pdf}
    \figsetgrpnote{
        Edge detection and TRGB selection band for target DDO71.
    }
    \figsetgrpend

    \figsetgrpstart
    \figsetgrpnum{7.2}
    \figsetgrptitle{DDO78}
    \figsetplot{f7_2.pdf}
    \figsetgrpnote{
        Edge detection and TRGB selection band for target DDO78.
    }
    \figsetgrpend

    \figsetgrpstart
    \figsetgrpnum{7.3}
    \figsetgrptitle{DDO82}
    \figsetplot{f7_3.pdf}
    \figsetgrpnote{
        Edge detection and TRGB selection band for target DDO84.
    }
    \figsetgrpend

    \figsetgrpstart
    \figsetgrpnum{7.4}
    \figsetgrptitle{ESO540-030}
    \figsetplot{f7_4.pdf}
    \figsetgrpnote{
        Edge detection and TRGB selection band for target ESO540-030.
    }
    \figsetgrpend

    \figsetgrpstart
    \figsetgrpnum{7.5}
    \figsetgrptitle{HS117}
    \figsetplot{f7_5.pdf}
    \figsetgrpnote{
        Edge detection and TRGB selection band for target HS117.
    }
    \figsetgrpend

    \figsetgrpstart
    \figsetgrpnum{7.6}
    \figsetgrptitle{IC2574-SGS}
    \figsetplot{f7_6.pdf}
    \figsetgrpnote{
        Edge detection and TRGB selection band for target IC2574-SGS.
    }
    \figsetgrpend

    \figsetgrpstart
    \figsetgrpnum{7.7}
    \figsetgrptitle{KDG73}
    \figsetplot{f7_7.pdf}
    \figsetgrpnote{
        Edge detection and TRGB selection band for target KDG74.
    }
    \figsetgrpend

    \figsetgrpstart
    \figsetgrpnum{7.8}
    \figsetgrptitle{KKH37}
    \figsetplot{f7_8.pdf}
    \figsetgrpnote{
        Edge detection and TRGB selection band for target KKH37.
    }
    \figsetgrpend

    \figsetgrpstart
    \figsetgrpnum{7.9}
    \figsetgrptitle{M81-DEEP}
    \figsetplot{f7_9.pdf}
    \figsetgrpnote{
        Edge detection and TRGB selection band for target M81-DEEP.
    }
    \figsetgrpend

    \figsetgrpstart
    \figsetgrpnum{7.10}
    \figsetgrptitle{NGC0300}
    \figsetplot{f7_10.pdf}
    \figsetgrpnote{
        Edge detection and TRGB selection band for target NGC0300.
    }
    \figsetgrpend

    \figsetgrpstart
    \figsetgrpnum{7.11}
    \figsetgrptitle{NGC2403-HALO-6}
    \figsetplot{f7_11.pdf}
    \figsetgrpnote{
        Edge detection and TRGB selection band for target NGC2403-HALO-5.
    }
    \figsetgrpend

    \figsetgrpstart
    \figsetgrpnum{7.12}
    \figsetgrptitle{NGC2976-DEEP}
    \figsetplot{f7_12.pdf}
    \figsetgrpnote{
        Edge detection and TRGB selection band for target NGC2976-DEEP.
    }
    \figsetgrpend

    \figsetgrpstart
    \figsetgrpnum{7.13}
    \figsetgrptitle{NGC3077-PHOENIX}
    \figsetplot{f7_13.pdf}
    \figsetgrpnote{
        Edge detection and TRGB selection band for target NGC3077-PHOENIX.
    }
    \figsetgrpend

    \figsetgrpstart
    \figsetgrpnum{7.14}
    \figsetgrptitle{NGC3741}
    \figsetplot{f7_14.pdf}
    \figsetgrpnote{
        Edge detection and TRGB selection band for target NGC3741.
    }
    \figsetgrpend

    \figsetgrpstart
    \figsetgrpnum{7.15}
    \figsetgrptitle{NGC4163}
    \figsetplot{f7_15.pdf}
    \figsetgrpnote{
        Edge detection and TRGB selection band for target NGC4164.
    }
    \figsetgrpend

\figsetend

%% file: figset_tip.tex
\figsetstart
\figsetnum{8}
\figsettitle{TRGB fitting results}

    \figsetgrpstart
    \figsetgrpnum{8.1}
    \figsetgrptitle{DDO71}
    \figsetplot{f8_1.pdf}
    \figsetgrpnote{
        Results of XDGMM TRGB fitting for target DDO71.
    }
    \figsetgrpend

    \figsetgrpstart
    \figsetgrpnum{8.2}
    \figsetgrptitle{DDO78}
    \figsetplot{f8_2.pdf}
    \figsetgrpnote{
        Results of XDGMM TRGB fitting for target DDO78.
    }
    \figsetgrpend

    \figsetgrpstart
    \figsetgrpnum{8.3}
    \figsetgrptitle{DDO82}
    \figsetplot{f8_3.pdf}
    \figsetgrpnote{
        Results of XDGMM TRGB fitting for target DDO83.
    }
    \figsetgrpend

    \figsetgrpstart
    \figsetgrpnum{8.4}
    \figsetgrptitle{ESO540-030}
    \figsetplot{f8_4.pdf}
    \figsetgrpnote{
        Results of XDGMM TRGB fitting for target ESO540-030.
    }
    \figsetgrpend

    \figsetgrpstart
    \figsetgrpnum{8.5}
    \figsetgrptitle{HS117}
    \figsetplot{f8_5.pdf}
    \figsetgrpnote{
        Results of XDGMM TRGB fitting for target HS117.
    }
    \figsetgrpend

    \figsetgrpstart
    \figsetgrpnum{8.6}
    \figsetgrptitle{IC2574-SGS}
    \figsetplot{f8_6.pdf}
    \figsetgrpnote{
        Results of XDGMM TRGB fitting for target IC2574-SGS.
    }
    \figsetgrpend

    \figsetgrpstart
    \figsetgrpnum{8.7}
    \figsetgrptitle{KDG73}
    \figsetplot{f8_7.pdf}
    \figsetgrpnote{
        Results of XDGMM TRGB fitting for target KDG73.
    }
    \figsetgrpend

    \figsetgrpstart
    \figsetgrpnum{8.8}
    \figsetgrptitle{KKH37}
    \figsetplot{f8_8.pdf}
    \figsetgrpnote{
        Results of XDGMM TRGB fitting for target KKH37.
    }
    \figsetgrpend

    \figsetgrpstart
    \figsetgrpnum{8.9}
    \figsetgrptitle{M81-DEEP}
    \figsetplot{f8_9.pdf}
    \figsetgrpnote{
        Results of XDGMM TRGB fitting for target M81-DEEP.
    }
    \figsetgrpend

    \figsetgrpstart
    \figsetgrpnum{8.10}
    \figsetgrptitle{NGC0300}
    \figsetplot{f8_10.pdf}
    \figsetgrpnote{
        Results of XDGMM TRGB fitting for target NGC0300.
    }
    \figsetgrpend

    \figsetgrpstart
    \figsetgrpnum{8.11}
    \figsetgrptitle{NGC2403-HALO-6}
    \figsetplot{f8_11.pdf}
    \figsetgrpnote{
        Results of XDGMM TRGB fitting for target NGC2403-HALO-5.
    }
    \figsetgrpend

    \figsetgrpstart
    \figsetgrpnum{8.12}
    \figsetgrptitle{NGC2976-DEEP}
    \figsetplot{f8_12.pdf}
    \figsetgrpnote{
        Results of XDGMM TRGB fitting for target NGC2976-DEEP.
    }
    \figsetgrpend

    \figsetgrpstart
    \figsetgrpnum{8.13}
    \figsetgrptitle{NGC3077-PHOENIX}
    \figsetplot{f8_13.pdf}
    \figsetgrpnote{
        Results of XDGMM TRGB fitting for target NGC3077-PHOENIX.
    }
    \figsetgrpend

    \figsetgrpstart
    \figsetgrpnum{8.14}
    \figsetgrptitle{NGC3741}
    \figsetplot{f8_14.pdf}
    \figsetgrpnote{
        Results of XDGMM TRGB fitting for target NGC3741.
    }
    \figsetgrpend

    \figsetgrpstart
    \figsetgrpnum{8.15}
    \figsetgrptitle{NGC4163}
    \figsetplot{f8_15.pdf}
    \figsetgrpnote{
        Results of XDGMM TRGB fitting for target NGC4163.
    }
    \figsetgrpend

\figsetend

%% file: 4_results.tex
\section{Results} \label{sec:results}

\subsection{Apparent TRGB magnitudes and colors}

In this section we compare the TRGB apparent magnitudes and colors we have measured using the techniques developed in this paper to those used in \citetalias{2012ApJS..198....6D}. All revised apparent magnitudes and errors are reported in \autoref{tab:apparent}.

\begin{figure}[ht] 
\epsscale{1.1}
\plotone{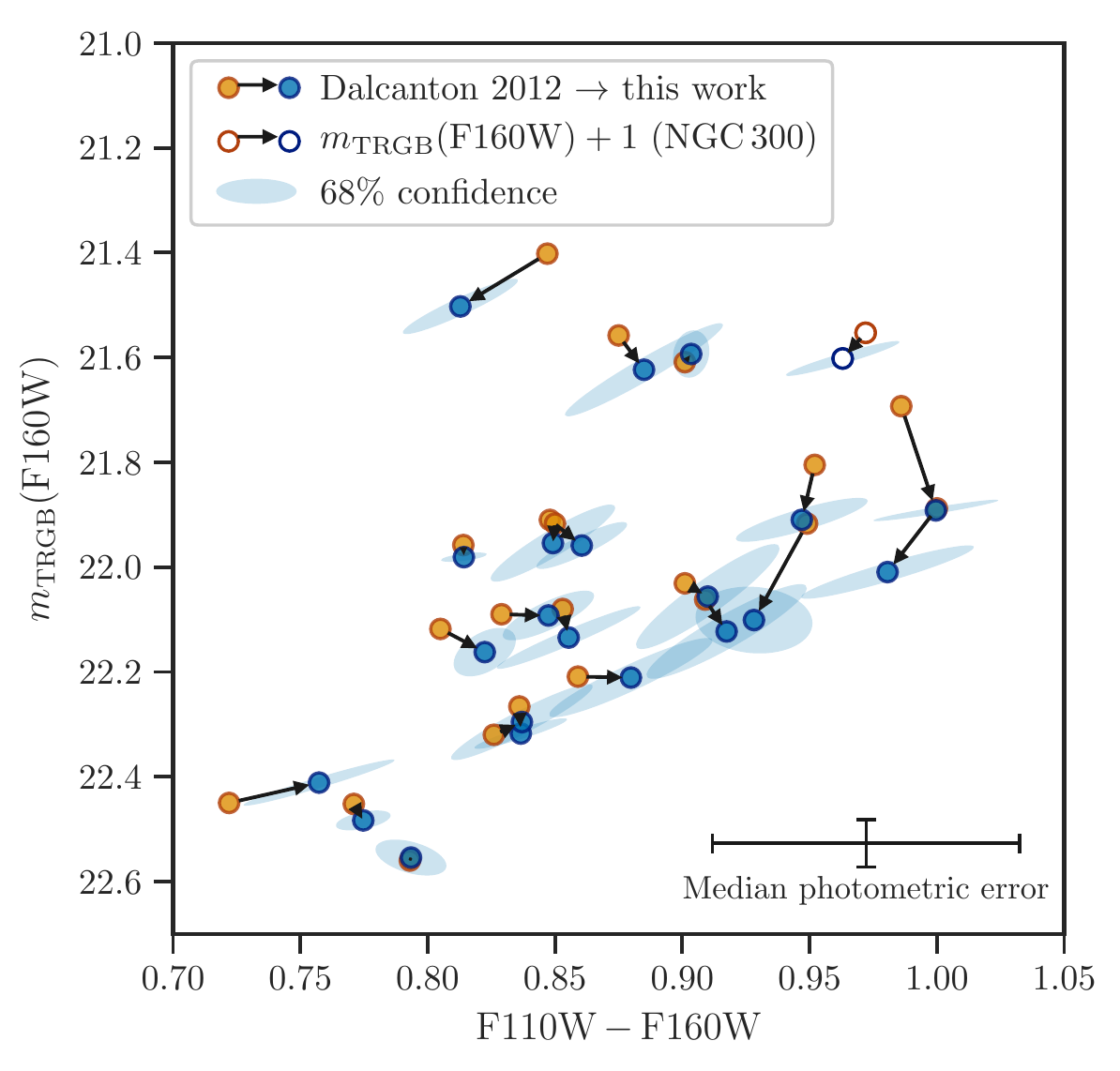}
\caption{Comparison of the revised tip F160W apparent magnitudes and F110W -- F160W colors from this work (blue points) to those of \citetalias{2012ApJS..198....6D} (orange points).  Arrows indicate per-target correspondence between \citetalias{2012ApJS..198....6D} and the new measurements. The blue ellipses show the 68\% confidence regions on the measurements of this paper from XDGMM fitting, and the bottom right errorbars indicate the median photometric uncertainties in color and magnitude for an individual star. For NGC$\,300$ (unfilled points) we plot $M_{\rm F160W} + 1$ rather than $M_{\rm F160W}$, as it is $\sim\!1$ mag brighter than the remainder of the sample. 
On average, our mean color-magnitude tip results are redder and slightly fainter than  \citetalias{2012ApJS..198....6D}. 
}
\label{fig:nir_apparent_changes}
\end{figure}

First, \autoref{fig:nir_apparent_changes} compares the change in apparent F160W magnitude and F110W--F160W color between this work (blue points) and \citetalias{2012ApJS..198....6D} (orange points) for each target in our sample.
The 68\% confidence intervals are shown for our measurements and demonstrate that the difference between this work and \citetalias{2012ApJS..198....6D} is almost always larger than our measurement uncertainties, albeit, as shown in the lower right, most are within the color-magnitude photometric error circle for an individual source at the tip.

\begin{figure}[ht]
\epsscale{1.1}
\plotone{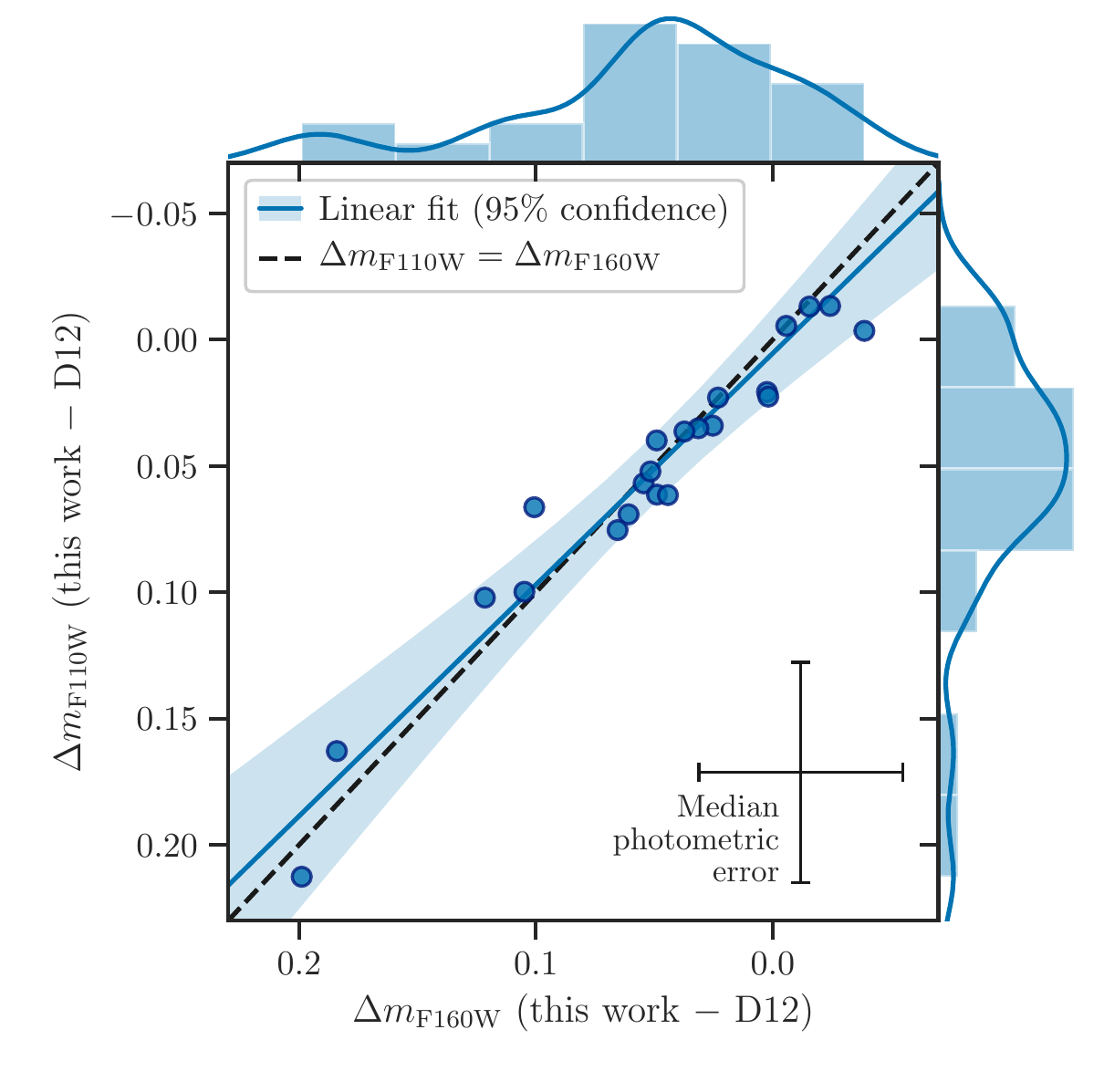}
\plotone{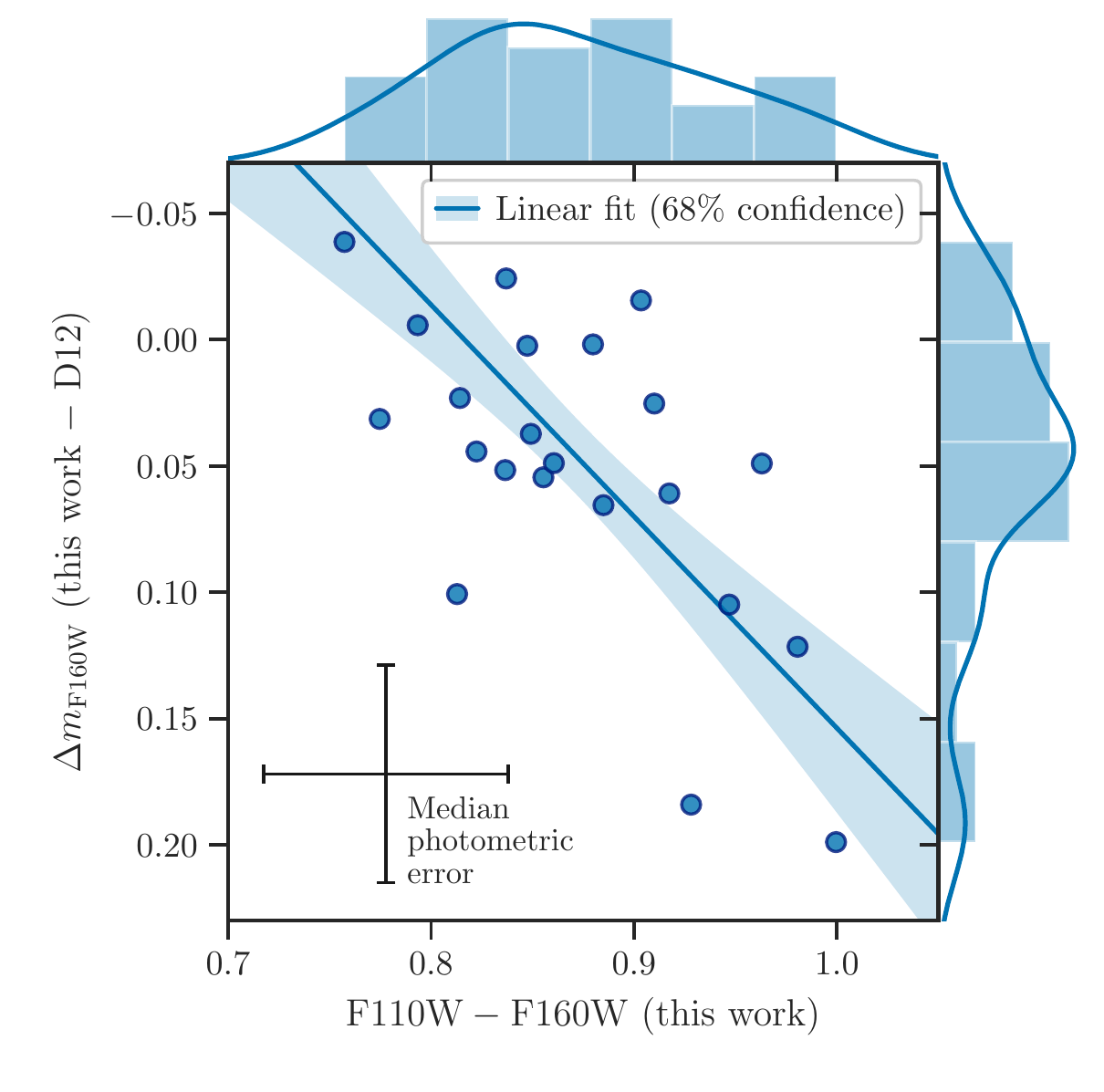}
\caption{Top: changes in the $M_{\rm F160W}$ tip magnitudes compared to changes in the $M_{\rm F110W}$ TRGB magnitudes between this work and \citetalias{2012ApJS..198....6D}.
Bottom: changes in the $M_{\rm F160W}$ tip magnitudes between this work and \citetalias{2012ApJS..198....6D} against the NIR color measured in this work.
Both panels show histograms with overlaid kernel density estimates of the marginal distributions of the quantities on each axis, linear fits with shaded confidence intervals, and scale bars with typical photometric uncertainties.
}
\label{fig:dF110W_dF160W_correlation}
\end{figure}

The origin of these offsets can be determined by comparing the individual differences between the photometry. 
\autoref{fig:dF110W_dF160W_correlation} compares the relative change between the measurements of this work and that of \citetalias{2012ApJS..198....6D} for the F110W (x-axis) and F160W (y-axis). 
For both $\Delta m_{\rm F160W}$ and $\Delta m_{\rm F110W}$ (defined as this work minus \citetalias{2012ApJS..198....6D}), the median difference is approximately +0.05 mag; histograms are shown in \autoref{fig:dF110W_dF160W_correlation} on each axis. 
Interestingly, the offsets are highly correlated; in the upper panel, a one-to-on line is shown in black-dashed line with a fit to the results given in blue solid, and the 95\% confidence interval (shown in the shaded region) encompasses the one-to-one line.

\begin{figure*}[ht]
\plotone{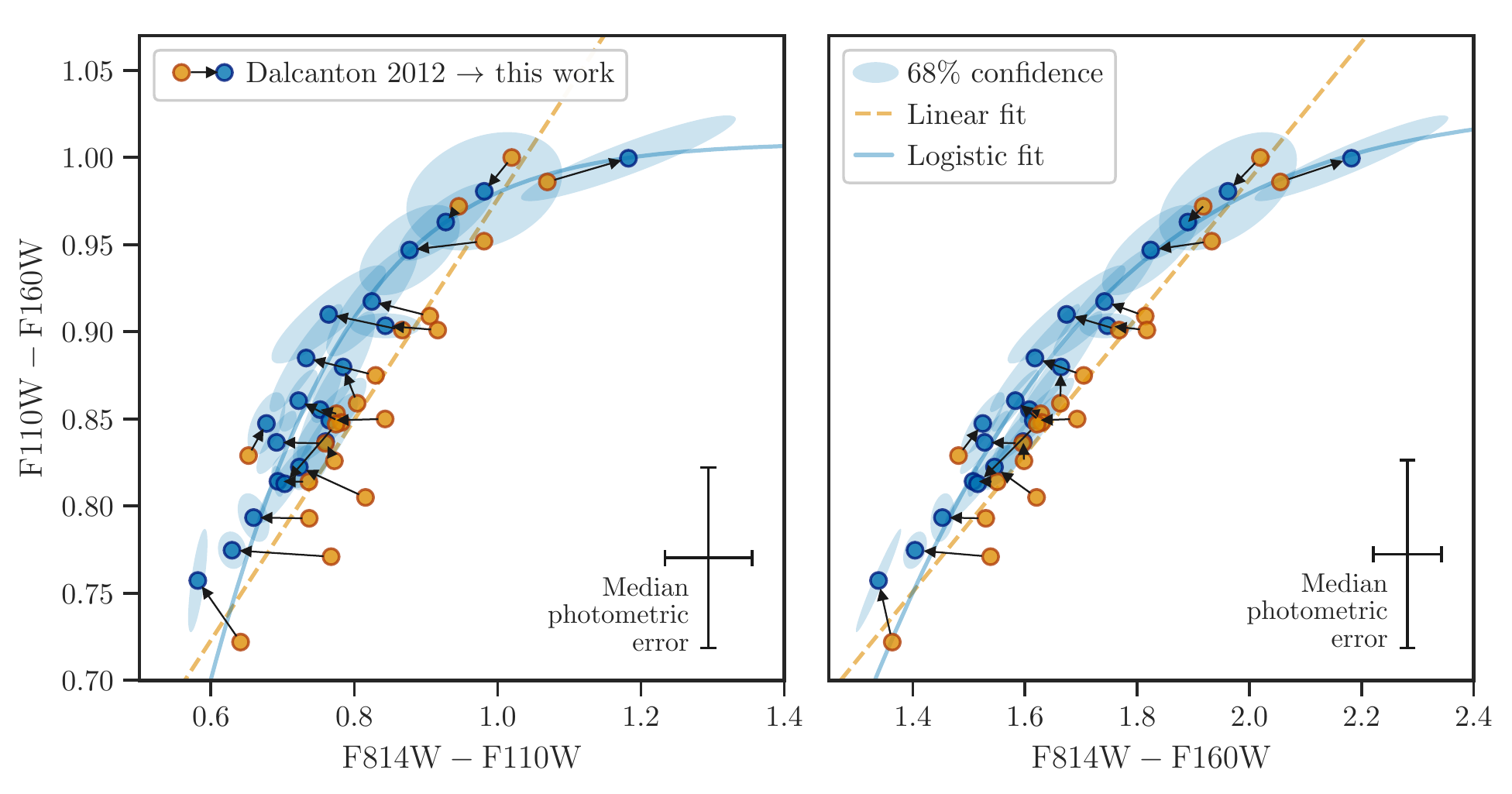}
\caption{Comparison of \citetalias{2012ApJS..198....6D} color-color relation to the TRGB color-color relation determined in this work, with F110W -- F160W against F814W--F110W (left) and F814W--F160W (right). As in \autoref{fig:nir_apparent_changes}, orange points are from \citetalias{2012ApJS..198....6D}, blue points are from this work,  arrows connect the corresponding results, and the blue shading indicates our two-dimensional uncertainties from XDGMM. We show a linear fit to the \citetalias{2012ApJS..198....6D} values (as done in that work) and a generalized logistic fit to values from this work to highlight the changes in morphology in our new color-color relations.}
\label{fig:color_color}
\end{figure*}

\autoref{fig:color_color} displays the F814W--F110W (left) and F814W--F160W (right) to F110W--F160W color-color diagrams for the results of this work (blue) compared to that of \citetalias{2012ApJS..198....6D} (orange). 
Relative to \citetalias{2012ApJS..198....6D}, the measurements from this work move the color-color relations to the left in this diagram -- bluer in F814W--F110W and F814W--F160W and slightly redder in F110W--F160W. 

In \autoref{fig:color_color} we provide reference lines to highlight the color behavior, using a linear function for the \citetalias{2012ApJS..198....6D} photometry and a logistic function 
for our new photometry. (We caution that these fitting relations should not be taken as physically meaningful.)

\input{table_apparent.tex}

\subsection{The TRGB color-absolute magnitude relation}

To derive the color dependence of the NIR TRGB absolute magnitude, we must adjust the apparent magnitudes in \autoref{fig:nir_apparent_changes} by the appropriate distance modulus for each galaxy.

We first present a revised NIR color-absolute magnitude relation adopting the same distances as in \citetalias{2012ApJS..198....6D}, and then explore the use of the most up-to-date stellar models to derive revised distance moduli and absolute magnitudes. 

\subsubsection{Adopting D12 distances}
The distance moduli used in \citetalias{2012ApJS..198....6D} were detemined using the F814W TRGB, which enables a fully self-consistent study of the TRGB across bandpasses. With the exception of NGC\,7793, these distances were originally published in \citet{2009ApJS..183...67D}, whereas the distance for NGC\,7793 is from \citet{2003A&A...404...93K}, which also uses the F814W TRGB. 
The absolute calibration of the F814W (ground-based $I$-band) TRGB at \textless\ 5\% precision is unclear
\citep{2017ApJ...835...28J, 2018SSRv..214..113B, 2019ApJ...882...34F, 2019ApJ...886...61Y, 2019ApJ...886L..27R, 2020ApJ...891...57F}.
Historically, it has been assumed to be a constant value of approximately $M^I_{\rm TRGB} \sim -4.05$ mag \citep{1993ApJ...417..553L, 1997MNRAS.289..406S}. However, this magnitude is only anticipated to be roughly constant for uniformly old and metal-poor populations \citep{2018SSRv..214..113B, 2017A&A...606A..33S, 2006essp.book.....S}; more specifically, [M/H] \textless{} $-0.5$ dex and \textgreater{} 4 Gyr. 

The stellar populations in the  \citetalias{2012ApJS..198....6D} sample, however, span a wide range of ages and metallicities that preclude the assumption of a single value for the TRGB F814W luminosity.
Rather than adopting a single value for $M_{\rm TRGB}^{\rm F814W}$, \citet{2009ApJS..183...67D} used the mean optical colors of stars within 0.2 mag of the apparent F814W TRGB to choose fiducial \citet{2008PASP..120..583G} isochrones with corresponding colors. \citet{2009ApJS..183...67D} then determined the predicted F814W TRGB absolute magnitude for each galaxy from the isochrone sets,  subtracted that from their measured F814W TRGB apparent magnitudes, and corrected for foreground extinction to obtain their distance moduli.

\begin{figure}[ht]
    \plotone{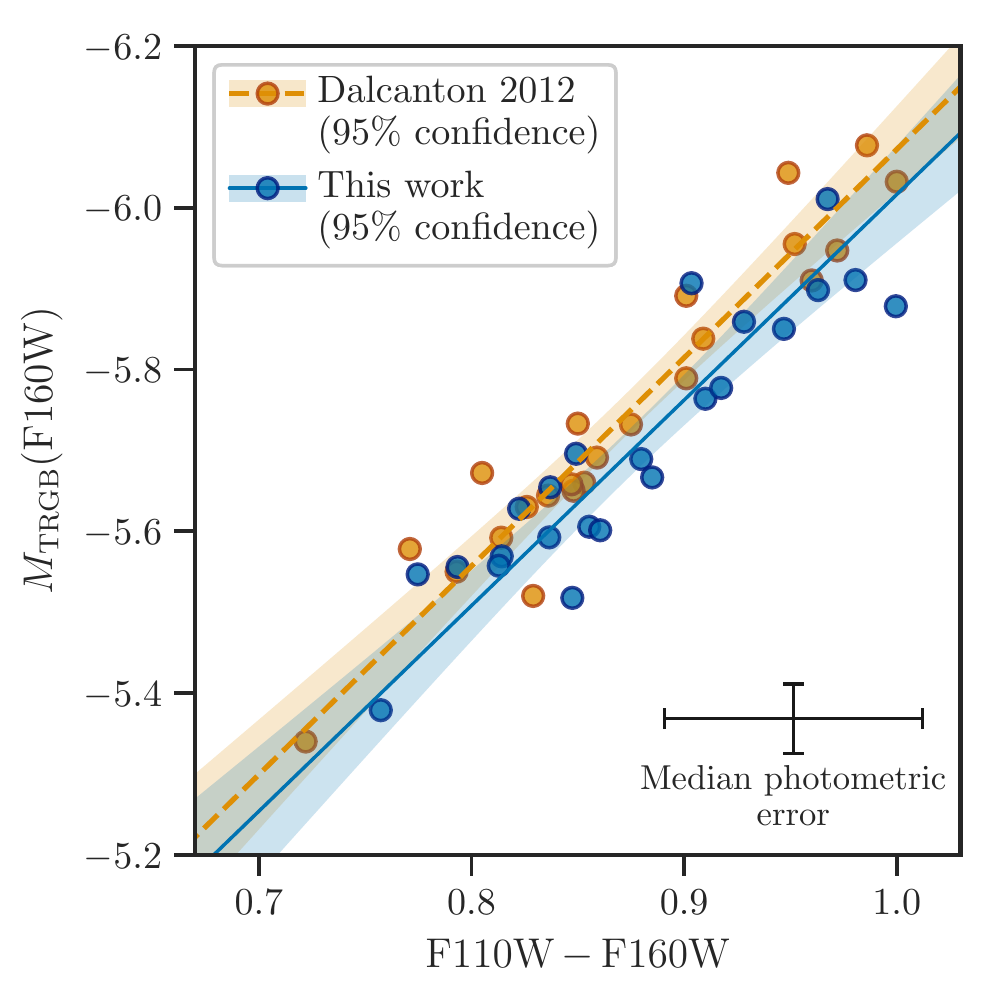}
    \caption{Comparison of revised NIR color-absolute magnitude relations to \citetalias{2012ApJS..198....6D} with revised absolute magnitudes derived using the same distances as in \citetalias{2012ApJS..198....6D}. Blue points are values from the current work and orange points are from \citetalias{2012ApJS..198....6D}. The corresponding color-coded lines show linear fits to each dataset and the shaded regions show 95\% confidence intervals. Again, we see that our results are slightly redder and slightly fainter than \citetalias{2012ApJS..198....6D} using their distances. 
    This color-color relation is distance-independent.}
\label{fig:d12_absolute_changes}
\end{figure}

We calculate NIR TRGB absolute magnitudes by subtracting the \citetalias{2012ApJS..198....6D} distance moduli from our apparent tip magnitudes, as reported in \autoref{tab:abs_d12}.
The resulting NIR absolute magnitudes and color-magnitude relation are compared to that of \citetalias{2012ApJS..198....6D} in \autoref{fig:d12_absolute_changes}. 

We fit a linear relation to the absolute F160W magnitudes and F110W--F160W colors determined in this work using orthogonal distance regression \citep[ODR,][]{odr}, and find:
\begin{equation}\label{eq:d12_recalib}
  M_{\rm TRGB}^{\rm F160W} = -2.541({\rm F110W}-{\rm F160W})-3.475
\end{equation}

The uncertainties on our slope and zeropoint are 0.057 mag color$^{-1}$ and 0.050 mag, respectively. Compared to the equivalent fit from \citetalias{2012ApJS..198....6D} (their eq.~1),
\begin{equation}\label{eq:d12_calib}
  M_{\rm TRGB}^{\rm F160W} = -2.576({\rm F110W}-{\rm F160W})-3.496,
\end{equation}
we find an 0.02 mag fainter zero-point (\textless{} 1\% in distance) and a change in the slope of less than 0.04 mag color$^{-1}$, both of which are well within our uncertainties.
As expected, the difference in the zeropoint is roughly equivalent to the differences in measured TRGB photometry observed in \autoref{fig:nir_apparent_changes}.

\input{table_abs_d12.tex}

\subsubsection{Recalibrating Distances to Recent Models} \label{sec:recalib_d12dist}

Both the physical isochrones and the filter transformations described in \citet{2008PASP..120..583G} have undergone many revisions in the intervening years \citep{2012MNRAS.427..127B, 2017ApJ...835...77M}, and thus the F814W TRGB zeropoints adopted in \citetalias{2009ApJS..183...67D} and \citetalias{2012ApJS..198....6D} may no longer be appropriate.
Here we apply a similar distance estimation method as in \citetalias{2009ApJS..183...67D} to our revised measurements, using \replaced{isochrones}{synthetic photometry} from the \deleted{most recent} model suites \deleted{from} PARSEC \added{v.~1.2S\footnote{\url{http://stev.oapd.inaf.it/cgi-bin/cmd_3.3}}} \citep{2012MNRAS.427..127B, 2017ApJ...835...77M} and MIST \added{v.~1.2\footnote{\url{http://waps.cfa.harvard.edu/MIST/index.html}}} \citep{2016ApJ...823..102C}, both of which are used routinely for stellar populations work.
\added{We retrieved the synthetic photometry directly from the cited web services.}
For both sets, we use isochrones with ages spanning 8 to 14 Gyr with $\log({\rm age})$ spacing of 0.05 dex. The PARSEC metallicities span $-2.2 \leq [{\rm Fe/H}] \leq 0$ dex with a spacing of 0.1 dex, whereas the MIST metallicities span $-2.0 \leq [{\rm Fe/H}] \leq 0$ dex with a spacing of 0.25 dex.
\added{Both model suites use scaled-solar abundances, albeit with slightly different calibrations ($Z_{\odot} = 0.0152$ and $Y_{\odot} = 0.275556$ for PARSEC, and $Z_{\odot} = 0.0142$ and $Y_{\odot} = 0.2703$ for MIST).} 
\replaced{For each model, we calculate colors and adopt absolute magnitudes for the TRGB as tagged in the model sets.}{We use the evolutionary phase tags in each model set to select the predicted TRGB at each age/metallicity combination.}

\begin{figure*}[ht]
\plotone{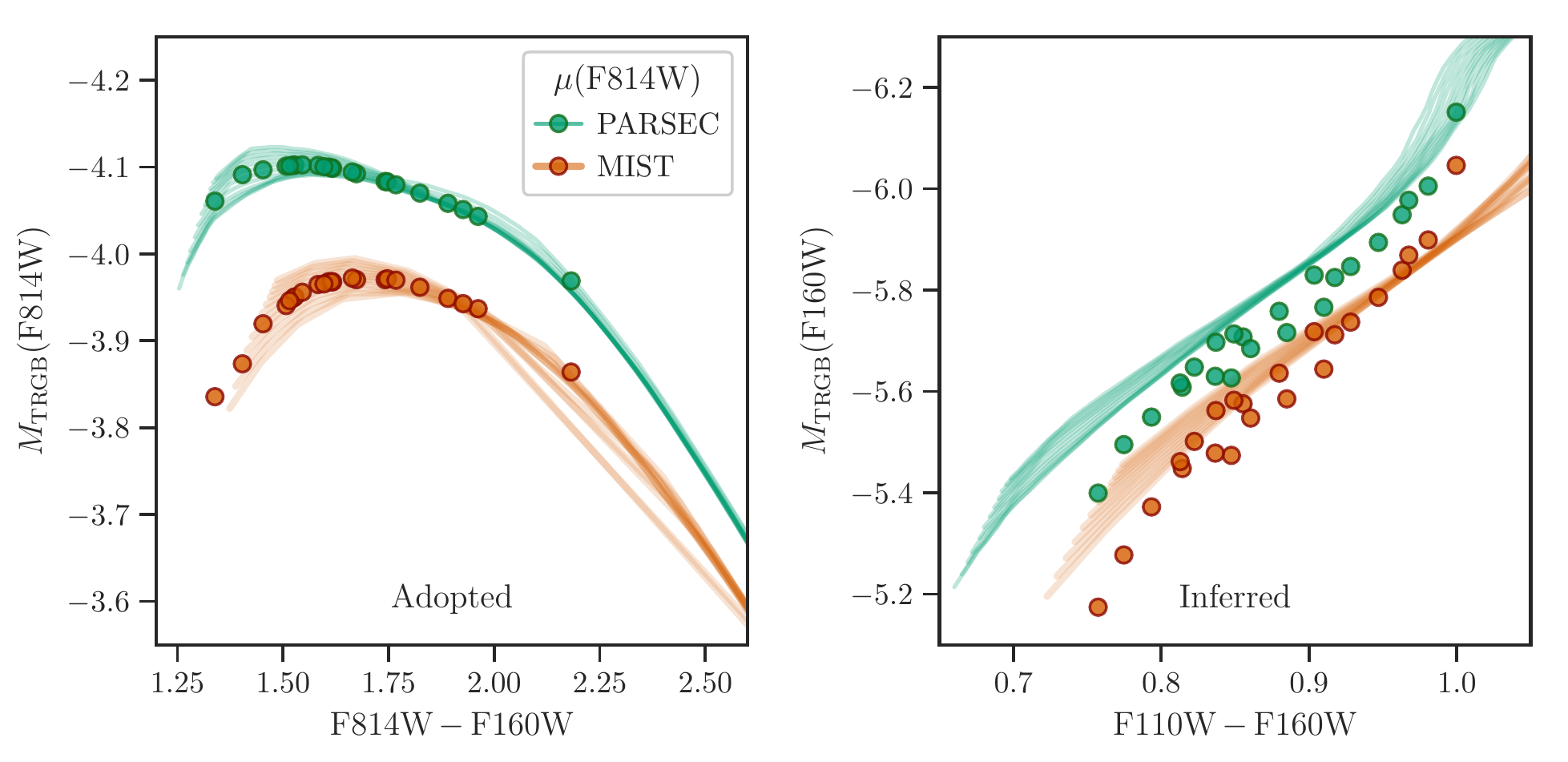}
\caption{
Revised absolute tip magnitudes using distance moduli calibrated to $M_{\rm F814W}$ versus F814W--F160W derived from \replaced{theoretical isochrones}{synthetic photometry} from the MIST (orange) and PARSEC (green) model suites. Each solid line represents a set of theoretical tip star colors and absolute magnitudes at a single age.
In the left panel, we tie the observed colors to an absolute magnitude in either isochrone set to determine $\mu_{\rm F814W}$.
On the right, we use $\mu_{\rm F814W}$ to determine $M_{\rm TRGB}^{\rm F160W}$ and plot against our F110W--F160W color; we find that these measurements are systematically offset from the corresponding isochrone models. }
\label{fig:newdistances_F814W}
\end{figure*}

For each of these model sets, we estimate new sets of distance moduli using two color-magnitude combinations: (i) F814W vs.\ F814W--F160W and (ii) F160W vs.\ F110W--F160W.
For most of our color measurements, there are multiple isochrones with TRGB colors that fall within the measurement uncertainties, each with slightly different absolute TRGB magnitudes.
We calculate a fiducial tip absolute magnitude for each galaxy by taking the weighted mean of the isochrone absolute magnitudes, where the weights are defined by a Gaussian with a center at the measured tip color and a width from the color-uncertainty. 
Derived distance moduli for all color-magnitude combinations and model sets are reported in \autoref{tab:abs_new}.

\begin{figure*}[ht]
\plotone{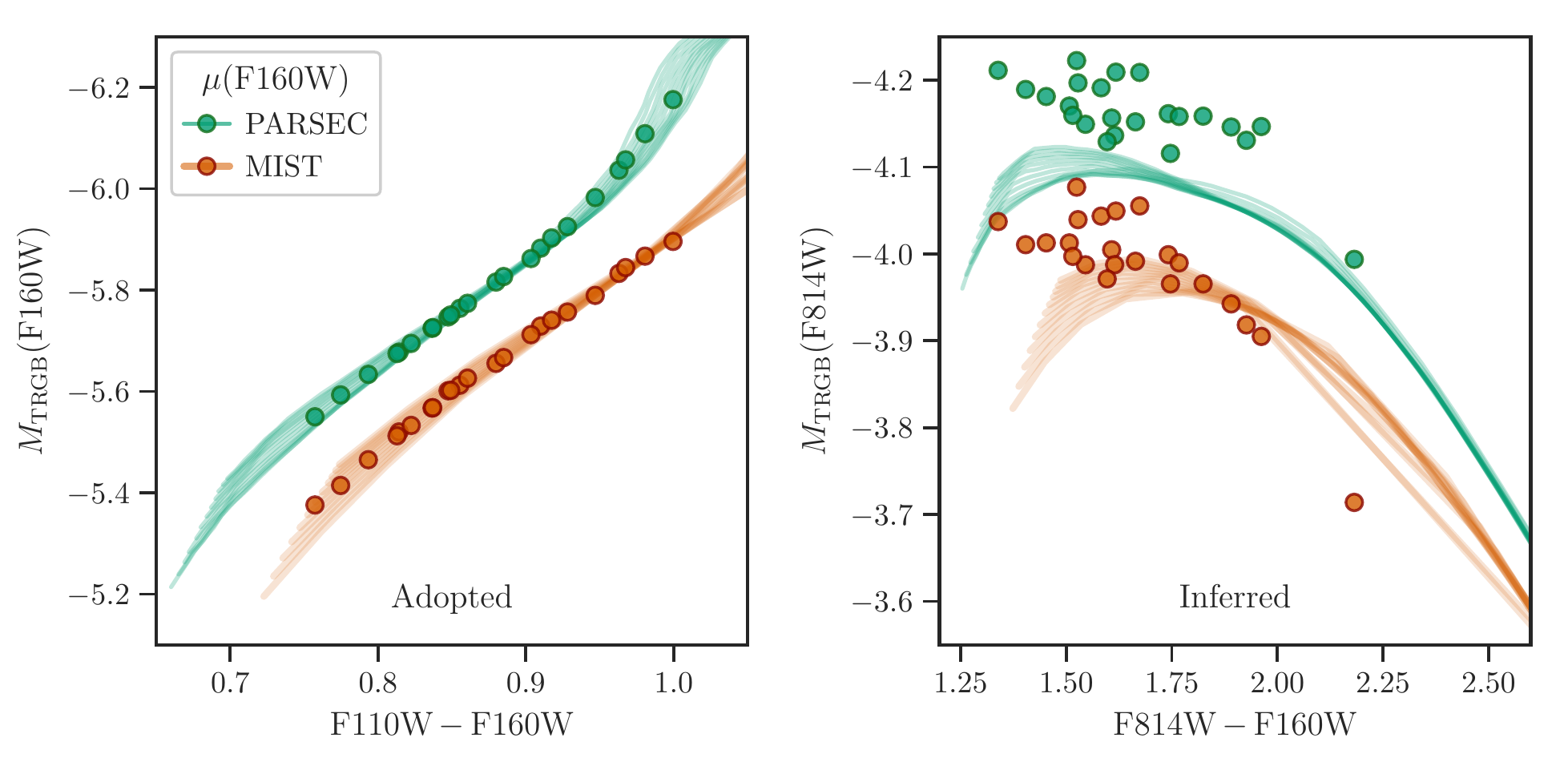}
\caption{We invert the demonstration of \autoref{fig:newdistances_F814W}. 
In the left panel, we tie the observed colors to an absolute magnitude in either isochrone set to determine $\mu_{\rm F160W}$.
On the right, we use $\mu_{\rm F160W}$ to compute $M_{\rm TRGB}^{\rm F814W}$ and plot against our F814W--F160W color; we again find that our measurements are systematically offset from the corresponding isochrone models.
Taken with \autoref{fig:newdistances_F814W}, this suggests that distance moduli calibrated to models in one band will systematically mispredict the corresponding tip behavior in other bands. }
\label{fig:newdistances_F160W}
\end{figure*}

The results of this procedure are shown in \autoref{fig:newdistances_F814W} where the left panel shows the adopted values of $M_{\rm TRGB}$(F814W)--(F814W-F160W) and the right panel shows the inferred values of $M_{\rm TRGB}$(F160W)--(F110W-F160W).
In \replaced{either}{each} panel, \replaced{each unique color-metallicity isochrone}{sets of synthetic photometry at single ages (8-13 Gyr), with metallicities spanning $-2.0 < [Fe/H] < -0.25$~dex,} are plotted as transparent lines, with PARSEC models plotted in green and MIST models in orange.
Overall, the mag-color behavior of the two sets is qualitatively similar, but the absolute magnitudes differ by $\Delta M_{\rm F814W}~\sim~0.15$ mag, with PARSEC being brighter than MIST for the same color ($\sim$8\% in distance).

The fits of our data to the \added{predicted} $M_{\rm TRGB}$(F814W)--(F814W-F160W) \deleted{model} distributions are shown as the points in the panels of \autoref{fig:newdistances_F814W}. 
From this, we determine a distance modulus to each galaxy, which we denote as $\mu_{\rm F814W}$. 
In the right panel, we use $\mu_{\rm F814W}$ to translate $m_{\rm TRGB}$(F160W) to $M_{\rm TRGB}$(F160W), and compare these values to the same isochrone sets used to derive $\mu_{\rm F814W}$.
The PARSEC-based distances place the observed NIR TRGB $\sim0.05$ mag fainter than \replaced{models}{predicted}, but they do trace the same underlying variations with color. In contrast, the MIST\replaced{isochrones}{-derived values} are less offset overall in magnitude, but show shape deviations that become particularly pronounced at red colors.

\autoref{fig:newdistances_F160W} repeats this process in reverse by determining a distance modulus, $\mu_{\rm F160W}$, based off of the $M_{\rm TRGB}$(F160W)--(F110W-F160W) model predictions (left panel) and then comparing the absolute $M_{\rm TRGB}$(F814W)--(F814W-F160W) empirical relationship to the models in the right panel. 
In this case, different behavior is observed: for both model sets, $M_{\rm TRGB}$(F814W) from our data are too bright by $\sim$0.1~mag at the blue (low-metallicity) end, and there is a slight color offset of $\sim$0.05~mag. 

Overall \autoref{fig:newdistances_F814W} and \autoref{fig:newdistances_F160W} suggest that the isochrone predictions are inconsistent both with each other and, when used internally to predict multi-band magnitudes, with our measurements. \added{We will further discuss potential reasons for these apparent inconsistencies in \autoref{ssec:model_discussion}.}

\subsubsection{New Distance Moduli Compared to D12}

\begin{figure}[ht]
\plotone{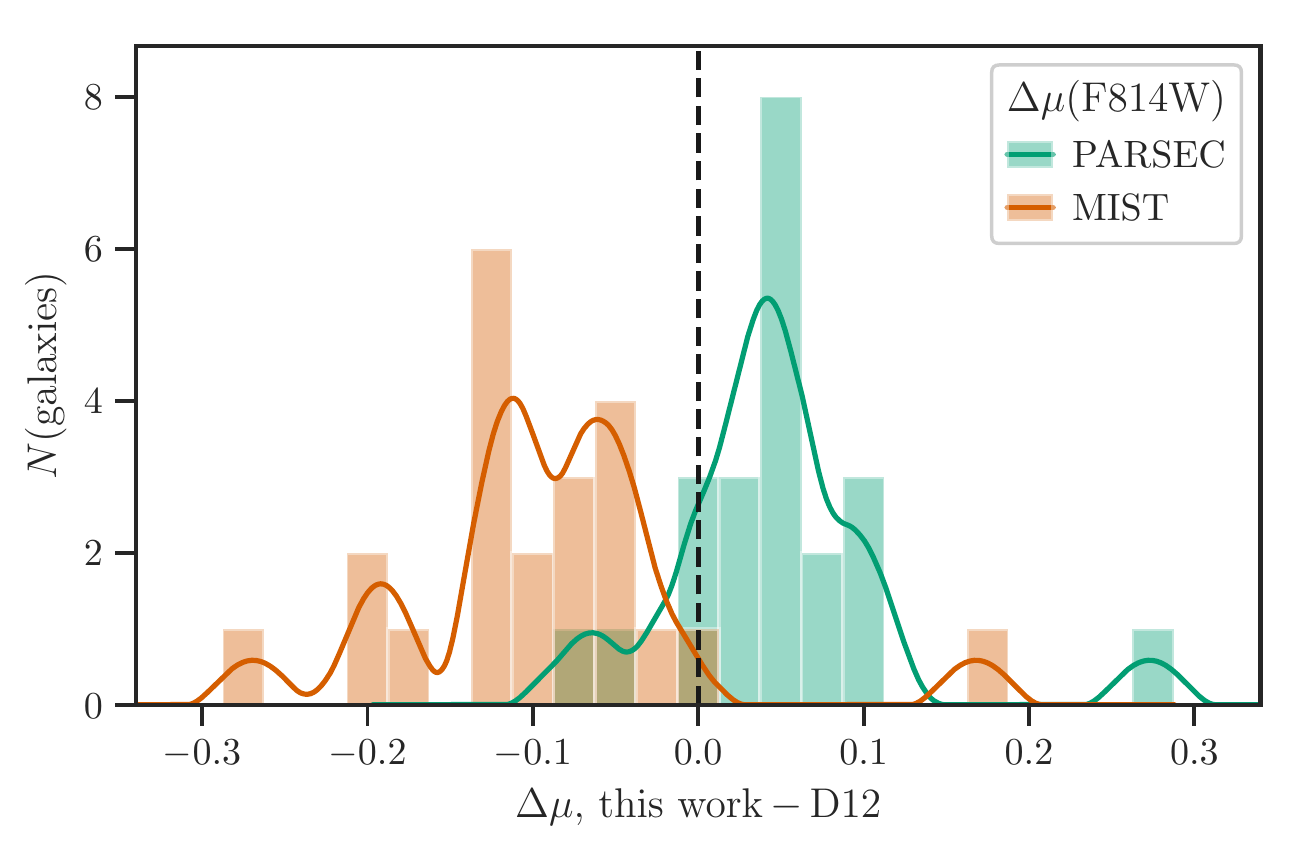}
\plotone{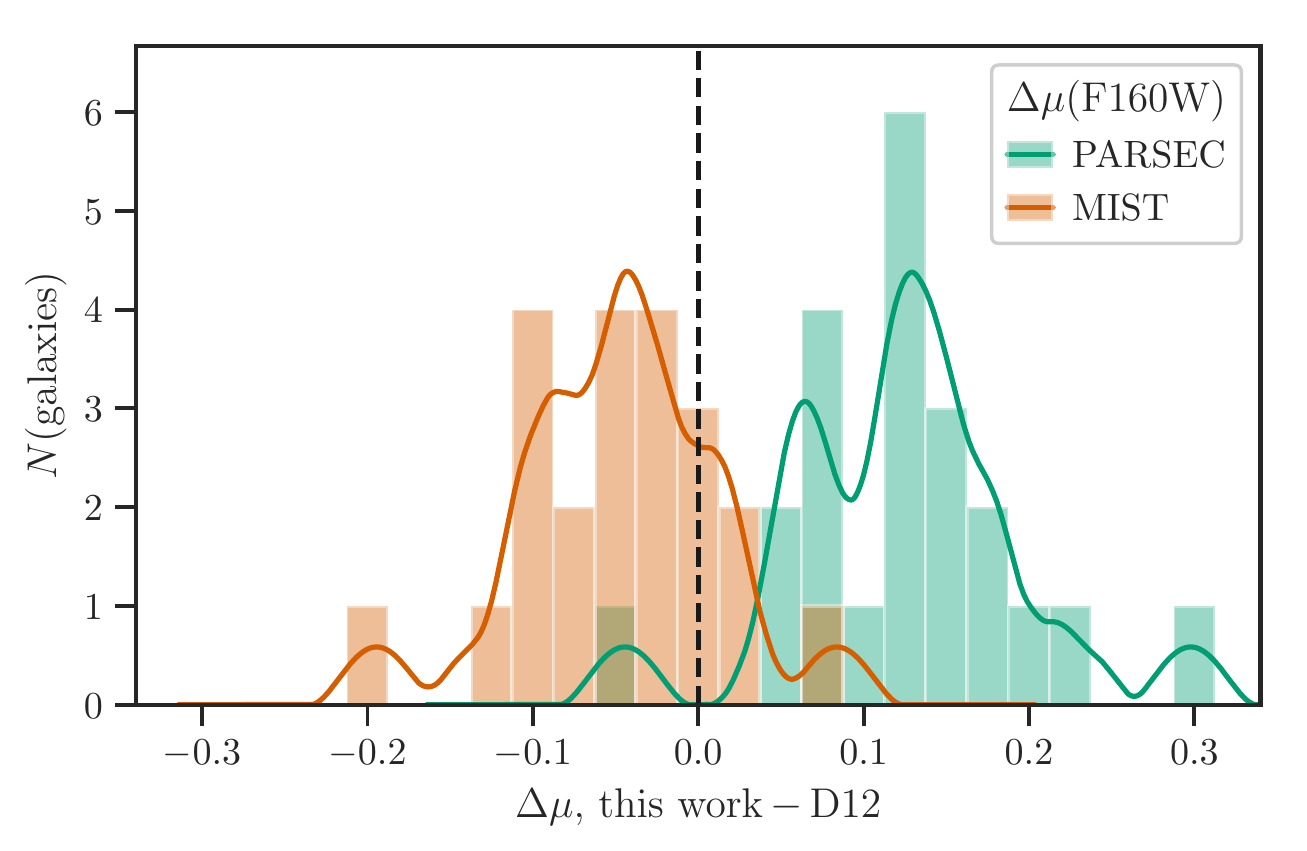}
\caption{Histograms and overlaid biweight kernel density estimates of $\Delta \mu$, where $\Delta \mu$ is the difference between distance moduli derived in this work and the distance moduli used in \citetalias{2012ApJS..198....6D}. The top panel shows $\Delta \mu$ using distance moduli calibrated to $M_{\rm F814W}$ and the bottom shows $\Delta \mu$ using distance moduli calibrated to $M_{\rm F160W}$.
These differences are much larger than those measured using the same distances (\autoref{fig:d12_absolute_changes}) and thus can be interpreted as differences between model predictions of tip magnitudes. }
\label{fig:delta_dmod}
\end{figure}

\autoref{fig:delta_dmod} compares the distance moduli determined in the previous subsection to those from \citetalias{2012ApJS..198....6D}.
The top panel compares the distances calibrated to F814W (the same filter in either case) and the bottom panel compares the distances from F160W. 
No difference ($\Delta\mu=0$ mag) is indicated by the vertical dashed line. 
In both cases, the PARSEC-based calibration is systematically larger than in  \citetalias{2012ApJS..198....6D} by median values of $0.043(\pm0.069)$ mag in F814W and $0.126(\pm 0.071)$ mag in F160W, corresponding to $2(\pm3)$\% and $6(\pm3)$\% greater distances, respectively. The MIST calibrations, on the other hand, are systematically smaller, with median differences of $-0.096(\pm0.084)$ mag in F814W and $-0.045(\pm0.063)$ mag in F160W ($4(\pm4)$\% and $2(\pm3)$\% closer respectively).

\input{table_new_distances.tex}

%% file: table_apparent.tex
\begin{table*}[!ht]
\caption{Apparent TRGB magnitudes}
\hskip-0.5in
\resizebox{1\textwidth}{!}{
\begin{tabular}{lcccccccccr}
\hline
\hline
 & \multicolumn{3}{c}{F814W} & \multicolumn{3}{c}{F110W} & \multicolumn{3}{c}{F160W} & \\
Target &  $m$ &  $\sigma_{\rm fit}$ &  $\sigma_{\rm phot}$ &  $m$ &  $\sigma_{\rm fit}$ &  $\sigma_{\rm phot}$ &  $m$ &  $\sigma_{\rm fit}$ &  $\sigma_{\rm phot}$ &  $N_\star$ \\
\hline
DDO71           &           23.742 &                           0.002 &                            0.015 &           22.990 &                           0.022 &                            0.047 &           22.134 &                           0.040 &                            0.048 &      1478 \\
DDO78           &           23.730 &                           0.005 &                            0.032 &           22.966 &                           0.050 &                            0.042 &           22.056 &                           0.066 &                            0.042 &      1948 \\
DDO82           &           23.864 &                           0.005 &                            0.044 &           23.040 &                           0.041 &                            0.051 &           22.123 &                           0.060 &                            0.051 &      4525 \\
ESO540-030      &           23.617 &                           0.007 &                            0.036 &           22.940 &                           0.023 &                            0.039 &           22.092 &                           0.031 &                            0.043 &       631 \\
HS117           &           23.845 &                           0.011 &                            0.036 &           23.154 &                           0.009 &                            0.045 &           22.318 &                           0.019 &                            0.040 &       592 \\
IC2574-SGS      &           23.875 &                           0.005 &                            0.040 &           23.091 &                           0.031 &                            0.062 &           22.211 &                           0.050 &                            0.057 &      5227 \\
KDG73           &           23.887 &                           0.021 &                            0.031 &           23.258 &                           0.023 &                            0.046 &           22.483 &                           0.024 &                            0.052 &       312 \\
KKH37           &           23.542 &                           0.004 &                            0.032 &           22.819 &                           0.020 &                            0.044 &           21.959 &                           0.030 &                            0.048 &       902 \\
M81-DEEP        &           24.074 &                           0.103 &                            0.013 &           22.892 &                           0.021 &                            0.039 &           21.892 &                           0.024 &                            0.035 &       551 \\
NGC0300         &           22.493 &                           0.043 &                            0.017 &           21.565 &                           0.009 &                            0.022 &           20.602 &                           0.022 &                            0.022 &      1350 \\
NGC2403-HALO-6  &           23.340 &                           0.020 &                            0.027 &           22.497 &                           0.036 &                            0.027 &           21.593 &                           0.036 &                            0.028 &       378 \\
NGC2976-DEEP    &           23.734 &                           0.055 &                            0.027 &           22.857 &                           0.016 &                            0.044 &           21.910 &                           0.028 &                            0.040 &      1771 \\
NGC3077-PHOENIX &           23.972 &                           0.080 &                            0.016 &           22.990 &                           0.015 &                            0.044 &           22.010 &                           0.034 &                            0.041 &      1136 \\
NGC3741         &           23.488 &                           0.004 &                            0.034 &           22.795 &                           0.005 &                            0.043 &           21.981 &                           0.006 &                            0.043 &       798 \\
NGC4163         &           23.241 &                           0.007 &                            0.030 &           22.508 &                           0.039 &                            0.038 &           21.623 &                           0.059 &                            0.039 &      2429 \\
NGC7793-HALO-6  &           23.868 &                           0.007 &                            0.051 &           23.029 &                           0.046 &                            0.032 &           22.101 &                           0.042 &                            0.038 &       866 \\
SCL-DE1         &           24.007 &                           0.028 &                            0.015 &           23.348 &                           0.034 &                            0.049 &           22.554 &                           0.030 &                            0.052 &       454 \\
SN-NGC2403-PR   &           23.416 &                           0.076 &                            0.067 &           22.457 &                           0.030 &                            0.100 &           21.489 &                           0.032 &                            0.079 &      1641 \\
UGC4305         &           23.569 &                           0.004 &                            0.034 &           22.803 &                           0.034 &                            0.050 &           21.954 &                           0.049 &                            0.049 &      4883 \\
UGC4459         &           23.708 &                           0.004 &                            0.034 &           22.985 &                           0.028 &                            0.044 &           22.162 &                           0.031 &                            0.048 &      1292 \\
UGC5139         &           23.893 &                           0.003 &                            0.021 &           23.133 &                           0.031 &                            0.047 &           22.296 &                           0.048 &                            0.050 &      2015 \\
UGC8508         &           23.018 &                           0.005 &                            0.027 &           22.315 &                           0.021 &                            0.032 &           21.503 &                           0.035 &                            0.034 &      1402 \\
UGCA292         &           23.750 &                           0.021 &                            0.028 &           23.168 &                           0.023 &                            0.044 &           22.411 &                           0.035 &                            0.048 &       177 \\
\hline
\end{tabular}
}

\label{tab:apparent}
\end{table*}

%% file: table_abs_d12.tex
\begin{table*}[!ht]
\caption{Absolute TRGB magnitudes from D12 distances}
\hskip-1cm
\resizebox{1\textwidth}{!}{
\begin{tabular}{lccccccc}
\hline
\hline
Target &   $\mu$ (D12) &  $M_{\rm F814W}$ &  $\sigma_{\rm F814W}$ &  $M_{\rm F110W}$ &  $\sigma_{\rm F110W}$ &  $M_{\rm F160W}$ &  $\sigma_{\rm F160W}$ \\
\hline
DDO71           &     $27.740$ &         $-3.998$ &               $0.016$ &         $-4.750$ &               $0.052$ &         $-5.606$ &               $0.062$ \\
DDO78           &     $27.820$ &         $-4.090$ &               $0.033$ &         $-4.854$ &               $0.065$ &         $-5.764$ &               $0.078$ \\
DDO82           &     $27.900$ &         $-4.036$ &               $0.044$ &         $-4.860$ &               $0.066$ &         $-5.777$ &               $0.079$ \\
ESO540-030      &     $27.610$ &         $-3.993$ &               $0.037$ &         $-4.670$ &               $0.046$ &         $-5.518$ &               $0.053$ \\
HS117           &     $27.910$ &         $-4.065$ &               $0.037$ &         $-4.756$ &               $0.045$ &         $-5.592$ &               $0.044$ \\
IC2574-SGS      &     $27.900$ &         $-4.025$ &               $0.040$ &         $-4.809$ &               $0.069$ &         $-5.689$ &               $0.076$ \\
KDG73           &     $28.030$ &         $-4.143$ &               $0.037$ &         $-4.772$ &               $0.051$ &         $-5.547$ &               $0.057$ \\
KKH37           &     $27.560$ &         $-4.018$ &               $0.032$ &         $-4.741$ &               $0.049$ &         $-5.601$ &               $0.056$ \\
M81-DEEP        &     $27.770$ &         $-3.696$ &               $0.104$ &         $-4.878$ &               $0.045$ &         $-5.878$ &               $0.043$ \\
NGC0300         &     $26.500$ &         $-4.007$ &               $0.046$ &         $-4.935$ &               $0.024$ &         $-5.898$ &               $0.031$ \\
NGC2403-HALO-6  &     $27.500$ &         $-4.160$ &               $0.034$ &         $-5.003$ &               $0.045$ &         $-5.907$ &               $0.046$ \\
NGC2976-DEEP    &     $27.760$ &         $-4.026$ &               $0.061$ &         $-4.903$ &               $0.047$ &         $-5.850$ &               $0.049$ \\
NGC3077-PHOENIX &     $27.920$ &         $-3.948$ &               $0.082$ &         $-4.930$ &               $0.047$ &         $-5.910$ &               $0.053$ \\
NGC3741         &     $27.550$ &         $-4.062$ &               $0.035$ &         $-4.755$ &               $0.043$ &         $-5.569$ &               $0.044$ \\
NGC4163         &     $27.290$ &         $-4.049$ &               $0.031$ &         $-4.782$ &               $0.055$ &         $-5.667$ &               $0.070$ \\
NGC7793-HALO-6  &     $27.960$ &         $-4.092$ &               $0.052$ &         $-4.931$ &               $0.056$ &         $-5.859$ &               $0.057$ \\
SCL-DE1         &     $28.110$ &         $-4.103$ &               $0.031$ &         $-4.762$ &               $0.060$ &         $-5.556$ &               $0.060$ \\
SN-NGC2403-PR   &     $27.500$ &         $-4.084$ &               $0.101$ &         $-5.043$ &               $0.105$ &         $-6.011$ &               $0.086$ \\
UGC4305         &     $27.650$ &         $-4.081$ &               $0.034$ &         $-4.847$ &               $0.060$ &         $-5.696$ &               $0.069$ \\
UGC4459         &     $27.790$ &         $-4.082$ &               $0.034$ &         $-4.805$ &               $0.052$ &         $-5.628$ &               $0.057$ \\
UGC5139         &     $27.950$ &         $-4.057$ &               $0.022$ &         $-4.817$ &               $0.056$ &         $-5.654$ &               $0.069$ \\
UGC8508         &     $27.060$ &         $-4.042$ &               $0.027$ &         $-4.745$ &               $0.039$ &         $-5.557$ &               $0.049$ \\
UGCA292         &     $27.790$ &         $-4.040$ &               $0.035$ &         $-4.622$ &               $0.049$ &         $-5.379$ &               $0.060$ \\
\hline
\end{tabular}
}

\label{tab:abs_d12}
\end{table*}

%% file: table_new_distances.tex
\begin{table*}[!ht]
\caption{New distance moduli}
\hskip-2cm 
\resizebox{1\textwidth}{!}{
\begin{tabular}{lcccccccccccc}
\hline
\hline
 & \multicolumn{4}{c}{PARSEC} & \multicolumn{4}{c}{MIST} \\
 & \multicolumn{2}{c}{F814W} & \multicolumn{2}{c}{F160W} & \multicolumn{2}{c}{F814W} & \multicolumn{2}{c}{F160W} \\
Target &  $\mu$ &  $\sigma$ &  $\mu$ &  $\sigma$  &  $\mu$ &  $\sigma$ &  $\mu$ &  $\sigma$ \\
\hline
DDO71           &                          27.842 &                              0.065 &                          27.899 &                              0.093 &                        27.710 &                            0.065 &                        27.747 &                            0.093 \\
DDO78           &                          27.822 &                              0.093 &                          27.939 &                              0.100 &                        27.700 &                            0.093 &                        27.785 &                            0.100 \\
DDO82           &                          27.948 &                              0.101 &                          28.026 &                              0.109 &                        27.835 &                            0.101 &                        27.864 &                            0.109 \\
ESO540-030      &                          27.719 &                              0.072 &                          27.839 &                              0.080 &                        27.566 &                            0.072 &                        27.694 &                            0.080 \\
HS117           &                          27.948 &                              0.072 &                          28.042 &                              0.075 &                        27.796 &                            0.072 &                        27.885 &                            0.075 \\
IC2574-SGS      &                          27.969 &                              0.094 &                          28.027 &                              0.115 &                        27.847 &                            0.094 &                        27.866 &                            0.115 \\
KDG73           &                          27.978 &                              0.072 &                          28.077 &                              0.090 &                        27.761 &                            0.072 &                        27.898 &                            0.090 \\
KKH37           &                          27.643 &                              0.072 &                          27.733 &                              0.087 &                        27.506 &                            0.072 &                        27.585 &                            0.087 \\
M81-DEEP        &                          28.043 &                              0.162 &                          28.068 &                              0.070 &                        27.938 &                            0.162 &                        27.788 &                            0.070 \\
NGC0300         &                          26.551 &                              0.079 &                          26.639 &                              0.047 &                        26.441 &                            0.079 &                        26.435 &                            0.047 \\
NGC2403-HALO-6  &                          27.423 &                              0.061 &                          27.456 &                              0.060 &                        27.311 &                            0.061 &                        27.306 &                            0.060 \\
NGC2976-DEEP    &                          27.804 &                              0.098 &                          27.893 &                              0.079 &                        27.696 &                            0.098 &                        27.699 &                            0.079 \\
NGC3077-PHOENIX &                          28.015 &                              0.124 &                          28.118 &                              0.083 &                        27.909 &                            0.124 &                        27.877 &                            0.083 \\
NGC3741         &                          27.590 &                              0.065 &                          27.659 &                              0.075 &                        27.429 &                            0.065 &                        27.501 &                            0.075 \\
NGC4163         &                          27.340 &                              0.079 &                          27.450 &                              0.091 &                        27.209 &                            0.079 &                        27.290 &                            0.091 \\
NGC7793-HALO-6  &                          27.948 &                              0.091 &                          28.026 &                              0.077 &                        27.838 &                            0.091 &                        27.858 &                            0.077 \\
SCL-DE1         &                          28.104 &                              0.064 &                          28.188 &                              0.094 &                        27.926 &                            0.064 &                        28.020 &                            0.094 \\
SN-NGC2403-PR   &                          27.467 &                              0.158 &                          27.546 &                              0.155 &                        27.358 &                            0.158 &                        27.334 &                            0.155 \\
UGC4305         &                          27.668 &                              0.084 &                          27.705 &                              0.099 &                        27.537 &                            0.084 &                        27.557 &                            0.099 \\
UGC4459         &                          27.810 &                              0.074 &                          27.857 &                              0.087 &                        27.664 &                            0.074 &                        27.695 &                            0.087 \\
UGC5139         &                          27.993 &                              0.075 &                          28.022 &                              0.099 &                        27.858 &                            0.075 &                        27.864 &                            0.099 \\
UGC8508         &                          27.119 &                              0.064 &                          27.178 &                              0.069 &                        26.964 &                            0.064 &                        27.015 &                            0.069 \\
UGCA292         &                          27.811 &                              0.071 &                          27.961 &                              0.091 &                        27.586 &                            0.071 &                        27.787 &                            0.091 \\
\hline
\end{tabular}
}
\tablecomments{Quoted errors are the quadrature sum of the photometric and fitting errors.}

\label{tab:abs_new}
\end{table*}

%% file: 5_discussion.tex
\section{Discussion} \label{sec:discussion}
In this section, we discuss the advantages and limitations of our adopted methods to trace the TRGB across multiple wavelengths.
Once established, we then discuss more fundamental limitations to our investigation, \added{which include knowledge of the absolute magnitude of the TRGB, details of the physical models underlying the isochrone suites, and possible systematics that are difficult to disentangle with the data at hand}. 

\subsection{Advantages of the MCR-TRGB method} \label{ssec:pros}
MCR-TRGB \added{simultaneously} measures the distributions of a pre-selected group of stars across an arbitrary number of color-magnitude combinations. 
As a result, \replaced{we ensure}{the method ensures} self-consistency in the measured color-magnitude behavior as it is determined from the same underlying set of stars. 
\added{In contrast}, traditional edge detection is done on a per-filter basis, and it is not guaranteed to detect the tip \deleted{at} using precisely the same stars across color-magnitude combinations. 
This \added{limitation} is particularly \replaced{true}{important} in the cases of steeply-sloped tips where the corresponding color-baseline changes significantly relative to the color-uncertainty.

Fitting the full color-magnitude covariance \deleted{matrix} \replaced{allows for characterization of the}{has the further benefit of characterizing} spread of colors and magnitudes both intrinsically, e.g., within a galaxy, and experimentally, e.g., with respect to our photometric precision. 
This differs from Sobel-based edge detection where it is difficult to properly account for measurement uncertainties in both magnitude and color.
Moreover, Sobel-based edge detection \replaced{has limitations}{does not automatically account} for intrinsic color-magnitude variation \added{across a given TRGB} and, as a result, the Sobel-edges are generalized to a mean color \replaced{and then use a 2-dimensional calibration}{on a per-filter basis}. 

Unlike the $T$-magnitude system of \citet{2009ApJ...690..389M}, which rectifies the photometry to an assumed TRGB slope, MCR-TRGB relies only on the assumption that there is one filter in which the tip magnitude has a weak enough color dependence enough to make an initial selection of tip star candidates.
This permits us to use the well-established Sobel method to find a ``flat" edge to define candidate tip stars and then utilize those stars in regimes where the tip is more difficult to detect using Sobel-based methods. 
This provides a fundamental advantage toward revealing the underlying intrinsic behavior of \replaced{the tip}{TRGB stars} to construct self-consistent color-magnitude calibrations.

\deleted{As we discuss in the following subsection, MCR-TRGB is perhaps non-ideal for generic distance measurement. However,} We posit that MCR-TRGB is a more effective tool to define the underlying color-magnitude calibrations for local, well-studied, and well-observed galaxies than relying on techniques more suited for distant galaxies. Stated differently, if your goal is to provide the best characterization of the behavior of tip stars across multiple bands, MCR-TRGB will perform better than standard Sobel-edge detection. It also provides a fully empirical basis to explore the ultimate precision of the TRGB as distance-measurement tool, from which we can gain understanding of both systematic and statistical biases for more distance measurements where there is a lower ability to probe these terms with available data.

\subsection{Limitations of MCR-TRGB} \label{ssec:cons}

Our method requires multi-wavelength data and relatively good data quality, which makes it less generally applicable to all distance measurement applications. 
More specifically, MCR-TRGB requires fairly stringent initial rejection of potential contaminants, which may not be feasible for all datasets due to photometric uncertainties or the complexity of the underlying stellar populations. 
Moreover, the \replaced{optical-to-NIR}{multiwavelength} tracing \added{of individual TRGB stars} may also be infeasible for many contexts where the acquisition of \replaced{optical or NIR}{multiband} imaging is too expensive. 
Thus, as just discussed, we \replaced{with to frame}{consider} MCR-TRGB\added{'s most significant role is} as a tool to define the underlying systematics affiliated with TRGB-based distances as the community explores different color-magnitude regimes.

\added{The} XDGMM \added{algorithm used in MCR-TRGB} loses its ability to resolve the shape of an intrinsic distribution \replaced{in the regime where}{when} the typical uncertainties on the input data points are comparable to the full range of the input data. 
This method \replaced{is therefore less advantageous}{loses its advantages} for low signal-to-noise photometry, filter combinations with very short color baselines, and simple stellar populations with little color spread near the TRGB. 
(See \autoref{sec:artdata_tests} for further discussion.)

\added{Another limitation of XDGMM is its assumption of Gaussianity.} \replaced{As previously mentioned}{In practice}, the intrinsic distribution of TRGB stars within a given magnitude range is \deleted{in reality} far from Gaussian. 
More complex models of this distribution should be investigated, ideally on high-precision photometry of systems with well-populated RGB sequences. 
For example, a multi-component Gaussian mixture could be of use in distinguishing remaining contaminants, such as a low-density AGB ``background", from the TRGB population. \added{Alternate fitting methods, such as Gaussian process regression, should also be considered.}
We reserve tests of this nature for future exploration \replaced{in}{using} local galaxies \added{with properties and observations} that are better matched \replaced{to performing this test}{to the requirements of drawing conclusions from such tests}.


\subsection{Absolute Calibration} \label{ssec:model_discussion}

\added{
In section~\ref{sec:recalib_d12dist}, Figures \ref{fig:newdistances_F814W} and \ref{fig:newdistances_F160W} introduce a puzzle with regards to using synthetic photometry as the absolute calibration for the TRGB. 
Inspection of Figures \ref{fig:newdistances_F814W} and \ref{fig:newdistances_F160W} reveal two concerns for the isochrone sets: 
 (i) a systematic offset between PARSEC and MIST is present for the absolute magnitudes regardless of filter, such that PARSEC is consistently brighter than MIST, and 
 (ii) there is a relative filter-to-filter offset between the models' predictions and our measurements, such that the absolute NIR magnitudes of our measurements derived by adopting the models' optical predictions do not correspond to the models' NIR predictions, and vice versa.
Understanding these differences requires considering how stellar interior models, which predict stellar structure, are mapped to stellar atmospheres used to construct the synthetic photometry shown in Figures \ref{fig:newdistances_F814W} and \ref{fig:newdistances_F160W}. 
An excellent discussion of this process is given by \citet{2014MNRAS.444..392C}.}

\added{Before examining the models more closely, we note that such offsets should not be surprising when viewed in the context of the larger literature.
Even in the well-studied F814W/$I$-band, there is a current debate in the value of the absolute magnitude of the TRGB, which is central to determining $H_{0}$ \citep[e.g., see][and references therein]{2019ApJ...882...34F,2019ApJ...886...61Y,2019ApJ...886L..27R,2020ApJ...891...57F}. 
Even the most detailed and careful calibrations have total uncertainties at the 0.05~mag level due to various systematic terms. The range of recently used F814W absolute values spans $\sim\!0.10$~mag \citep[as reviewed by][]{2018SSRv..214..113B}.
These discrepancies in the absolute magnitude of the tip can propagate into stellar models depending on exactly how the isochrone sets cross-check their own absolute scales.
As reviewed by \citet{2018SSRv..214..113B}, such discrepancies also affect the RR Lyrae and horizontal branches, which imparts uncertainty on the absolute scale for globular clusters \citep[as are explored in detail by][]{2014MNRAS.444..392C}.
Therefore, no theoretical predictions can be expected immune to the downstream effects of systematics in the empirical distance scale.} 

\added{
In this subsection, we first present a preliminary comparison of our results to the empirical TRGB relation derived for the (F606W--F814W)-$M_{\rm{F814W}}$ color-magnitude plane by \citet{2017ApJ...835...28J}, and discuss its implications with regard to assessing differences in the models' behavior.
We then consider two aspects of the synthetic photometry that might contribute to the discrepancies in Figures \ref{fig:newdistances_F814W} and \ref{fig:newdistances_F160W}:} (i) differences in adopted stellar atmospheres, which affect the conversion from bolometric luminosity to observed fluxes in specific filters, \added{discussed in section \ref{sssec:bcs}}, and (ii) differences in the underlying stellar evolution physics, \added{discussed in section \ref{sssec:physics}}. 
\deleted{The discrepancies we note in Figures \ref{fig:newdistances_F814W} and \ref{fig:newdistances_F160W} may be a manifestation of either or both of these differences.
}

\subsubsection{Comparison to empirical optical results} \label{sssec:jl}

\begin{figure*}[ht]
\plotone{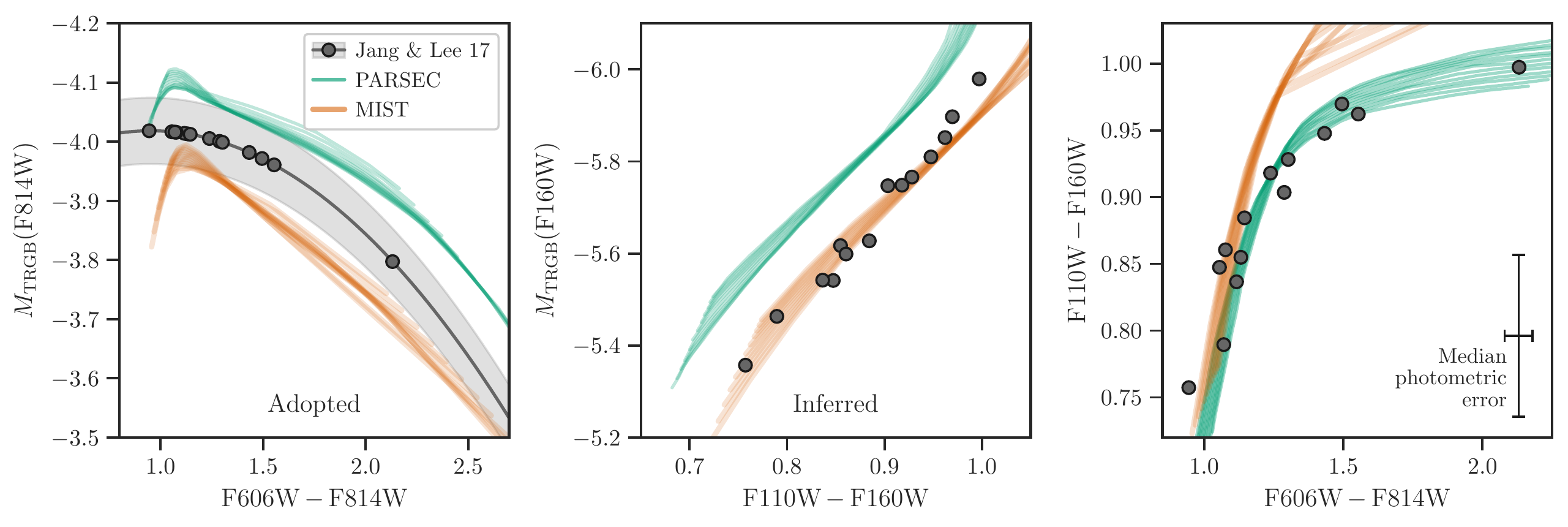}
\caption{As in \autoref{fig:newdistances_F814W}, we infer absolute F160W magnitudes via adopted F814W absolute magnitudes. Here the F814W absolute magnitudes are calibrated to the F814W vs.~F606W--F814W $QT$ relation presented by \citet{2017ApJ...835...28J} for the subset of our sample with F606W coverage.
We overplot MIST (orange) and PARSEC (green) synthetic photometry for comparison.
Left: The $QT$ relation with adopted $M_{\rm{F814W}}$ values.
Center: Inferred $M_{\rm{F160W}}$ vs.~ F110W--F160W.
Right: Distance-independent F110W--F160W vs.\ F606W--F814W color-color plot.}
\label{fig:jl_comparison}
\end{figure*}

\deleted{
Although fully empirical, model-independent calibrations of the F814W TRGB do exist (for example, \citet{2017ApJ...835...28J}, among others)
they are all tied to specific optical colors.
The inhomogeneity of our observations in optical filters other than F814W would require that we apply at least three different calibrations to bring all of our data onto an empirically-based absolute scale. 
This adds yet another layer of uncertainty when trying to understand the multi-wavelength behavior of the TRGB at precision.
Thus, our study on the absolute magnitude of the TRGB was fundamentally limited by the lack of model-independent TRGB distances. (However, we emphasize that color-color results are fully distance independent.)
}

\added{
The discrepancies between our measurements and both sets of synthetic photometry in Figures \ref{fig:newdistances_F814W} and \ref{fig:newdistances_F160W} make it unclear which model should be preferred, if either. As an alternative, we turn to the empirical F814W TRGB calibration presented by \citet{2017ApJ...835...28J}. Like the majority of existing empirical F814W/$I$ calibrations, it is based on a specific optical color baseline (F606W--F814W), which precludes us from adopting it for our entire sample, as F814W is the only optical filter common to all our targets. 
However, the subset of our sample with F606W observations (14 out of 23 targets) allows us to make a preliminary comparison as a benchmark against theoretical calibrations.\footnote{A more detailed analysis that incorporates recently revised distances to the two absolute-scale zeropoint anchors used by \citet{2017ApJ...835...28J} is currently in preparation.}
}

\added{\citet{2017ApJ...835...28J} employ a quadratic functional form (the $QT$ system) for the $M_{\rm{F814W}}$ vs.\ (F606W--F814W) relation, which they calibrate over an extensive color range of $0.8 < {\rm F606W}-{\rm F814W} < 3$ mag.
We show the results of adopting the $QT$ relation for the galaxies in our sample with F606W coverage in \autoref{fig:jl_comparison}.}

\added{\autoref{fig:jl_comparison} shows that the \citeauthor{2017ApJ...835...28J} calibration falls squarely between the PARSEC and MIST predictions in F814W.
Adopting the distances from this calibration, we then plot the inferred absolute magnitude in the F160W band in the center panel.
The resulting F160W TRGB absolute magnitudes agree quite well with the MIST predictions in the NIR.
We explore this apparent color mismatch further in the right panel, which compares the optical F606W--F814W colors to the NIR F110W--F160W colors for the 14 galaxies with F606W data.
Although the center panel of \autoref{fig:jl_comparison} appears to favor MIST predictions in the NIR at all but the reddest colors, the optical-IR color-color behavior strongly favors PARSEC.}

\subsubsection{Bolometric corrections} \label{sssec:bcs}

\begin{figure*}[ht]
\plotone{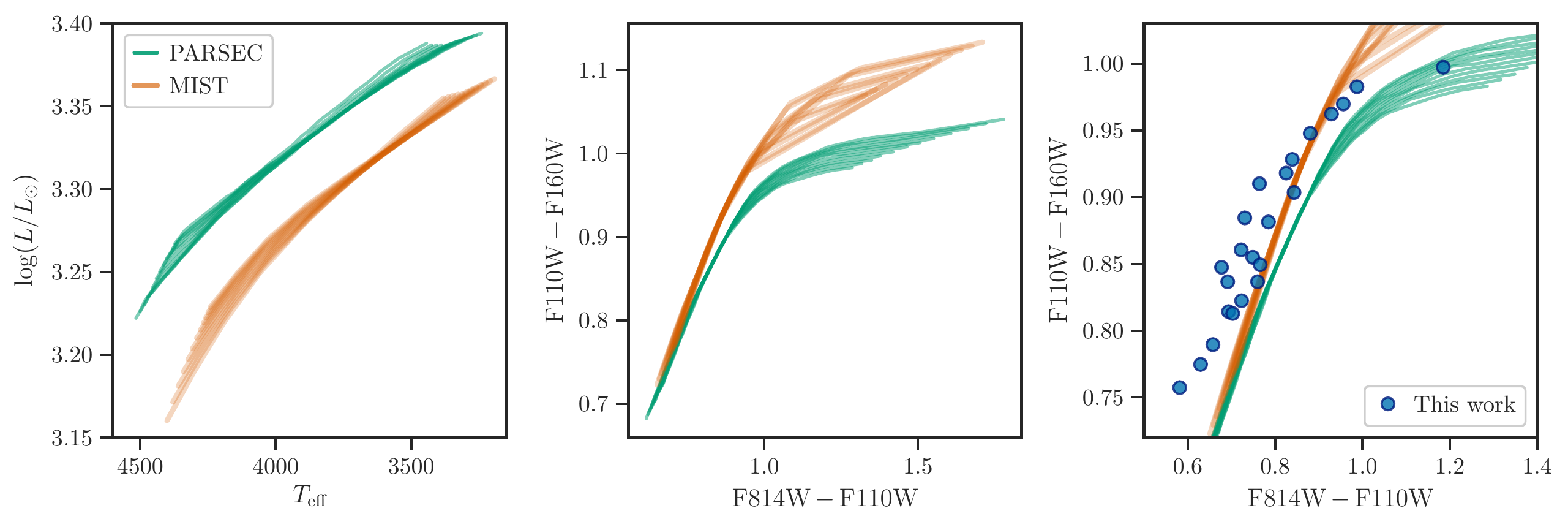}
\caption{Left: Bolometric luminosity vs.\ effective temperature for the MIST (orange) and PARSEC (green) model tip stars used in this work (ages 8~-~14 Gyr and $-2 \leq \rm{[Fe/H]} \leq -0.25$ dex). PARSEC luminosities are found to be consistently $\sim$1.1 times brighter than MIST at the same age and metallicity. Center: (F110W--F160W) vs.\ (F814W--F110W) color-color plots for MIST and PARSEC tip star predictions.
Right: Color-color results of this work overlaid on MIST and PARSEC predictions.}
\label{fig:models_are_inconsistent}
\end{figure*}

\added{We make a direct comparison of the models' physical predictions in the left panel of} \autoref{fig:models_are_inconsistent}, which compares the MIST and PARSEC model TRGB in temperature-luminosity space \deleted{(left) and F110W--F160W vs.\ F814W--F110W colors (center)} \added{for the same range of ages and metallicities as in Figures \ref{fig:newdistances_F814W} and \ref{fig:newdistances_F160W}}. 
\deleted{We use the internal evolution tags in each model set to identify the tip stars.}
We see that the models are offset from each other in $T_{\rm eff}$ and $\log(L/L_{\odot})$, indicating differences in the underlying stellar structure, \replaced{with PARSEC stars having $\sim$10\% higher luminosities at the same $T_{\rm eff}$}{with PARSEC running approximately $\sim$10\% more luminous and 50-150 K warmer than MIST at the same age and metallicity}. 
(For comparison, \citet{2018ApJ...860..131C} find uncertainties on the absolute $T_{\rm eff}$ scale of $\pm100$~K due to boundary conditions.)
The PARSEC predictions also show a slightly larger spread in $T_{\rm eff}$ than MIST over the same range of age and metallicity. 

\added{We attempt to isolate filter-to-filter differences between the models' predictions by examining their color-color behavior. The middle panel of \autoref{fig:models_are_inconsistent} shows} the F814W--F110W to F110W--F160W color-color behavior of the two model sets \deleted{is shown}.
There is a divergence on the order of 0.1~mag for stars with F814W--F110W \textgreater{} 1 mag, indicating significant differences in the model atmospheres of cooler stars. Close inspection of the bluer side shows that although the color-color relations have similar slopes, they are slightly offset from one another at the same metallicity.

\added{We compare these color-color predictions to our measurements in the right panel of \autoref{fig:models_are_inconsistent}, which overlays the color-color results derived via the MCR-TRGB technique. 
Unlike Figures \ref{fig:newdistances_F814W} and \ref{fig:newdistances_F160W}, the color-color relations are distance-independent, which alleviates concerns about the absolute scale of the measurements.
The observed TRGB is systematically redder than predicted in F110W-F160W and/or systematically bluer in F814W--F110W (which appears to effectively rule out unaccounted-for extinction as a source of disagreement).
Comparison to the right panel of \autoref{fig:jl_comparison}, which shows good agreement between observations and PARSEC predictions for F606W--F814W and F110W--F160W colors, suggests that PARSEC's predicted TRGB colors are overall accurate in the optical and IR independently, but that there may be offsets in the relative cross-calibration between the two wavelength regimes in either the stellar atmosphere models or in our data.
More specifically, if PARSEC's predicted F110W and F160W absolute magnitudes were shifted to be $\sim\!0.1$ mag dimmer (or if our NIR measurements were 0.1 mag brighter) with no changes to the optical, the predicted and observed F814W--F110W colors would be brought into alignment, with no change to the F110W--F160W or F606W--F814W colors. 
}

\replaced{Atmospheric differences are manifested in the}{The stellar atmospheres used to generate synthetic photometry are encoded as} bolometric corrections, which transform bolometric luminosities into filter-specific quantities. 
The bolometric corrections depend \added{primarily} on $\log({\rm g})$, $T_{\rm eff}$, and \replaced{[M/H]}{both [Fe/H] and [$\alpha$/H]}. \deleted{even if both PARSEC and MIST used the same set of corrections,} \added{While the differences in predicted bolometric luminosities shown in the left panel of} \autoref{fig:models_are_inconsistent} suggest \replaced{they}{that PARSEC and MIST} would still \replaced{have}{predict} different absolute magnitudes for a tip star of a given age and metallicity, \added{bolometric corrections and the stellar $T_{\rm{eff}}$ scale are likely to be a source of some of the filter-to-filter offsets we observe}.

\added{As of this writing, the PARSEC web service (CMD v.~3.3) uses {\tt PHOENIX} \citep{2012EAS....57....3A} bolometric corrections for stars with $T_{\rm{eff}} < 5500$~K ($\sim\!1000$~K warmer than the warmest TRGB stars), and MIST uses {\tt{ATLAS12}} \citep{2014dapb.book...39K}.
\citet{2019A&A...632A.105C} have explored how the {\tt PHOENIX} \citep{2012EAS....57....3A} and {\tt ATLAS9} \citep{2014dapb.book...39K} model atmospheres affect predicted colors and magnitudes of PARSEC isochrones in optical and IR passbands.
Their Figure 2 shows that the {\tt PHOENIX} bolometric corrections produce RGB colors that are biased red by up to 0.1 mag in $V-I$, which translates to an artificially bright TRGB in the NIR consistent with what we see in Figures~\ref{fig:newdistances_F814W} and \ref{fig:newdistances_F160W}.
Although \citet{2019A&A...632A.105C} claim that the {\tt PHOENIX} bolometric corrections are preferable for giants because they are computed with spherical geometry, \citet[][section 3.2.3]{2018MNRAS.476..496F} find that the {\tt PHOENIX} bolometric corrections cannot reproduce the observed RGB colors in 47~Tuc, which they term the ``RGB-too-red" problem.
While we have been unable to locate any similar such studies of MIST's predictions for the RGB, \citet{2018MNRAS.476..496F} caution that {\tt{ATLAS12}} atmospheres may be unreliable for $T_{\rm{eff}} < 4000$~K, which may explain the divergence we see between the MIST and PARSEC predictions at F814W--F110W \textgreater{} 1 in the center panel of \autoref{fig:models_are_inconsistent}.
}

\replaced{Similar to what we see in \autoref{fig:newdistances_F814W} and \autoref{fig:newdistances_F160W},}{Similar evidence for the importance of bolometric corrections for the optical TRGB was explored by}
\citet{2017A&A...606A..33S}, \added{who} directly compared the predicted absolute magnitudes of the $I$-band TRGB from the BaSTI models \citep{2013A&A...558A..46P} using four sets of bolometric corrections (see their fig.~8). 
They see differences at the $\sim$10\% level \deleted{comparable to what we see here} when applying different sets of bolometric corrections to models using the same underlying physics, \added{comparable to the amplitude of discrepancies we see here}.
\added{Although this investigation focused on the optical, their finding of $\sim\!0.1$ mag discrepancies aligns with the scale of the adjustments needed to bring our measurements and the models' predictions into alignment.}

\added{While further quantitative investigation is clearly required, we conclude that bolometric corrections are a likely source of a substantial part of the optical-IR discrepancies we observe in Figures~\ref{fig:newdistances_F814W} and \ref{fig:newdistances_F160W}.
}

\subsubsection{Physical properties of TRGB stars} \label{sssec:physics}

\added{The left-most panel of} \autoref{fig:models_are_inconsistent} suggests that there are currently real differences in the predicted stellar structure at the TRGB due to the different physical assumptions between the models, even before differences in atmospheres or bolometric corrections are included.
Our comparison of PARSEC and MIST \added{broadly} agrees with the conclusions from a more detailed model-focused study by \citet{2017A&A...606A..33S}, who investigated both the physical and computational factors \replaced{that contributed}{contributing} to differences between the \added{predicted} TRGB luminosities of two sets of stellar models (BaSTI and GARSTEC).
\added{\citet{2017A&A...606A..33S} were able to produce identical predictions of tip stars' physical properties from two different model suites only when certain physical processes, such as neutrino energy loss and electron screening, as well as some numerical criteria such as integration timestep, were implemented consistently between stellar evolution codes (the full set of which they term ``concordance physics").
While a comparable investigation of such sources of difference between MIST and PARSEC is well outside the scope of this work, we find it reasonable to conclude that some aspects of their physical differences are likely due to limitations in our current understanding of certain ``cutting edge'' topics in stellar astrophysics, and may also be in part due to differing computational approaches.
Tip stars, in addition to being both cool and luminous, are at an evolutionary transition point, and so may be especially sensitive to these details.}

\added{Another possible source of disagreement between the models' predictions and our data is elemental abundances, including the helium fraction $Y$ and $\alpha$ enhancement, the latter of which is of particular concern at [Fe/H] $\lesssim -1$ dex (F110W--F160W $\lesssim 0.9$~mag for PARSEC).
At present, neither MIST nor PARSEC have publicly available $\alpha$-enhanced models, although they are slated to be included in future releases of both \citep{2016ApJ...823..102C, 2018MNRAS.476..496F}.
\citet{2017A&A...606A..33S} report that, at least after the adoption of their concordance physics, $\alpha$ enhancement produces what they consider to be negligible effects on the TRGB bolometric luminosity ($< 1$\% difference between [$\alpha$/Fe] = 0.4 and [$\alpha$/Fe] = 0 at fixed [M/H]), $T_{\rm{eff}}$ ($< 2$\% difference), and predicted $V\!I\!J\!K$ magnitudes ($< 0.01$~mag difference at constant color).
Similarly, they find that a change in the helium fraction $Y$ of 0.01 (approximately the range over which estimates of the primordial helium mass fraction vary) has no more than a 1\% effect on TRGB temperatures and luminosities.
Nonetheless, as forthcoming versions of PARSEC will offer options for variable $\alpha$ enhancement and helium abundance \citep{2018MNRAS.476..496F} as well as updated bolometric corrections\footnote{The {\tt PHOENIX} bolometric corrections currently employed in the PARSEC web service only take total $Z$ into account in their color transformations \citep{2018MNRAS.476..496F}, which is problematic for understanding potential photometric effects of varying abundance ratios. {\tt ATLAS12} model atmospheres for $\alpha$-enhanced PARSEC isochrones down to $T_{\rm{eff}} = 4000$~K are currently in development \citep{2018MNRAS.476..496F, 2019A&A...632A.105C}.}, we anticipate that a direct analysis of these quantities' impacts on predictions of TRGB behavior as they pertain to this work will be both easily achievable and informative.
}


\subsection{TP-AGB contamination}

While we believe that our method of determining $P({\rm RGB})+$ is overall effective at rejecting contaminating populations, such as red supergiants and the bulk of the AGB, there may be some amount of remaining contamination, particularly from thermally-pulsing AGB (TP-AGB) stars, which we briefly discuss here.

Although TP-AGB stars are generally intrinsically brighter than the TRGB, extinction from circumstellar dust can substantially impact their observed magnitudes and bring them closer to luminosities typical of the upper RGB. However, as they are also heavily reddened, their colors are inconsistent with RGB stars \citep[][see their fig.~8]{2017ApJ...851..152B}, so it is likely that our method of TRGB candidate selection successfully rejected most, if not all, of these stars.

TiO absorption is another factor that may bring certain TP-AGB stars, particularly M-type stars at high metallicity, closer in luminosity to the TRGB \citep{2019ApJ...879..109B}. While the bulk of our targets are low-metallicity dwarf galaxies, this may be an issue for some of the larger galaxies in our sample, such as M81.

Finally, TP-AGB stars may cross the TRGB when they reach the minimum point in their pulsation cycle. In this case, the fact that our NIR data were taken several years later than our optical data is an advantage; TP-AGB stars at their minimum in our optical observations are unlikely to be at their minimum in the NIR, and vice versa.

\subsection{Limited Empirical Constraints on TRGB Magnitude Stability}

We briefly discuss other physical concerns that may affect the colors and magnitudes of TRGB stars, including binarity, low amplitude pulsational variability, mass loss, \added{and planetary engulfment}.

First, the presence of a companion star may affect the intrinsic photometric properties of a TRGB star through mass transfer in a binary interaction. Preliminary investigations suggest that the photometric effects of the former phenomenon are overall secondary to the variation of TRGB magnitude with metallicity \citep[][private communication]{eldridge19}. 
Thus, we qualitatively conclude that binarity is likely to be a source of some amount of residual scatter in our measurements rather than a primary driver of TRGB variation across populations.

Second, as recent high-precision, high-cadence photometry has demonstrated, low-amplitude variability exists for many to most stellar types.
Pulsational variability was first proposed for stars on the upper RGB by \citet{2002MNRAS.337L..31I} and has since been observationally confirmed \citep{2004MNRAS.347..720I, 2005A&A...441.1117L, 2015MNRAS.448.3829W}. 
Variability may contribute some amount of uncertainty to TRGB measurements by effectively blurring the TRGB edge. However, there are few established constraints on relevant characteristics, such as typical periods, amplitudes, fractions of stars that exhibit variability, and dependence on stellar properties such as age and metallicity.
We have thus disregarded TRGB variability as a potential systematic in this work due to lack of empirical constraints. We expect that any overall effects are small compared to our dominant sources of uncertainty.

\deleted{Finally,} RGB stars are known to experience mass loss driven by chromospheric activity \citep{2007ApJ...667L..85O, 2012A&A...540A..32G, 2014ApJ...789...28P}, which may be amplified by either of the first two properties.
\added{\citet{2020arXiv200311499J} predict that variations in the mass loss parameter $\eta$ at the TRGB may affect individual stars' luminosities by over 5\%, although they estimate that the net effect on measured TRGB distances does not exceed 2\%, and that it is strongly metallicity-dependent.}
Additionally, although mass loss has been correlated with the blueshifting of optical and near-IR spectral lines such as H$\alpha$ and the calcium triplet \citep{2007A&A...476.1261M, 2015MNRAS.448.3829W}, the impact of this blueshifting on broadband photometry has not been quantified.

\added{\citet{2020arXiv200311499J} also consider planetary engulfment, wherein an RGB star consumes one or more planets in close orbit as it expands. They predict that the increased turbulence in the star's convective envelope, corresponding to an increase in mixing length, may result in a net decrease in TRGB luminosity by up to 5\% for a single star that has consumed a giant planet. However, they conclude that both detailed hydrodynamical simulations and further studies of planetary system formation are required to accurately constrain potential impacts of this phenomenon on the TRGB as a distance indicator.}

Again, we expect that these effects are overall \replaced{negligible for the purposes}{well within the uncertainties} of this work, but \replaced{should}{may need to} be taken under consideration in future high-precision TRGB studies.

%% file: 6_conclusions.tex
\section{Conclusions and Future Work} \label{sec:conclusions}

\subsection{Conclusions}

We have developed a method to measure TRGB magnitudes and colors in multiple filters simultaneously.
This method, MCR-TRGB, was designed to use a set of likely RGB stars, which were defined where traditional TRGB-detection methods using edge-detection can be employed reliably, to study the multi-wavelength behavior of the TRGB using those same stars.
We applied MCR-TRGB to a re-reduction of optical+NIR \HST{} data originally presented in \citetalias{2012ApJS..198....6D}; these new reductions use the optical observations, which have higher spatial resolution and are generally more complete at the TRGB, to produce more complete and precise photometry in the infrared-bands.
When using the same distances as \citetalias{2012ApJS..198....6D}, we find only minor adjustments to the color-magnitude behavior of the IR-TRGB.
However, the \citetalias{2012ApJS..198....6D} absolute magnitudes were determined relative to color-magnitude predictions from stellar models. 
Thus, we compared three different absolute-magnitude calibrations of the measured TRGB magnitudes, one using the same distance moduli as in \citetalias{2012ApJS..198....6D}, and two using distance moduli derived from the predicted TRGB absolute magnitudes from two commonly used isochrone sets (PARSEC and MIST). 
We find that the isochrone-based absolute calibrations are inconsistent with each other at the $\sim\!0.1$ mag level, consistent with previous work in this domain, and that both sets of isochrones are internally inconsistent with our measurements of the TRGB magnitudes and colors when \deleted{both} optical and infrared measurements are \added{used} together.
We further caution that adoption of model-based absolute calibrations for the TRGB, a conservative ~10\% systematic included in the calibrations \added{for the TRGB} based on differences between the isochrone sets \added{at the TRGB}. 
\added{We find that these tensions persist even with the application of a state-of-the-art empirical calibration. From examining the distance-independent color-color behavior of our data against model predictions, we conclude that bolometric corrections and the underlying stellar $T_{\rm{eff}}$ scale are likely to explain a large part of the inconsistences we have found.}

\subsection{Future work}

An empirical absolute TRGB calibration in WFC3/IR bandpasses remains elusive. 
A fully model-independent calibration, as in \citet{2017ApJ...835...28J}, is clearly necessary.
However, there are a limited number of systems that are distant enough that their apparent TRGB magnitudes are easily observable with \HST{}, but nearby enough to have distances that are well-constrained by other means.

Another limitation of our study is the lack of precise independent distances to the galaxies in this work.
Indeed, the majority of these systems {\it only} have distances determined from the TRGB itself. 
Some of these galaxies are within a volume for variable-star based distances with \emph{HST}, though we are cautious about their precision given the metallicity dependence of such relations and the difficulty of inferring stellar metallicities for galaxies at these distances \citep[for RR Lyrae see][]{2018SSRv..214..113B}.

Our initial goal in this work was to relate empirical results on the multi-wavelength TRGB to the physical characteristics of the underlying stellar populations, such as age and metallicity.
We found that goal challenging due to the internal mismatches we observe in the \replaced{models}{isochrone sets}, given that the aforementioned physical parameters are ultimately inferred via comparison to those from \replaced{stellar models}{isochrone sets} once a distance, also often \replaced{model}{isochrone-}dependent, is assumed.

%% file: app_tests.tex
\section{Tests on artificial data} \label{sec:artdata_tests}

\twocolumngrid %

In this section, we diagnose potential biases and systematics induced by our technique by applying the above methods to simulated CMDs with known theoretical TRGB magnitudes. We use the results of this analysis to determine bias corrections to our final tip magnitudes and to refine the uncertainties on our measurements. \\

We first generate a set of idealized (i.e., error-free) photometry of artificial RGB sequences with MATCH based on the PARSEC model suite.
\citet{2017A&A...606A..33S} demonstrated that metallicity is the primary driver of variation in TRGB colors and magnitudes for old ages, so we hold all parameters except metallicity constant.
We use a Chabrier IMF with a slope of 1.3, a binary fraction of 0.3, and a constant star formation rate with an age range of 100 Myr to 14 Gyr.
We vary metallicity between $-2.0$ \textless{} [Fe/H] \textless{} $-0.5$ dex with a spacing of 0.1 dex, which covers a color range representative of our data.
We constrain the output magnitudes to $\sim$1 mag brighter than the TRGB to $\sim$2.5 mag dimmer for each filter: $-5.0$ \textless{} F814W \textless{} $-1.5$ mag, $-6.0$ \textless{} F110W \textless{} $-3.0$ mag, and $-7.0$ \textless{} F160W \textless{} $-3.5$ mag.

We choose to model a constant SFR rather than a single-age population for two reasons. First, most of our data \citepalias[see star-formation histories in][]{2012ApJS..198....6D} show evidence for stellar sequences from young populations, and as such, mono-age populations are not realistic representations of our data. Second, for a mono-age population, the full color or magnitude range of the TRGB is small enough to be comparable to the typical errors on the data points. In this limit, XDGMM cannot resolve the underlying distribution of TRGB color and magnitude, and thus its performance cannot be tested.

We add fiducial photometric errors to the simulated stellar population using the aggregate of all of our AST results, which were determined for each galaxy in our dataset in \autoref{ssec:ASTs}.
We do not incorporate a net photometric bias or photometric incompleteness, both of which are negligible at the TRGB for the majority of our sample.

For all tests, we randomly subsample the artificial data to reach a desired total number of RGB stars less than 1 magnitude fainter than the fiducial-TRGB ($N_\star^{T+1}$) in the filter used to measure the Sobel edge. 
We also adjust all magnitudes by a value randomly generated from a Gaussian with a mean of 0 mag and standard deviation set to the star's photometric uncertainty.

\begin{figure*}[ht]
\epsscale{0.9}
\plotone{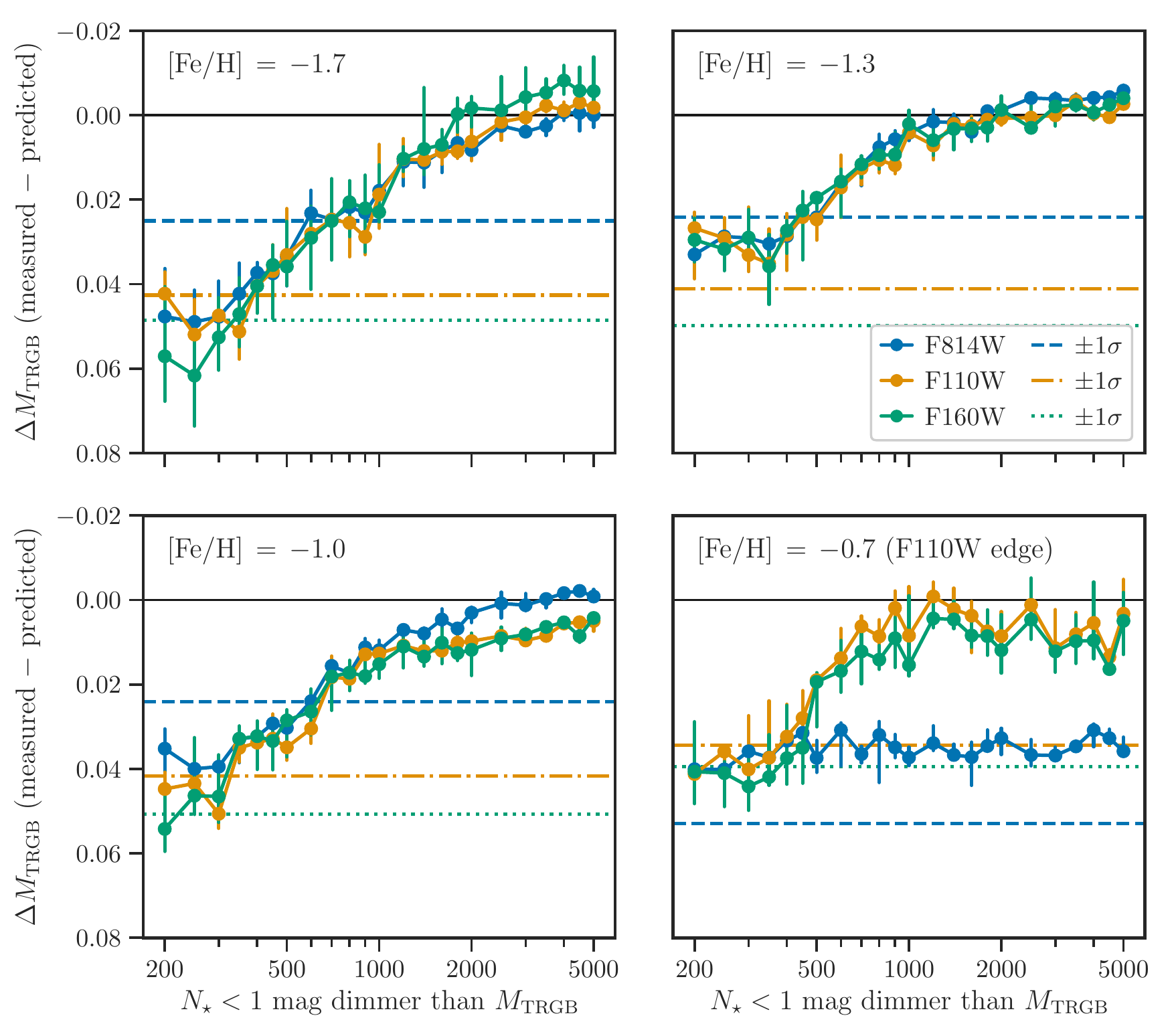}
\caption{Differences between measured and theoretical TRGB magnitude versus\ $N_\star^{T+1}$ (the number of stars within 1 mag fainter of the TRGB), for [Fe/H] = $-1.7$ dex (top left), $-1.3$ dex (top right), $-1.0$ dex (bottom left), and $-0.7$ dex (bottom right). The error bars on the points show the interquartile range of the results. The various dashed horizontal lines show the mean per-filter uncertainties on the tip fitting results, which we define as the quadrature sum of the XDGMM fitting uncertainty and the mean photometric uncertainty of the tip stars.
The bias in the XDGMM edge measurement is typtically within the 1-$\sigma$ TRGB uncertainty, with the exception of poorly populated CMDs ($N_\star^{T+1}$\textless$\sim$500 stars). }
\label{fig:offsets_nstars}
\end{figure*}

\subsection{Luminosity function sampling} \label{ssec:nstars}

Here we test our method against $N_\star^{T+1}$ over a range of $200 \leq N_\star^{T+1} \leq 5000$ stars, which spans the $N_\star^{T+1}$ values for the majority of the galaxies in our sample. At each $N_\star^{T+1}$ we run 20 end-to-end XDGMM tip fitting iterations and calculate the offsets $\Delta M_{\rm TRGB} \equiv M_{\rm TRGB}$(measured) -- $M_{\rm TRGB}$(true) for each filter. We repeat these tests at four different metallicities ([Fe/H = $\{-1.7, -1.3, -1.0, -0.7\}$ dex) to check for possible color-dependent effects. 

\autoref{fig:offsets_nstars} shows the median per-filter differences in the measured versus theoretical TRGB values against $N_\star^{T+1}$ for each of the four metallicities. 
The error bars on the points in \autoref{fig:offsets_nstars} show the interquartile range of the results.
The dashed horizontal lines are color-coded to match the filter and show the range of the per-filter mean uncertainty on the tip-fitting results; we define the uncertainty as the quadrature sum of the XDGMM fitting uncertainty and the median photometric error of the tip stars.

For all but [Fe/H] = $-0.7$ dex, the results are largely consistent: the offsets $\Delta M_{\rm TRGB}$ start out around 0.04 mag in the most undersampled case, increase approximately with $N_\star^{T+1}$ for $N_\star^{T+1} \lesssim 1000$ stars in all filters, and then begin to level off near $\Delta M_{\rm TRGB} \sim 0.01$ mag. These results are broadly similar to what is seen for traditional edge detection methods. \citet{2009ApJ...690..389M} found that Sobel edge detection is prone to bias when the RGB luminosity function is undersampled, and that a sample of $N_\star^{T+1} \gtrsim 500$ stars is required for edge detection to function accurately.

For [Fe/H] = $-0.7$ dex, where the edge detection and initial tip star selection are done in F110W rather than F814W, we see behavior similar to the lower metallicities in the NIR, but not in F814W, which hovers at 0.04 mag throughout. This foreshadows a possible selection effect when using F110W for the primary edge detection, which we investigate further in \autoref{ssec:metallicity}.

\subsection{Metallicity} \label{ssec:metallicity}

\begin{figure}[ht]
\epsscale{1.1}
\plotone{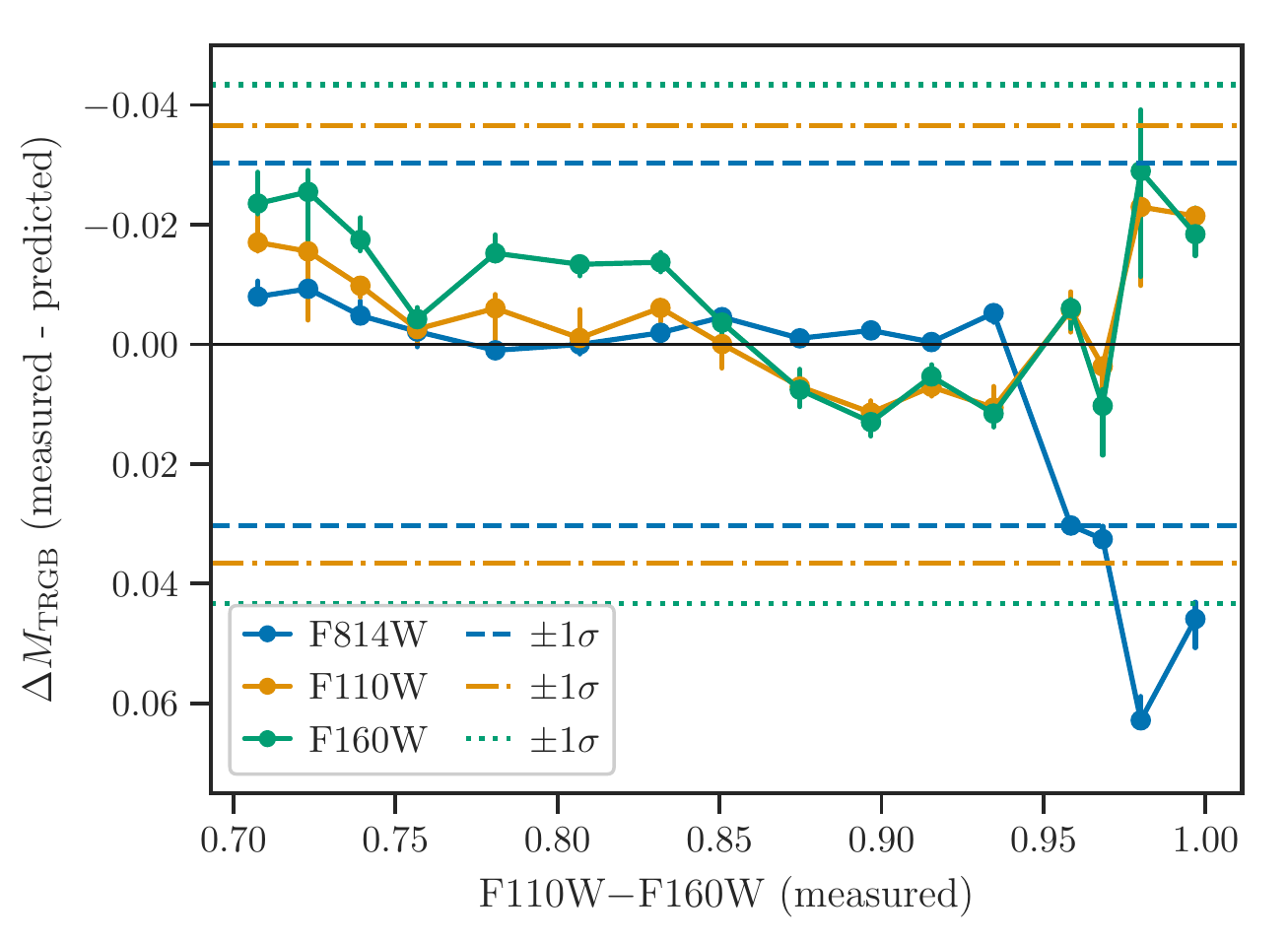}
\caption{The difference between measured and theoretical TRGB values against the median measured NIR tip color for the ensemble of simulated datasets. 
Results for F814W, F110W, and F160W are shown in blue, orange, and green, respectively. 
The horizontal lines show $\pm1\sigma$, where $\sigma$ is the mean quadrature sum of the photometric and fitting errors for each filter
This bias is within the TRGB detection uncertainty in almost all cases.}
\label{fig:offsets_feh_ircolor}
\end{figure}

For each metallicity in our artificial dataset, we run 20 end-to-end tip fitting iterations with $N_\star^{T+1} = 4000$ stars, in the regime where sampling effects are minimal.
\autoref{fig:offsets_feh_ircolor} shows the median per-filter differences in the measured versus the theoretical TRGB values against the measured IR-TRGB color; the IR-TRGB color increases approximately monotonically with metallicity in the artificial data.

The jump in offset values at F110W--F160W \textgreater{} 0.95 mag corresponds to the switch from using F814W to using F110W for the Sobel edge detection. We hypothesize that this jump is due to a difference in which stars are selected as candidate tip stars.
At high metallicity, stars that have similar magnitudes in F110W may have a large range of magnitudes in F814W due to the increasingly steep TRGB-color slope in F814W.
The stars that are the brightest in F814W are relatively dim in F110W and so may not fall within the F110W tip star selection window and, thus, our measured F814W tip magnitudes are skewed faint relative to the predicted values.

\subsection{Photometric uncertainties} \label{ssec:photerr}

\begin{figure}[ht]
\epsscale{1.1}
\plotone{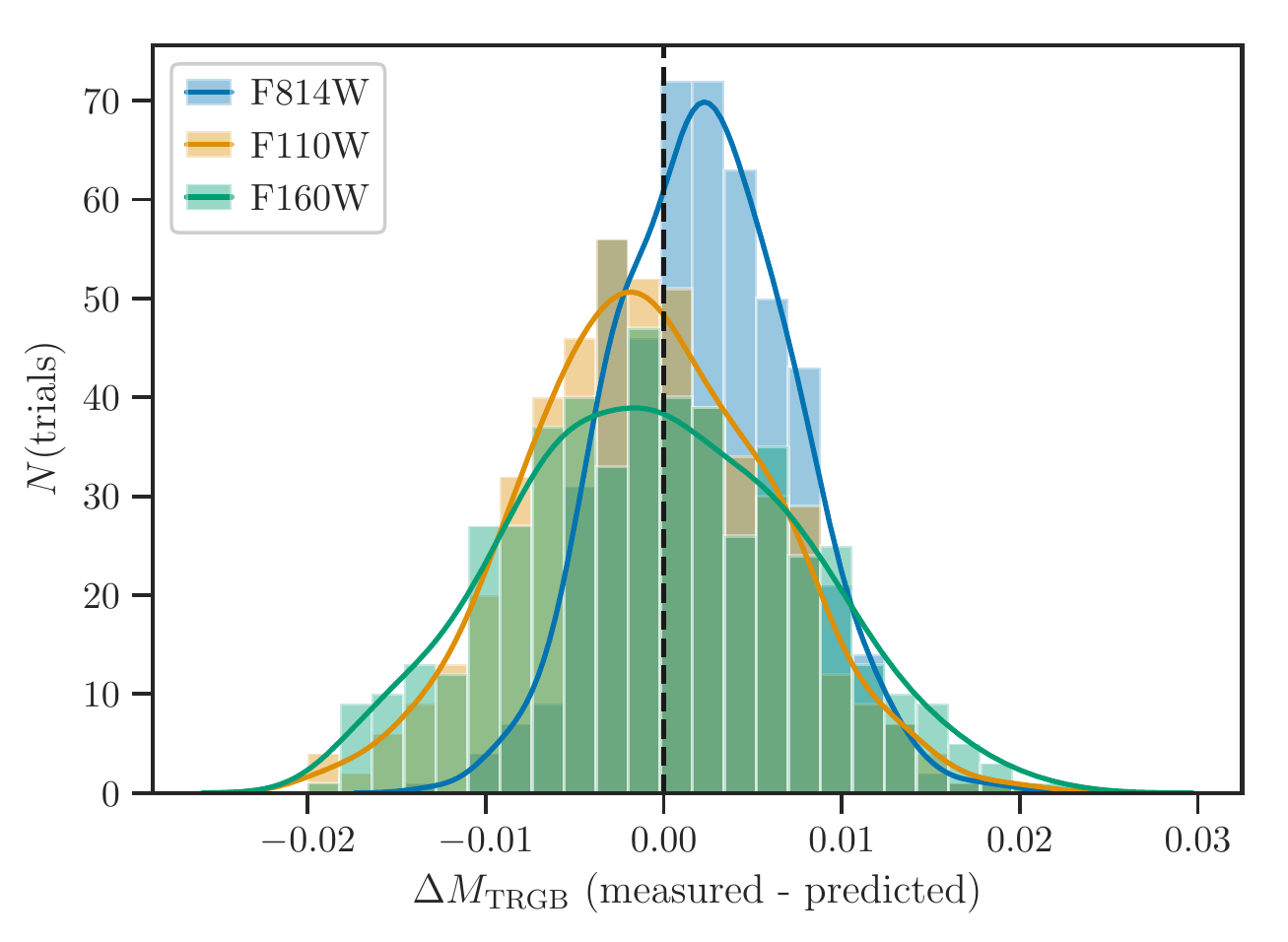}
\caption{Histograms of the difference between the measured and predicted $M_{\rm TRGB}$ for the with $N_\star^{T+1} = 2000$ stars and [Fe/H] = $-1.0$ dex artifical dataset for 500 trials, where each trial modifies the stellar magnitude randomly in proportion to its photometric error. The blue histogram is for F814W, the orange is for F110W, and the green is for F160W.
As expected, the widths of all histograms are smaller than the reported photometric uncertainties at the tip.}
\label{fig:offsets_photerr}
\end{figure}

To isolate the impact of photometric uncertainties on our TRGB-measurements, we use the artifical dataset with $N_\star^{T+1} = 2000$ stars and [Fe/H] = $-1.0$ dex. 
For each of 500 trials, we vary the input magnitudes by a different random value drawn from a Gaussian whose standard deviation is each star's photometric uncertainty.
The results of this test are shown in \autoref{fig:offsets_photerr}. 
We find that for all filters the standard deviation of the resulting distribution of offsets is roughly a third of the typical photometric uncertainty at the tip. 

\subsection{XDGMM vs.\ Sobel edge detection}

\begin{figure}[ht]
\plotone{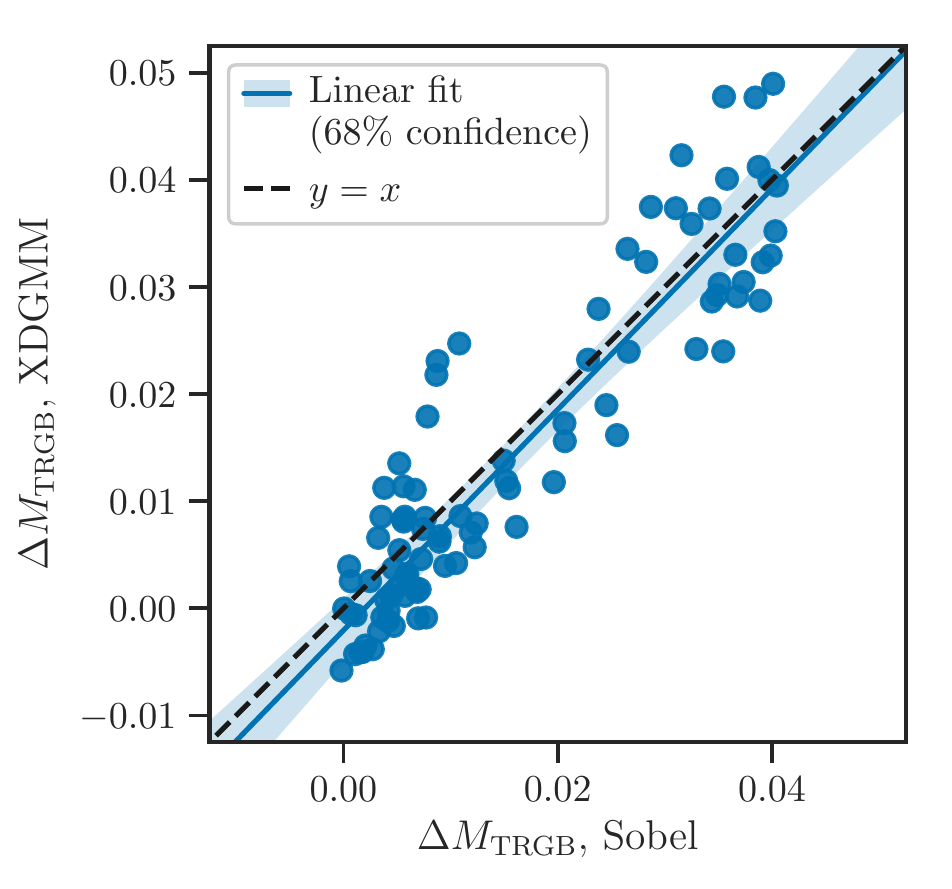}
\plotone{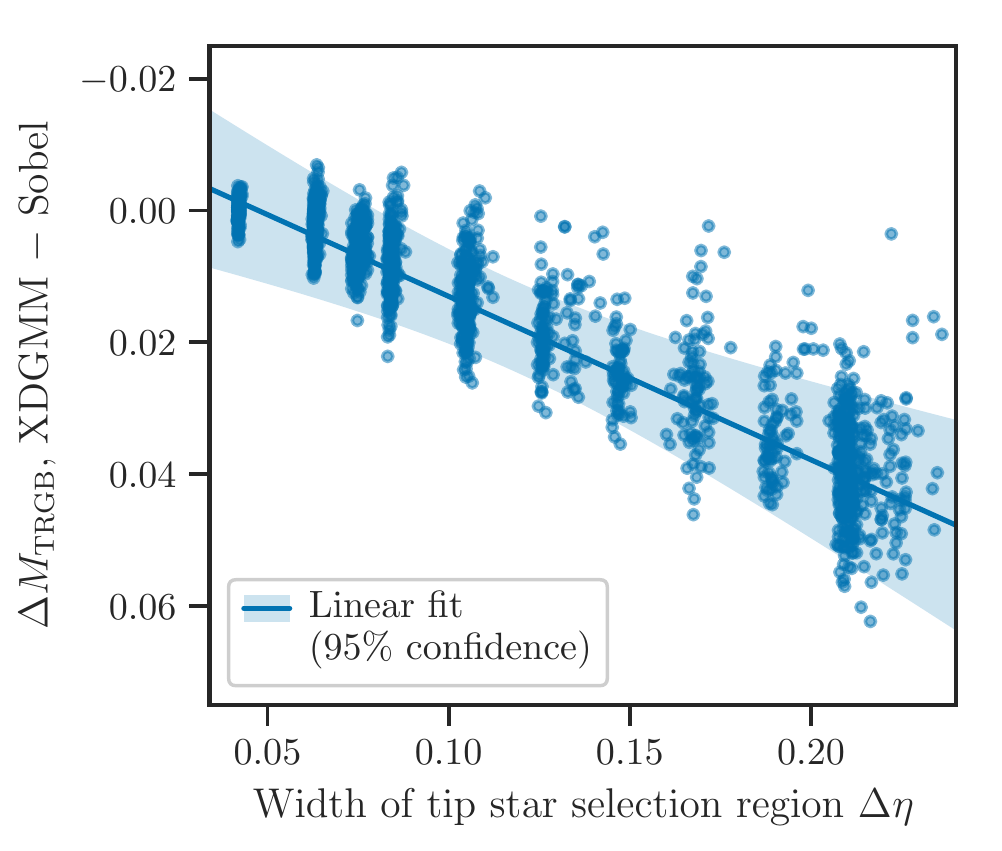}
\caption{Top: The median difference between XDGMM-fitted and theoretical tip magnitudes versus the median difference between the Sobel edge magnitude and theoretical tip magnitude in either F814W or F110W. Each point is the median result for one set of trials with fixed metallicity and $N_\star^{T+1}$. The solid line and flanking filled region show a linear fit to the data and its 68\% confidence interval, whereas the dashed line shows a one-to-one relation.
Bottom: Difference between the XDGMM-fitted mean and Sobel edge magnitude versus $\Delta \eta$, which is the width in magnitudes of the tip star selection region in the luminosity function. Each point represents the offset for a single trial with fixed metallicity and $N_\star^{T+1}$. The solid line and flanking filled region show a linear fit to the data and its a 95\% confidence interval, respectively.}
\label{fig:dsobel_dxdgmm}
\end{figure}

Here we investigate the behavior of XDGMM tip fitting relative to the standard method of Sobel edge detection using the same set of trials as in \autoref{ssec:nstars}.
As our method for XDGMM tip fitting itself uses Sobel edge detection to set the color-magnitude center of the initial tip star selection window, we can make a fully self-consistent comparison of the Sobel edge magnitudes to the XDGMM mean magnitudes for detections in F814W and F110W. (We do not perform edge detection on F160W in our method and so do not make the comparison.)

The top panel of \autoref{fig:dsobel_dxdgmm} shows the relation between the median difference between the XDGMM-fitted mean and the theoretical tip versus the median difference between the Sobel edge and theoretical tip.
The linear fit to the data is consistent within $1\sigma$ with a one-to-one relation, indicating that the methods produce overall consistent tip mangitudes.

The bottom panel of \autoref{fig:dsobel_dxdgmm} shows the relation between the width of the tip star selection region, $\Delta \eta$, and the difference between the XDGMM- and Sobel-derived tip magnitudes, $\Delta M_{\rm TRGB}$. The quantities are correlated, albeit with some scatter on the order of 0.01 mag, and are fit by the linear relation $\Delta M_{\rm TRGB}^{\rm X-S} = 0.25(\Delta \eta) - 0.01$ mag. 

\subsection{Adjustments to measurements}

In the previous subsections, a number of tests were performed to quantify the statistical and systematic uncertainties of the MCR method using artificial photometry. 
Here we match our observed galaxies to their artificial tests to determine both systematic terms that are applied in the form of bias corrections and statistical terms that are applied in the form of inflating the algorithmic uncertainties. 
The corrections are parameterized by two key observables: 
(i) how well populated the RGB is  as a proxy for the total mass, and
(ii) the F814W--F160W color as a proxy for the underlying stellar population properties. 
All such adjustments are summarized in \autoref{tab:bias} and if a given target does not appear in the table, then it did not require a modification.

\input{table_adjust.tex}

For each target, we determine the most appropriate sets of tests to use to determine the bias based on $N_\star^{T+1}$ and F814W--F160W color. The quoted adjustment values are adopted from the relevant set of trials, with uncertainties determined as the median and interquartile range of the offsets (measured -- predicted value) in each filter. We subtract the offsets from the measured TRGB apparent magnitude and add the associated uncertainty in quadrature to the fitting uncertainty. We also modify all relevant colors based on these adjustments.

The first four targets in \autoref{tab:bias} (KDG73, NGC2403-HALO-6, SCL-DE1, and UGCA292) all have $N_\star^{T+1} < 500$ stars.
(Although these targets do not all have the same colors, we found that differences between offsets were negligible at the relevant colors.)
For these we take the median and interquartile range of offsets for all trials with $N_\star^{T+1} < 500$ stars and ${\rm [Fe/H]} \leq -1.0$ dex.

The remaining targets (NGC0300, NGC2976-DEEP, NGC3077-PHOENIX, SN-NGC2403-PR, and M81-DEEP) use F110W as the edge detection filter. All but M81-DEEP have colors ${\rm F814W}\!-\!{\rm F160W} \sim 2$ mag, whereas M81-DEEP has ${\rm F814W}\!-\!{\rm F160W} \sim 2.25$ mag. We match these colors by adopting the median and interquartile range of offsets for trials with $-0.8 \leq {\rm [Fe/H]} \leq -0.7$ dex for all but M81-DEEP, and $-0.6 \leq {\rm [Fe/H]} \leq -0.5$ dex for M81-DEEP.

%% file: table_adjust.tex
\begin{table}[!ht]
\centering
\footnotesize
\caption{Applied bias corrections}

\begin{tabular}{>{\raggedright}p{2cm}rrr}
\hline
\hline
 & \multicolumn{3}{c}{Value($\pm$ error) subtracted from $m_{\rm TRGB}$} \\[-1em]
Target name     & \multicolumn{1}{c}{F814W} & \multicolumn{1}{c}{$\ $F110W} & \multicolumn{1}{c}{$\ $F160W} \\
\hline
KDG73           &  $0.04\pm0.02$ &  $0.04\pm0.02$ &  $0.04\pm0.02$ \\
NGC2403-HALO-6  &  $0.04\pm0.02$ &  $0.04\pm0.02$ &  $0.04\pm0.02$ \\
SCL-DE1         &  $0.04\pm0.02$ &  $0.04\pm0.02$ &  $0.04\pm0.02$ \\
UGCA292         &  $0.04\pm0.02$ &  $0.04\pm0.02$ &  $0.04\pm0.02$ \\
NGC0300         &  $0.03\pm0.01$ &                   &                   \\
NGC2976-DEEP    &  $0.03\pm0.01$ &                   &                   \\
NGC3077-PHOENIX &  $0.03\pm0.01$ &                   &                   \\
SN-NGC2403-PR   &  $0.03\pm0.01$ &                   &                   \\
M81-DEEP        &  $0.06\pm0.01$ &   $-0.02\pm0.02$ &   $-0.02\pm0.02$ \\
\hline
\end{tabular}
\label{tab:bias}
\end{table}